\documentstyle[11pt,aaspp,psfig]{article}
\def\gs{\mathrel{\raise0.35ex\hbox{$\scriptstyle >$}\kern-0.6em \lower0.40ex\hbox{{$\scriptstyle \sim$}}}}
\def\ls{\mathrel{\raise0.35ex\hbox{$\scriptstyle <$}\kern-0.6em \lower0.40ex\hbox{{$\scriptstyle \sim$}}}}

\begin{document}
\small

\title{A photometric and spectroscopic study of dwarf and giant galaxies 
in the Coma cluster --\\ III. Spectral ages and metallicities 
\footnote{Based on observations made with the William Herschel Telescope 
operated on the island of La Palma by the Isaac Newton Group in the 
Spanish Observatorio del Roque de los Muchachos of the Instituto de 
Astrofisica de Canarias.}}
\author{
Bianca M.\ Poggianti,$^{\!}$\altaffilmark{2}
Terry J.\ Bridges,$^{\!}$\altaffilmark{3}
Bahram Mobasher,$^{\!}$\altaffilmark{4}
Dave Carter,$^{\!}$\altaffilmark{5}
M. Doi,$^{\!}$\altaffilmark{6} 
M. Iye,$^{\!}$\altaffilmark{7} 
N. Kashikawa,$^{\!}$\altaffilmark{7} 
Y. Komiyama,$^{\!}$\altaffilmark{8}  
S. Okamura,$^{\!}$\altaffilmark{9,11} 
M. Sekiguchi,$^{\!}$\altaffilmark{10}
K. Shimasaku,$^{\!}$\altaffilmark{9}  
M. Yagi,$^{\!}$\altaffilmark{7}
N. Yasuda$^{\!}$\altaffilmark{7}
}
\smallskip

\affil{\scriptsize 2) Osservatorio Astronomico di Padova, vicolo dell'Osservatorio 5, 35122 Padova, Italy}
\affil{\scriptsize 3) Anglo-Australian Observatory, PO Box 296, Epping, NSW 1710, Australia}
\affil{\scriptsize 4) Space Telescope Science Institute, 3700 San Martin Drive, Baltimore, MD 21218, USA \\ Affiliated with the Space Sciences Department of the European Space Agency}
\affil{\scriptsize 5) Liverpool John Moores University, Astrophysics Research Institute, Twelve Quays House, Egerton Wharf, Birkenhead, Wirral, CH41 1LD, UK}
\affil{\scriptsize 6) Institute of Astronomy, School of Science, University of Tokyo, Mitaka, 181-0015, Japan}
\affil{\scriptsize 7) National Astronomical Observatory, Mitaka, Tokyo, 181-8588 Japan}
\affil{\scriptsize 8) Subaru Telescope, 650 North Aohoku Place, Hilo, HI 96720, USA}
\affil{\scriptsize 9) Department of Astronomy, University of Tokyo,
Bunkyo-ku, Tokyo 113-0033, Japan}
\affil{\scriptsize 10) Institute for Cosmic Ray Research, University of Tokyo,
Kashiwa, Chiba 277-8582, Japan}
\affil{\scriptsize 11) Research Center for the Early Universe, School of Science, University of Tokyo, Tokyo 113-0033, Japan}

\begin{abstract}
We present a detailed analysis of the spectroscopic catalog of galaxies
in the Coma cluster from Mobasher et al. (2001, Paper II of the series).
This catalog comprises $\sim$300 spectra of cluster members with absolute
magnitudes in the range $M_B=-20.5$ to $-14$ in two areas of $\sim 1\times 1.5$
Mpc towards the center and the South-West region of the cluster.
In order to study the star formation and metallicity properties of the
Coma galaxies as a function of their luminosity and environment,
spectral indices of the Lick/IDS system and equivalent widths of
the emission lines were measured in the range $\lambda=3600-6600$
\AA. 

In this paper the analysis is restricted to the 257 
galaxies with no emission lines in their spectra.  
The strength of the age-sensitive indices (such as $\rm
H\beta$, $\rm H{\gamma}_F$ and $\rm H\delta_F)$ is found to correlate
with galaxy magnitude over the whole magnitude range explored in
this study. Similarly, the metallicity-sensitive indices (such as $\rm
Mg_2$, $<$Fe$>$, $C_24668$) anticorrelate with magnitude.  By
comparing the observed indices with model grids based on the Padova
isochrones, we derive luminosity-weighted ages and metallicities.  We
present the distributions of ages and metallicities for galaxies in
various magnitude bins.  The mean metallicity decreases with 
galaxy magnitude and, at a given luminosity, appears to be generally
lower for galaxies in the South-West region of Coma as compared to the
center of the cluster.  A broad range of ages, from younger than 3 Gyr
to older than 9 Gyr, is found in galaxies of any magnitude. However,
systematic trends of age with luminosity are present among galaxies in
the central field, including a slight decrease of the mean age
for fainter galaxies.  Furthermore, in the central Mpc
of Coma, a large fraction of galaxies at {\it any} luminosity 
($50-60$\% of the giants, $>30$\% of the dwarfs) show no evidence 
in their central regions of star formation occurred at redshift $z<2$,
while the proportion of galaxies with significant star formation occurring 
at intermediate ($0.35<z<2$) and low ($z<0.35$) redshifts is found to
depend on galaxy luminosity. 

An additional surprising result is that the faint galaxies with young
luminosity-weighted ages appear 
to have a bimodal metallicity distribution that,
if confirmed, would point to a composite formation scenario involving
different physical processes. Coadding the spectra of these metal-rich and 
metal-poor galaxies separately supports the reality of the metallicity 
bimodality, although higher signal-to-noise spectra of the individual galaxies
will be needed to draw definite conclusions.

An anticorrelation between age and
metallicity is found to be present in galaxies of any given luminosity
bin and it is especially evident among the brightest subset with the
highest signal-to-noise spectra.

Finally, we present an interpretation of the index-magnitude relations
observed.  We show that the slopes of the indices/magnitude relations 
are the consequence of both age and metallicity trends with luminosity:
each such trend on its own would be sufficient to produce
relations similar to those observed.

\end{abstract}


\sluginfo
\newpage

\section{Introduction}

Dwarf galaxies are the most numerous type of galaxy in the local
Universe. In hierarchical models, they represent the ``building
blocks'' of more massive galaxies, and are expected to be the sites of
the earliest cosmic star formation (White \& Frenk 1991).  Their
observed numbers and properties at various redshifts are a strong test
of any theory of galaxy formation and evolution.

Furthermore, dwarfs are expected to be the type of galaxies that can
best teach us the influence of both internal and external conditions
on galaxy evolution.  Their structures and evolutionary histories are
predicted to be strongly affected by mass loss due to supernova-driven
outflows (Larson 1974, Dekel \& Silk 1986, Silk, Wyse \& Shields
1987), and their star formation is expected to be modulated also by
the external UV radiation field (Babul \& Rees 1992, Efstathiou 1992).

Dwarf galaxies are influenced by environment even more than
giant galaxies are. Dwarfs show the strongest morphology-density
relation: early-type dwarfs are found in clusters or clustered around
giant galaxies, while late-type dwarfs are the least clustered type of
galaxy.  In the Local Group, this morphological segregation could be
induced by the repeated action of tidal forces from the closest giant
galaxy (``tidal stirring'', Mayer et al. 2000).  In clusters like
Coma, physical processes peculiar to dense environments such as
interactions with a hot intracluster medium via pressure confinement,
shocks or ram pressure stripping (Silk, Wyse \& Shields 1987, Babul \&
Rees 1992, Murakami \& Babul 1999) and high-speed, repeated encounters
(``harassment'', Moore et al. 1996, 1998) are expected to have the
most dramatic effects on the dwarfs.  Many of the suggested processes
imply a morphological evolution from one dwarf type to another
(early-type to late-type dwarfs and vice-versa).

Surprisingly little is known about the stellar populations of 
non-starforming dwarf galaxies.  The two environments where such low
luminosity galaxies have been identified in significant numbers are
the Local Group and other groups and clusters of galaxies.  By far the
best studied environment in this respect is the Local Group, where
non-starforming dwarfs show evidence of varied and complex histories,
sometimes with multiple episodes of star formation (Mateo 1998, Grebel
1999). In fact, they vary widely in the star formation rate, the
length and the epoch of star formation episodes and the chemical
enrichment {\it even within the same morphological class}: ``no two
dwarf galaxies in the LG have the same star formation history''
(Grebel 1999).  Regardless of this variety of evolutionary histories,
the Local Group dwarf galaxies follow global relations between
absolute magnitude, mean metallicity and central surface brightness
(Caldwell 1999, Grebel \& Guhathakurta 1999).  Interestingly, dwarf
galaxy evolution in the Local Group appears to be determined both by
mass and by environmental effects (i.e. proximity to a massive spiral)
(Grebel 2000).

Spectroscopic studies of non-starforming dwarf galaxies outside the
Local Group have been limited so far to quite small samples, thus it
is still largely unknown whether low-luminosity cluster galaxies share
the diversity of star formation histories of dwarfs in the Local
Group, and whether they follow a similar metallicity-luminosity
relation.  A metallicity-luminosity relation valid for both bright and
dwarf ellipticals (the latter residing mostly in the Virgo and 
Fornax clusters) over the range $M_B=-14$ to $-22$ was presented by
Brodie \& Huchra (1991), who also noted that the scatter in this 
relation implies that additional parameters are involved.
Interestingly, ten dwarf ellipticals in the Fornax cluster were found
to cover a wide range of metallicities \it at a given luminosity \rm
(Held \& Mould 1994).

Broadband photometry first provided evidence for young or intermediate
age populations in some dwarf ellipticals (Caldwell 1983, Thuan 1985,
Caldwell \& Bothun 1987), and the poor correlation between the UBV
colors and the structural parameters was interpreted by Peterson \&
Caldwell (1993) as evidence for variations in the ages of dwarf
ellipticals.  Early spectroscopic studies of bright dwarf
ellipticals/spheroidals supported the photometric indications for
young or intermediate-age stellar populations in some of the dwarfs
(Zinnecker et al. 1985, Bothun \& Mould 1988, Gregg 1991, Held \&
Mould 1994).  Recently, six spheroidals in Virgo ($M_B=-17/-18$) have
been studied by Gorgas et al. (1997), who analyzed their Lick indices and
found four of them to be old and quite metal-poor ([Fe/H]$\sim -0.75$)
and two of them to be young and metal-rich.  Recent star formation in
a fraction of the early-type galaxies has been detected in Coma and
other clusters (Caldwell et al. 1993, Caldwell \& Rose 1997, Caldwell
\& Rose 1998), most of these galaxies being fainter than $M_B=-18.5$.

There are now several pieces of evidence indicating that the spread in
the average stellar ages of {\it low-luminosity} early-type galaxies in all
environments is larger than the age spread of the {\it luminous} early-type
galaxies (Bender, Burstein \& Faber 1993; Bressan et al. 1996; Worthey
1997; Worthey 1998; Kuntschner \& Davies 1998; Mehlert et al. 1998;
Vazdekis \& Arimoto 1999; Smail et al. 2001).  In the Coma cluster,
the scatter in age and metallicity has been found to increase with
decreasing mass/luminosity (Jorgensen 1999).  Concannon, Rose \&
Caldwell (2000) analyzed the index-velocity dispersion relation for
$\rm Mg_2$, $\rm H\beta$ and a Rose index of field and cluster E/S0
galaxies with a large magnitude range ($M_B=-22$ to -16) and found for
all three indices a relation whose scatter is greater for lower mass
galaxies. Their interpretation of this result is that less massive
galaxies have experienced a more varied SF history and the spread in
age increases towards lower velocity dispersions.

We are still far from a full understanding of the meaning of these
findings and their implications for theories of galaxy formation and
evolution.  The scarcity of information and the lack of systematic
studies regarding the evolutionary histories of dwarf galaxies
external to the Local group has hindered a detailed comparison with
the giant galaxies, whose ages and metallicities have been
investigated by a large number of studies.  Tight relations between
the strength of the Mg-absorption and the central velocity dispersion
(or luminosity) has long been known to exist for luminous ealy-type
galaxies (Faber 1973; Terlevich et al. 1981; Dressler 1984; Dressler
et al. 1987; Burstein et al.  1988; Guzman et al. 1992; Bender,
Burstein \& Faber 1993; Jorgensen, Franx \& Kj\ae rgaard 1996; Bender,
Ziegler \& Bruzual 1996; Ziegler \& Bender 1997; Bender et al. 1998;
Trager et al. 1998; Colless et al. 1999; Kuntschner et
al. 2001).  Correlations of other metal line indices with velocity
dispersion/luminosity have generally been found to be weak and to have a 
large scatter (Fisher, Franx \& Illingworth 1996; Jorgensen 1997,
1999; Trager et al. 1998; Terlevich et al. 1999).  Recently,
strong, significant correlations of metal indices other than magnesium
with velocity dispersion were measured in a sample of early-type
galaxies in the Fornax cluster (Kuntschner 2000; Kuntschner et
al. 2001).  Negative correlations of Balmer indices with the velocity
dispersion or luminosity have also been reported by a number of
authors (Gonz\'alez 1993; Jorgensen 1997; Trager et al. 1998; Terlevich
et al. 1999; Kuntschner 2000).  Shell and pair galaxies are known to
deviate from the $\rm H\beta-\sigma$ (both shell and pair galaxies)
and the $\rm Mg_2-\sigma$ (only shell galaxies) relations, most likely
due to secondary bursts of star formation (Longhetti et al. 2000).

The relations observed in luminous early-type galaxies have generally been
interpreted to be largely a sequence of metallicity versus
galaxy mass/luminosity (Forbes et al. 1998; Kobayashi \& Arimoto 1999;
Terlevich et al. 1999; Kuntschner 2000; Kuntschner et al. 2001),
although an issue still debated is whether the observed trends are
entirely due to a metallicity sequence or to a combination of both age
and metallicity effects (e.g. Jorgensen 1999, Trager et al. 2000b).
The most controversial issue in these studies are the star formation
histories of the early-type galaxies.  Ellipticals in the Fornax cluster
are mostly coeval and span a range in metallicity (Kuntschner \&
Davies 1998; Kuntschner 2000).  Similarly, a new sample of early-type
galaxies in cluster and group environments consists mostly of galaxies
with large ages spanning a range in metallicity (Kuntschner et
al. 2001).  In contrast, intermediate-age/young populations have been
found in a large fraction of non-cluster luminous ellipticals (Rose
1985; Gonz\'alez 1993; Forbes, Ponman \& Brown 1998).  The Gonz\'alez
dataset has been analyzed by a number of authors, all agreeing on the
existence of a large age spread in this sample (Faber et al. 1995;
Bressan et al. 1996; Tantalo et al. 1998; Kuntschner \& Davies 1998;
Trager et al. 2000a), except Maraston \& Thomas (2000) who argue that
the $\rm H\beta$ index of the majority of galaxies in the Gonz\'alez
sample (excluding the Balmer-strong lenticulars, intermediate-mass
field ellipticals and dwarf ellipticals) can be explained by a spread
in the metallicity of old stellar populations.

The different age ranges found in different samples of ellipticals
might be due to systematic effects as a function of environment,
with ellipticals in dense environments being older than those in
groups and in the field (Bower et al. 1990; Guzman et al. 1992; de
Carvalho \& Djorgovski 1992; Rose et al. 1994; Jorgensen 1997;
Kauffmann \& Charlot 1998; Kuntschner \& Davies 1998; Mobasher \&
James 2000; Trager et al. 2000b), although a large age spread has also
been found by Jorgensen (1999) in Coma E and S0 galaxies spanning a 5
magnitude range.  On the other hand, the luminosity-dependence of the
galactic properties has not been clarified yet. As will be
demonstrated by the present work, it is only meaningful to compare
galaxies of {\it similar luminosity/mass}. At least part of the
differences observed in the various datasets could be due to
differences in the luminosity range explored.

In addition, it is important to ask to what extent the star
formation history of a non-starforming galaxy is related to its Hubble
type.  In the Fornax cluster, luminous ellipticals are coeval (around
8 Gyr), while (less luminous) S0 galaxies display a significant age
spread (Kuntschner \& Davies 1998; Kuntschner 2000).  In apparent
constrast, Jorgensen (1999) finds no difference between the stellar
population properties of bright E's and S0's in Coma, and a similar
result is obtained for clusters at z=0.3-0.5 by Jones, Smail \& Couch
(2000). These results could be simultaneously interpreted if the 
dependence of galaxy age on morphology is relevant only for non-luminous
galaxies, as highlighted by the work of Smail et al. (2001) who find
that ellipticals of all magnitudes and luminous S0s in the cluster
A2218 at z=0.17 are coeval and trace a sequence of varying
metallicity, while faint S0 galaxies span a large range of
luminosity-weighted ages.

Obviously a key issue is understanding the relative roles of galaxy
mass (=luminosity, assuming an approximately constant
stellar mass-to-light ratio as for old populations), environmental conditions, 
and galaxy morphology, in determining
the evolutionary history of galaxies. This project aims at
investigating how many and which parameters govern the evolution of
low- and high-luminosity cluster galaxies, by obtaining
photometric and spectroscopic information for galaxies in the Coma
cluster.  For this purpose, the most valuable aspects of the present
survey are the large number of spectra, the broad magnitude range
covered, the diversity of environmental conditions surveyed
and the wide spectral coverage. This is the largest
spectroscopic survey of cluster dwarfs currently available and it is
complemented by a similarly large dataset for brighter galaxies up to
$M_B\sim -20.5$.  Given the paucity of spectroscopic studies of
low-luminosity cluster galaxies, in a sense we are venturing into an
unexplored realm.  The survey extends over more than 6 magnitudes, 
therefore providing an excellent opportunity to study the
spectroscopic properties as a function of galaxy luminosity in a
homogeneous dataset.  Two wide areas of $\sim 1$ $\times$ $1.5$ Mpc each
(toward the center and in the South-West region of the Coma cluster)
were chosen, to encompass very different environmental conditions in
terms of local galaxy density, distance from the cluster center, and
membership in a substructure. The wavelength range of the spectra
($\sim 3600-6600$ \AA) includes numerous absorption and emission
features, typically from [O{\sc ii}]$\lambda$3727 to $\rm H\alpha$,
hence it offers the chance to use several age and metallicity
indicators.

The main goal of this paper (Paper III of the series) is to study the
stellar population properties, deriving luminosity-weighted ages and
metallicities from the observed spectral indices.  After a brief
description of the spectroscopic catalog (\S2) and of the models
employed (\S3), we show the relations between the strength of the
spectral indices and galaxy luminosity (\S4.1), we derive
luminosity-weighted ages and metallicities from index-index diagrams
(\S4.2 and 4.3) and we present the age and metallicity distributions as a
function of galaxy magnitude, deriving the proportion of actively
starforming galaxies at various redshifts (\S4.3).  The link between
age and metallicity is discussed in \S4.4 and \S4.5, and in \S4.6 we
show how the observed index-magnitude relations are a
consequence of the systematic variations of age and metallicity with
luminosity. A summary of the main results can be found in \S5.

\section{Sample and spectral analysis}

We have carried out a spectroscopic survey of faint and bright
galaxies in two fields (each of size $32.5 \times 50.8 \, \rm
arcmin^2$) in the direction of the center of the Coma cluster (Coma1),
and in the south-west region (Coma3). The coordinates and extent of
these fields are presented in Mobasher et al. (2001, Paper II).  These
areas were imaged in the B and R bands with the Japanese mosaic CCD camera on
the William Herschel Telescope (Komiyama et al. 2001, Paper I).  A full
description of the selection of the spectroscopic targets,
observations and data reduction can be found in Mobasher et al. (Paper
II).  Here we only summarize the most important characteristics of the
sample and the data.

The spectra were taken with the WYFFOS multi-fibre spectrograph at the
William Herschel Telescope. They are centered on 5100 \AA $\,$ and
extend over more than $3000$ \AA $\,$ with a resolution $6-9$ \AA $\,$
FWHM (see Appendix A, dispersion $\sim 3 \, \rm \AA/pixel$).  
Each fibre has a diameter of 2.7 arcsec, thus
the spectra are sampling the central 1.3 kpc of each galaxy ($H_0=65
\, \rm km \, s^{-1} \, Mpc^{-1}$). For comparison, typical exponential
scalelengths of dwarf galaxies are in the range 0.8--1.3 kpc for
$M_B=-15$ to $-17$ (Ferguson \& Binggeli 1994).

In this paper we restrict our analysis to galaxies with velocities in
the range $4000<v<10000 \, \rm km \, s^{-1}$ (roughly corresponding to
$3\sigma$ cuts), which are considered members of the Coma cluster.
With this membership criterion, our spectroscopic catalog comprises a
total of 278 galaxies, of which 189 are in Coma1 and 89 in Coma3.
The R magnitude distributions of Coma members are presented in Fig.~1
for the Coma1 and the Coma3 fields separately.  This survey extends
over more than 6 magnitudes, corresponding to a B band absolute
magnitude in the range $M_B\sim-20.5$ to $-14$ mag
(the distance modulus adopted is 35.16).  For convenience,
in the following we will sometimes refer to ``dwarf'' and ``giant''
galaxies adopting a threshold $R_{3Kr}=16.3$\footnote{Magnitude over
an aperture of radius 3 times the Kron radius. Throughout the paper,
$R=R_{3Kr}$.}  corresponding approximately to $M_B=-17.3$; this
magnitude limit is expected to separate dwarf galaxies and
low-luminosity ellipticals from spiral galaxies and giant ellipticals
in most cases (Ferguson \& Binggeli 1994). Nevertheless, we stress that
this division should only be considered a way to separate the bright
and the faint subsamples and does not provide information about any
other galactic characteristic such as morphological type or central
surface brightness. According to our adopted division, the
spectroscopic catalog comprises 160 dwarf and 118 giant galaxies.

The selection criteria of the spectroscopic targets were different for
the faintest and brightest subsets ($R> 17\, \rm and \, R < 17$).  At the
bright end, the only criterion was membership in the Coma cluster,
as already established by previous spectroscopic studies in the
literature. At the faint end, a color criterion ($B-R>1$) was adopted;
with this color cut, only the most extreme starburst galaxies will be
excluded from the sample.  The majority of the faint objects were
selected to satisfy $B-R<2$ (0.5 mag redder than the mean
color-magnitude sequence), but a sample was also constructed with
redder galaxies to verify that the red color cut did not exclude a
significant number of cluster members. 
A full description of the selection
criteria can be found in Paper II (Mobasher et al. 2001).

In order to study the star formation (SF) and metallicity properties
of the Coma galaxies as a function of their luminosity and
environment, spectral indices and equivalent widths of the main lines
were measured.  Two types of measurements were performed: the indices
of the Lick/IDS system and the strength of eventual emission
lines. Furthermore all spectra were classified according to the
spectral classification scheme developed by the MORPHS group (Dressler
et al. 1999).  The Lick indices are employed in this paper to
determine the luminosity-weighted abundances and ages in spectra
without emission lines (257 Coma galaxies). We stress that this is a
{\it spectroscopic} criterion (only considering non-emission-line
galaxies), while no {\it morphological} criterion was applied. Hence
the sample employed in this paper can and {\it probably is} composed
of galaxies of different Hubble types, presumably ellipticals and S0s
but possibly also passive, non-starforming spirals such as those
observed in distant clusters (Poggianti et al. 1999).  The properties
of the galaxies with emission-line spectra (21 cluster members in
total) and the results of the MORPHS spectral classification will be
dealt with in future papers of the series.

\hbox{~}
\centerline{\psfig{file=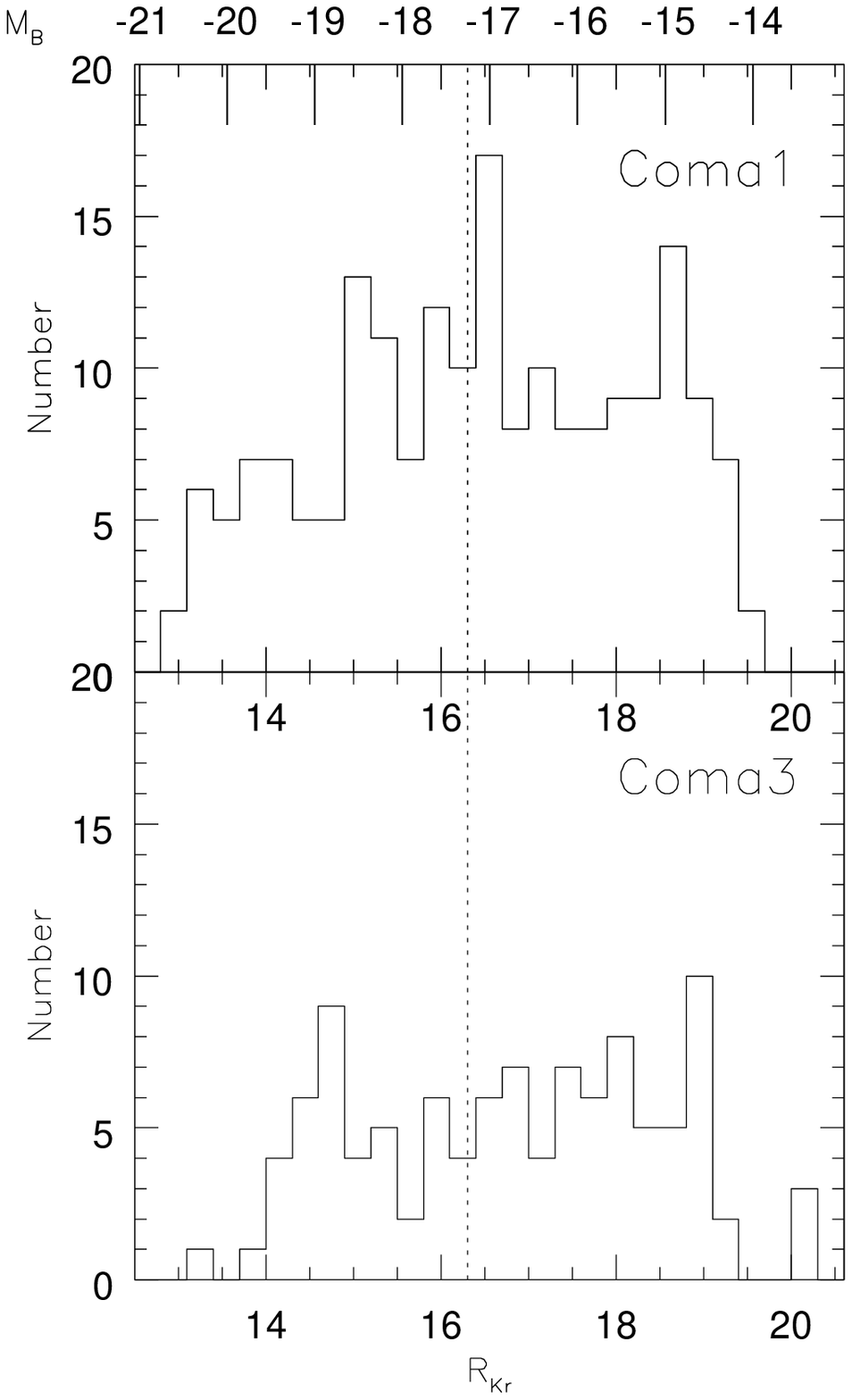,angle=0,width=5.0in}}
\noindent{\scriptsize
\addtolength{\baselineskip}{-3pt} 
\hspace*{0.1cm} Fig.~1. The R band magnitude distributions of Coma1 and Coma3
galaxies that are cluster members. The corresponding absolute B
magnitudes (at the top) have been found 
assuming a distance modulus to Coma
of 35.16 ($<v>=7000 \, \rm km \, s^{-1}$,
$H_0=65 \, \rm km \, s^{-1} \, Mpc^{-1}$) and B-R=1.6.
The dotted vertical line represents the adopted division between dwarfs and giants.
\addtolength{\baselineskip}{3pt}
}

\subsection{Lick indices}

The Lick/IDS system (Burstein et al. 1984, Worthey et al. 1994, Trager
et al. 1998) is the spectral index system that is most widely used to
investigate the abundances and star formation histories of galaxies
without current star formation.  It is a convenient method to study
practically all the main optical absorption lines in spectra of even
modest resolution and it has the advantage that great theoretical
effort has been devoted to interpret the results (see \S3).  The
definitions of the Lick indices are shown in Table~1 and are taken
from Trager et al. (1998) and Worthey \& Ottaviani (1997). In this
table we have flagged with an asterisk the numbers of those indices
that will be used in this paper.
All the indices listed in Table~1 have been measured in all our
spectra, but here we will restrict our analysis only to galaxies
with no detected emission lines. The presence of emission lines was
assessed by inspecting each spectrum and interactively searching for a
line in emission at the wavelengths of the principal features ([O{\sc
ii}]3727, $\rm H\beta$, [O{\sc iii}]4959,5007, $\rm H\alpha$, [N{\sc
ii}]6548,6584), hence we cannot exclude that low-level emission (not
apparent in a visual inspection) is present in some of our
``non-emission-line'' spectra.  We note, however, that
non-negligible emission (partially or totally filling-in, for example,
the $\rm H\beta$ line) would be expected to show up as an emission
line in $\rm H\alpha$ or [O{\sc ii}], which are the lines most
sensitive to even small amounts of photoionization; therefore,
significant contamination of the Balmer indices in our non-emission
line spectra seems unlikely. 

\begin{table}
{\scriptsize
\begin{center}
\centerline{\sc Table 1}
\vspace{0.1cm}
\centerline{\sc Index definitions}
\vspace{0.3cm}
\begin{tabular}{llcccc}
\hline\hline
\noalign{\smallskip}
$N_i$  & Name & Index Bandpass & Blue Bandpass & Red bandpass & Units \cr
\hline
\noalign{\smallskip}
01 & $\rm CN_1$    &     $4142.125-4177.125$ & 4080.125-4117.625 & 4244.125-4284.125 & mag \cr  
02 & $\rm CN_2$    &     $4142.125-4177.125$ & 4083.875-4096.375 & 4244.125-4284.125 & mag \cr
03 & Ca4227        &     $4222.250-4234.750$ & 4211.000-4219.750 & 4241.000-4251.000 & \AA \cr
04 & G4300         &     $4281.375-4316.375$ & 4266.375-4282.625 & 4318.875-4335.125 & \AA \cr
05 & Fe4383        &     $4369.125-4420.375$ & 4359.125-4370.375 & 4442.875-4455.375 & \AA \cr
06 & Ca4455        &     $4452.125-4474.625$ & 4445.875-4454.625 & 4477.125-4492.125 & \AA \cr
07 & Fe4531        &     $4514.250-4559.250$ & 4504.250-4514.250 & 4560.500-4579.250 & \AA \cr
08* & $\rm C_2$4668 &     $4634.000-4720.250$ & 4611.500-4630.250 & 4742.750-4756.500 & \AA \cr
09* & $\rm H_\beta$ &     $4847.875-4876.625$ & 4827.875-4847.875 & 4876.625-4891.625 & \AA \cr
10 & Fe5015        &     $4977.750-5054.000$ & 4946.500-4977.750 & 5054.000-5065.250 & \AA \cr
11 & $\rm Mg_1$    &     $5069.125-5134.125$ & 4895.125-4957.625 & 5301.125-5366.125 & mag \cr
12* & $\rm Mg_2$    &     $5154.125-5196.625$ & 4895.125-4957.625 & 5301.125-5366.125 & mag \cr
13 & Mg$b$         &     $5160.125-5192.625$ & 5142.625-5161.375 & 5191.375-5206.375 & \AA \cr
14 & Fe5270        &     $5245.650-5285.650$ & 5233.150-5248.150 & 5285.650-5318.150 & \AA \cr
15 & Fe5335        &     $5312.125-5352.125$ & 5304.625-5315.875 & 5353.375-5363.375 & \AA \cr
16 & Fe5406        &     $5387.500-5415.000$ & 5376.250-5387.500 & 5415.000-5425.000 & \AA \cr
17 & Fe5709        &     $5696.625-5720.375$ & 5672.875-5696.625 & 5722.875-5736.625 & \AA \cr
18 & Fe5782        &     $5776.625-5796.625$ & 5765.375-5775.375 & 5797.875-5811.625 & \AA \cr
19 & Na D          &     $5876.875-5909.375$ & 5860.625-5875.625 & 5922.125-5948.125 & \AA \cr
20 & $\rm TiO_1$   &     $5936.625-5994.125$ & 5816.625-5849.125 & 6038.625-6103.625 & mag \cr
21 & $\rm TiO_2$   &     $6189.625-6272.125$ & 6066.625-6141.625 & 6372.625-6415.125 & mag \cr
22 &$\rm {H\delta}_A$ &  $4083.500-4122.250$ & 4041.600-4079.750 & 4128.500-4161.000 & \AA \cr
23 &$\rm {H\gamma}_A$ &  $4319.750-4363.500$ & 4283.500-4319.750 & 4367.250-4419.750 & \AA \cr
24* &$\rm {H\delta}_F$ &  $4091.000-4112.250$ & 4057.250-4088.500 & 4114.750-4137.250 & \AA \cr
25* &$\rm {H\gamma}_F$ &  $4331.250-4352.250$ & 4283.500-4319.750 & 4354.750-4384.750 & \AA \cr
26* & $<\rm Fe>$    & (Fe5270+Fe5335)/2 & \multispan{2}{\hfil{}\hfil}& \AA \cr
27 & [MgFe]        & $\sqrt{\rm Mg{\it b} \, \cdot <Fe>}$  &\multispan{2}{\hfil{}\hfil}& \AA \cr
\noalign{\smallskip}
\noalign{\hrule}
\noalign{\smallskip}
\end{tabular}
\end{center}
}
\vspace*{-0.8cm}
\end{table}

In order to compare with the model predictions, it is
first necessary to take into account: (a) the different spectral
resolution of the WYFFOS spectra as compared to the Lick/IDS set-up;
(b) the internal velocity dispersion of each galaxy; (c) any 
systematic offsets in the indices mainly due to the fact that the Lick
stars were not flux calibrated. This was done as described in Appendix
A. The final, fully-corrected indices of the Lick system were computed
by a FORTRAN program that reads in the raw measurements (obtained
after applying the correction for spectral resolution described in 
point a) of Appendix A), computes and applies the velocity dispersion
corrections and finally applies the systematic offsets (point c) of
Appendix A).

To determine the random errors in our Lick indices, we followed the
approach of Gonz\'alez (1993), as explained and reproduced in Cardiel et
al.  (1998, C98).  Gonz\'alez has derived analytical formulae to
evaluate the random index errors, taking into account the data, the
statistical propagation of errors, and the variance in each pixel.
Cardiel et al. show that, for most situations, the formulae of
Gonz\'alez agree very well with numerical simulations.  We used the {\it
raw} (ie. non flux-calibrated) spectra, de-redshifted to zero
velocity, and accounted for both CCD gain and read-noise to calculate
the total variance per pixel (eqn. 9 of C98).  The errors in each of
the atomic and molecular indices were then calculated using
eqns. 5-7,11,12,13-18 of C98. Negative counts (representing poor sky
subtraction) were ignored if they constituted less than half the
pixels in a given bandpass (index or continuum); if more than half of
the pixels in a given bandpass were negative, then the absolute value
of the counts was used.

The correctness of the calibration of our indices onto the Lick/IDS
system has been verified with a number of tests as described in 
Appendix B: from them, we conclude that our index calibration onto the
Lick system has been successful and that -- within the limitations and
drawbacks intrinsic to the data and model uncertainties -- we can
determine ages and metallicities by comparing the galaxy measurements
and the model predictions.

\section{Models}

Many authors have used stellar population models to compute
Lick line indices for stellar populations of different ages and metallicities
(Worthey 1994, Weiss, Peletier \& Matteucci 1995, Buzzoni 1995, Vazdekis et 
al. 1996, Bressan, Chiosi \& Tantalo 1996, Maraston 1998,
Tantalo, Chiosi \& Bressan 1998, Maraston, Greggio \& Thomas 1999).
The models adopted in this paper are those of Worthey (1994) in the new WEB
version\footnote{http://astro.sau.edu/\~{}worthey/dial/dial\_a\_pad.html} 
that is based upon the Padova isochrone library (Bertelli et al. 1994).
The Padova-based models were prefered to the Yale version for two reasons.
First, the Yale version of the model did not cover the low-metallicity, 
young corner ([Fe/H]$<$-0.225, age $<8$ Gyr) which could be relevant in 
this study\footnote{In fact, this region of the parameter space turned out 
to be populated by a significant fraction of our dwarf galaxies, see later.}
while the Padova version has ages in the range 0.4-19.9 Gyr
and [Fe/H] between -1.7 and +0.4. Second, the Padova isochrones include
the existence of blue horizontal branches (HB) in old populations of 
low metallicities. A blue HB produces moderate Balmer line strengths, 
hence an accurate treatment of the HB is needed for a correct interpretation 
of these lines
(Poggianti \& Barbaro 1997, Lee, Yoon \& Lee 2000, Maraston \& Thomas 
2000). The Padova version reproduces well the increase in Balmer indices 
with decreasing metallicity observed in old populations (e.g. Galactic 
globular clusters).\footnote{As an example, the $\rm H\beta$ index of the 
models used in this paper is 2.6 for [Fe/H]=$-$1.7, 2.5 at [Fe/H]=$-$1.5 and 
2.4 at [Fe/H]=$-$1.3 (compare with Fig.~1 in Maraston \& Thomas 2000).}
We have used Worthey's WEB model interpolation engine to produce grids of
indices of Single Stellar Populations with a standard Salpeter IMF (x=2.35,
$M=0.6-120 M_{\odot}$)
for ages=0.4,0.5,0.7,0.8,1.0,1.5,2.0,3.0,4.0,5.0,6.0,7.0,8.0,9.0,10.0,11.0,12.0,13.0,14.0,15.0,16.0,17.0,18.0,19.0,19.99 Gyr 
and [Fe/H]=$-$1.7,$-$1.5,$-$1.3,$-$1.0,$-$0.7,$-$0.5,$-$0.25,0.0,0.25,0.39. 
Another complete set of predicted Lick indices is given by 
Vazdekis et al. (1996) for ages between 1 and 17.4 Gyr and six
metallicities between [Fe/H]=-1.7 and +0.4. This set is based upon the Padova
isochrones as well, and therefore has the same stellar evolutionary
background as the models used in this paper. 
Comparing the synthetic grids of the indices employed in \S4.2
for the same IMF, we found a general agreement 
between the Worthey and Vazdekis models. The only 
systematic difference was that Vazdekis's models reach
slightly lower values of $\rm H\beta$ ($\sim 0.2$ \AA $\,$ lower) at old ages
($> 14$ Gyr) and slightly higher values (0.01 mag) of $\rm Mg_2$ at 
low metallicity ($ \leq -0.7 $),
producing at most a $\sim 0.1$ difference in Z and an age difference
smaller than the uncertainty at old ages.
Both sets of models lead to the conclusions presented below, 
and we decided to use Worthey's models because they allowed us to produce 
a model grid as fine as desired, in the range 0.4 to 20 Gyr (but see
also Vazdekis's WEB page for an update of his models).

It is generally assumed that the calibrating stars used to derive the
Lick/IDS fitting functions adopted in the models have solar abundance
ratios; more likely, there is a variation of abundance ratio built
into the models, for example the calibrating globular cluster (low
metallicity) stars have [Mg/Fe] ratios above solar (Worthey 1998).
Under the (likely correct) assumption that {\it around solar
metallicity} the model ratios are indeed solar, it has been found that
luminous ellipticals have non-solar abundance ratios and that the
[Mg/Fe] ratio increases with the luminosity (Peletier 1989, Worthey et
al. 1992, Vazdekis et al. 1997, Worthey 1998 and references therein,
Trager et al. 2000a and 2000b, Kuntschner et al. 2001). Originally
referred to as a magnesium enhancement, this effect is now recognized
to be more precisely an iron depression (Vazdekis et al. 1997, Trager
et al. 2000a).

In this paper we do not attempt to vary the [Mg/Fe] ratio of the
models because (a) the variations of the ratio {\it intrinsic} to the
models are not well understood over the metallicity range of interest
here, and (b) our sample does not comprise extremely luminous galaxies
($M_B<-21$), and consists mostly of low luminosity objects which
are believed to be consistent with solar abundance [Mg/Fe] ratios
(e.g. Gorgas et al. 1997). However, we caution that solar abundance
ratios in low luminosity galaxies are usually deduced from the
agreement with models which probably {\it do not} have solar ratios at
low metallicities, as explained above.  
In the following, we will employ both iron and
magnesium indices and compare the results obtained.

\section{Results}

\subsection{Index results}

The distributions of the most relevant indices are shown in Fig.~2
for four magnitude bins and for Coma1 and Coma3 galaxies separately 
(empty and shaded histogram respectively). 
Three main results appear evident: 

a) For both fields, the mean value of the metallicity-sensitive
indices ($\rm Mg_2$, $\rm <Fe>$, $C_{2}4668$) {\it increases} for brighter
galaxies.  Similarly, the mean value of the age-sensitive (Balmer)
indices {\it decreases} with galaxy luminosity.

b) Within a given magnitude bin, in the Coma1 region there is generally
a higher proportion of galaxies with higher metallicity-indices 
and lower age-indices than in Coma3.

c) The width of the distributions increases towards fainter galaxies.

In the following we will mostly focus on the trends of the spectral
indices with magnitude for the whole sample (Coma1+Coma3), pointing
out the differences between Coma1 and Coma3 only when luminosity and
environmental effects need to be disentangled.  The spatial dependence
of the galaxy properties will be presented in a future paper of the
series.

\hbox{~}
\centerline{\hspace{1.3in}\psfig{file=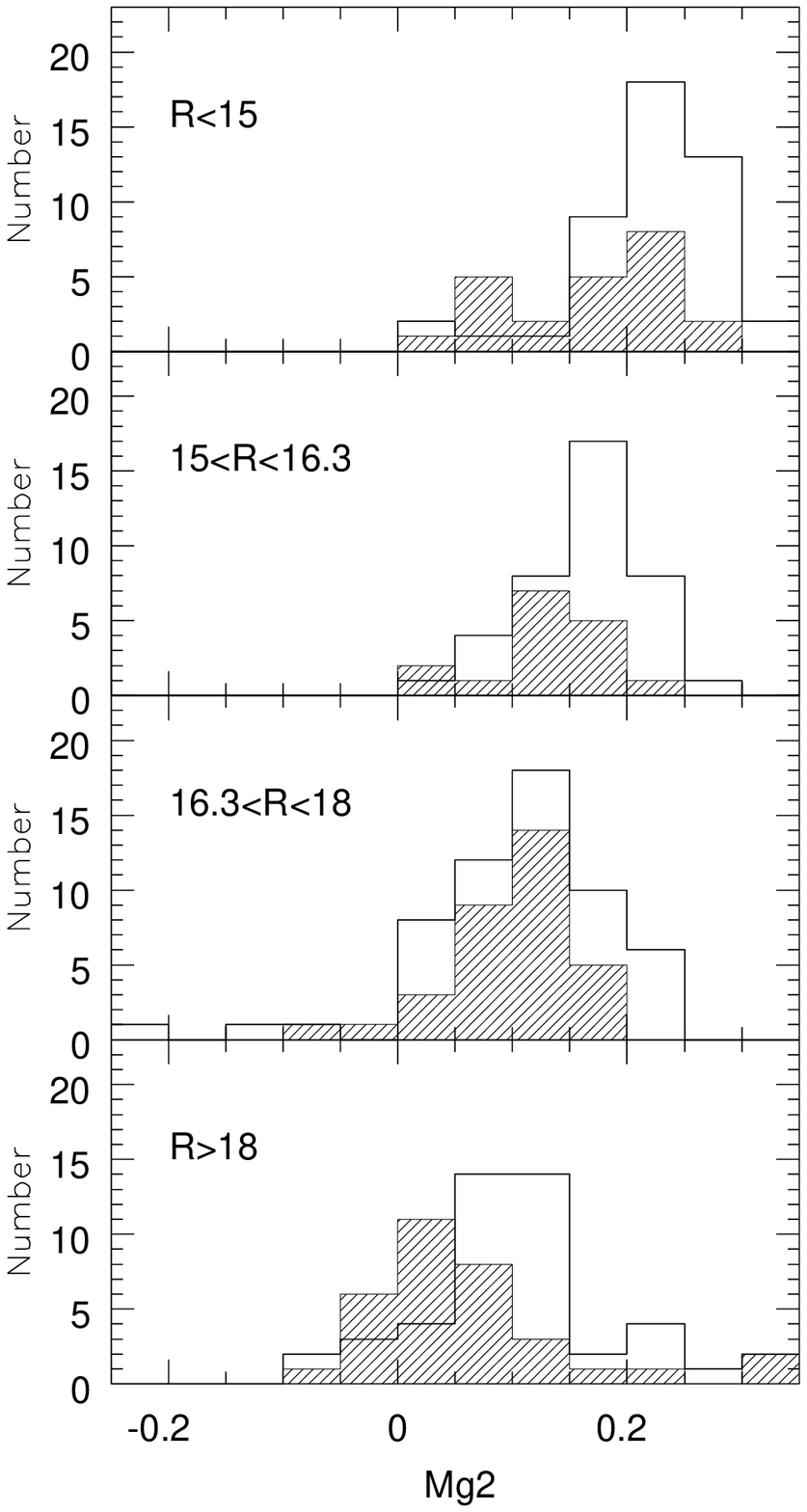,angle=0,width=4in}
\hspace{-2in}\psfig{file=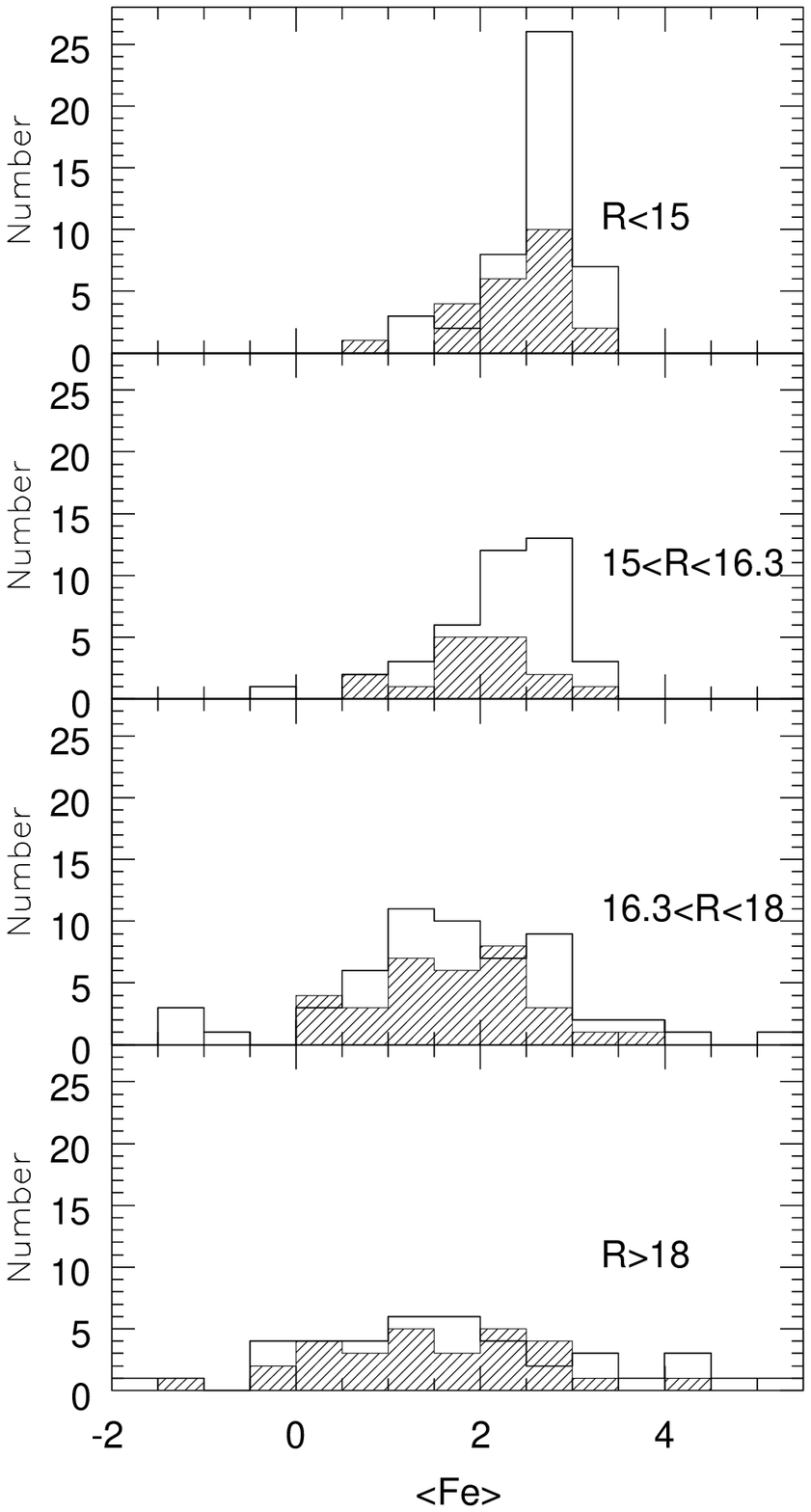,angle=0,width=4in}
\hspace{-2in}\psfig{file=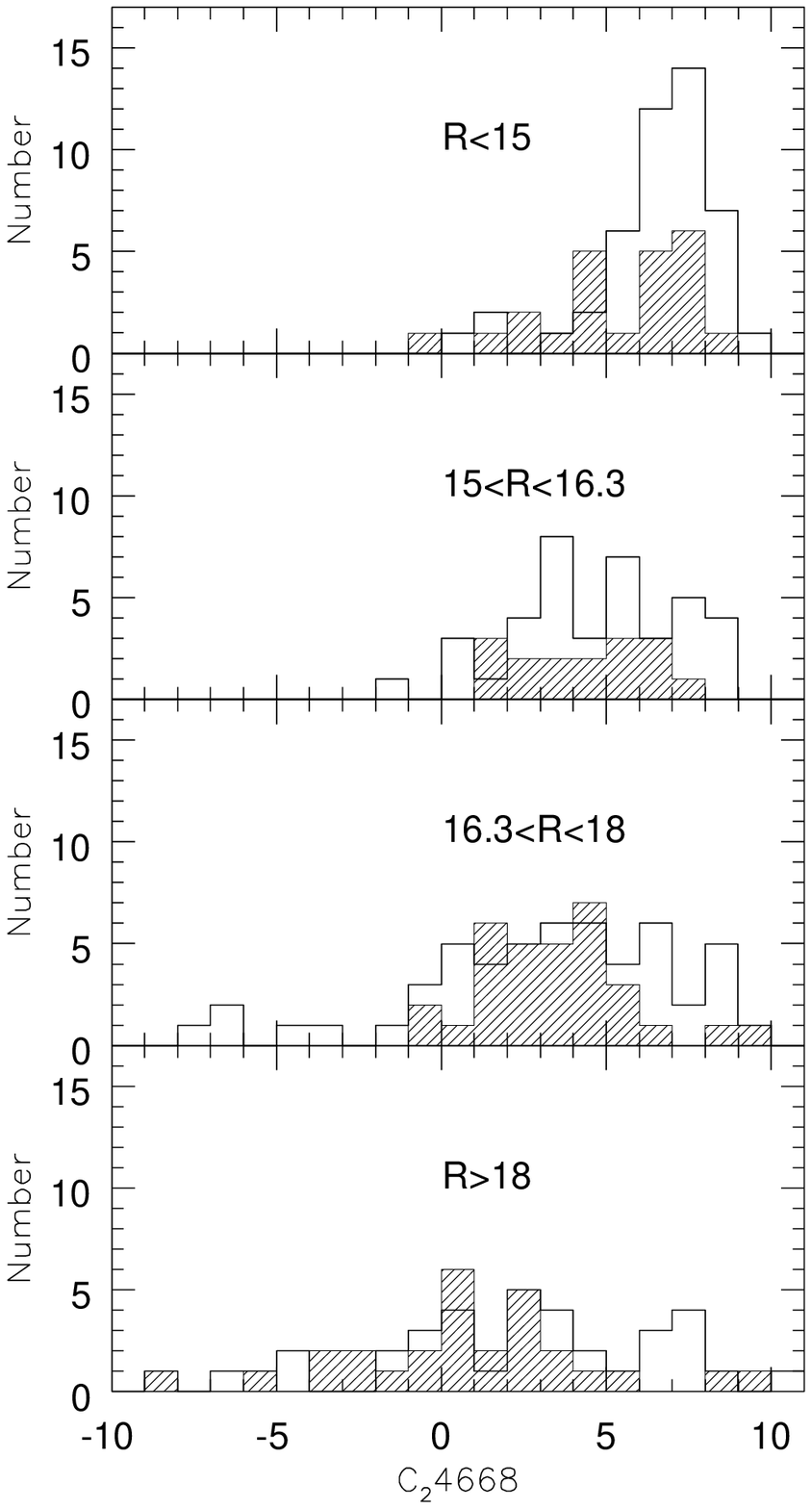,angle=0,width=4in}
}
\centerline{\hspace{1.3in}\psfig{file=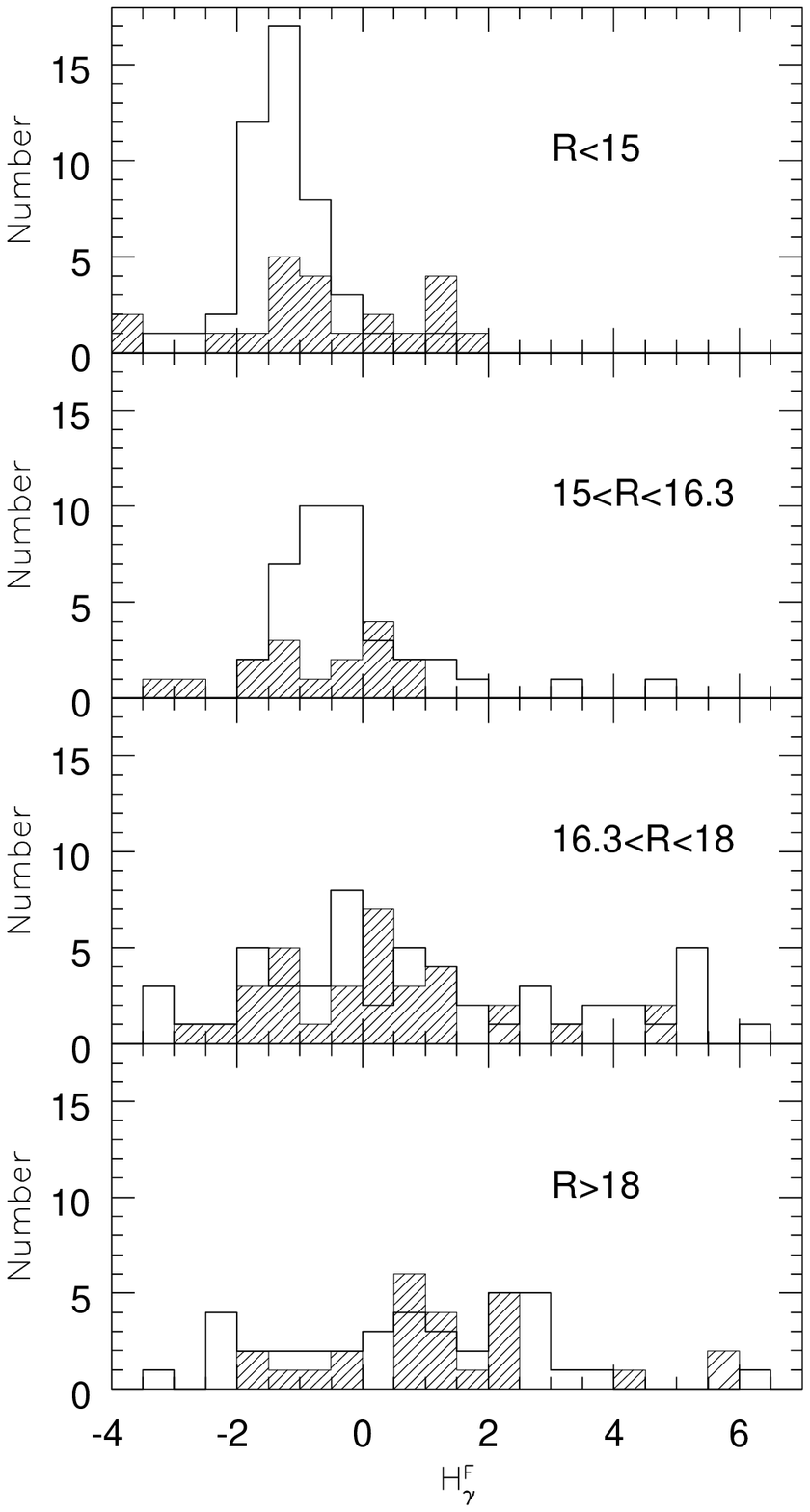,angle=0,width=4in}
\hspace{-2in}\psfig{file=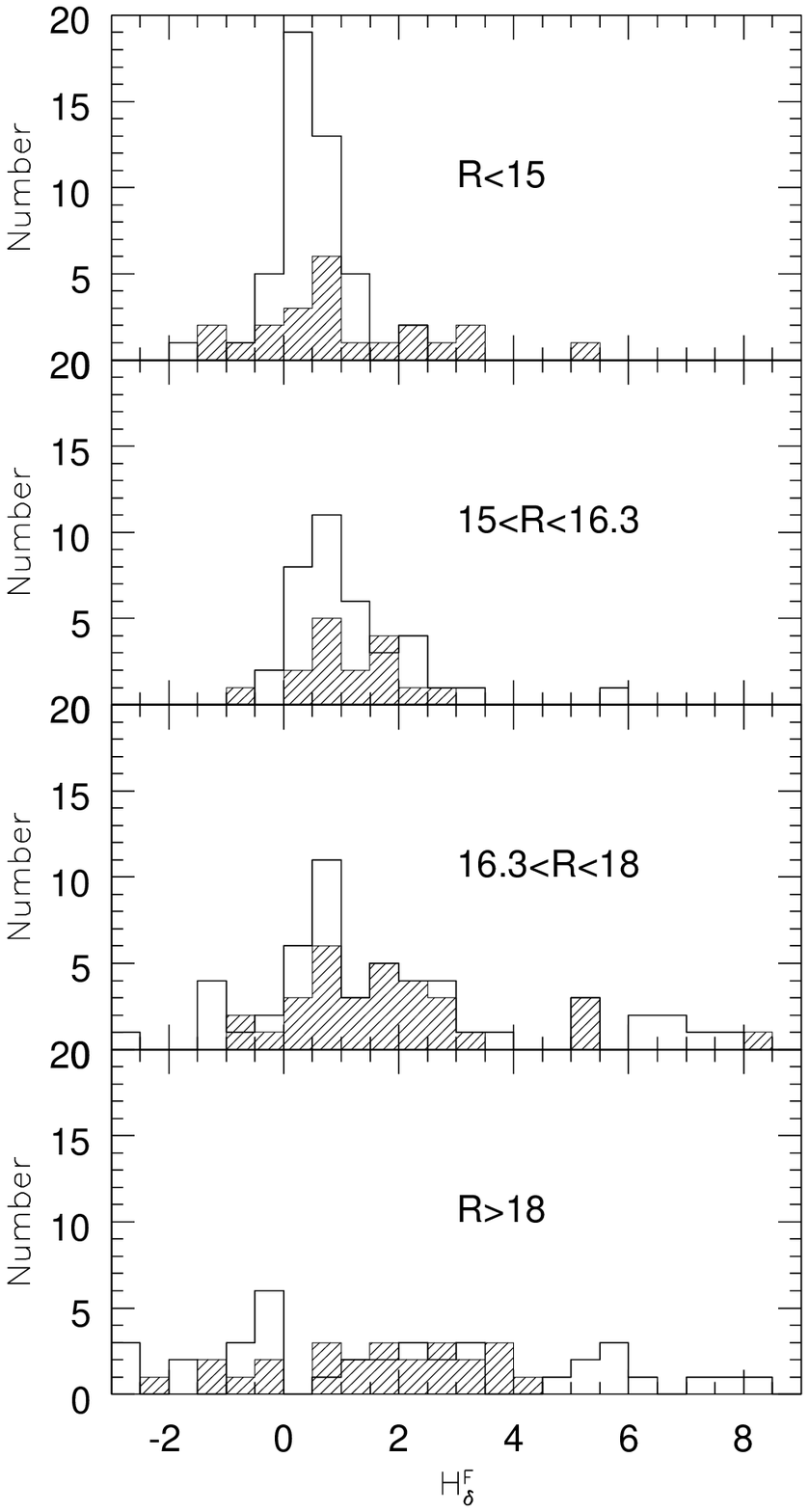,angle=0,width=4in}
\hspace{-2in}\psfig{file=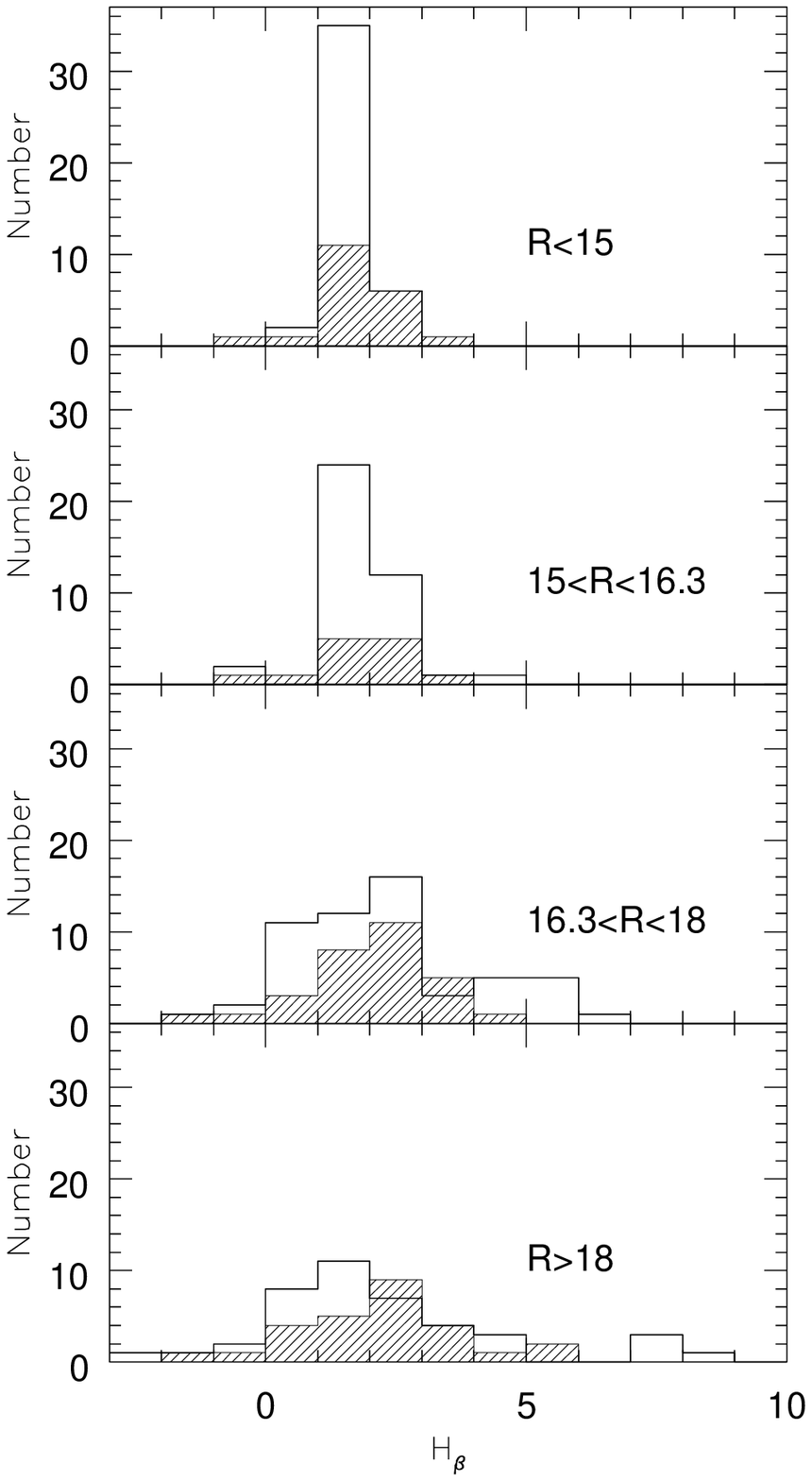,angle=0,width=4in}
}
\noindent{\scriptsize
\addtolength{\baselineskip}{-3pt} 
\hspace*{0.1cm} Fig.~2.\ Index distributions of Coma1 (empty histogram)
and Coma3 (shaded histogram) galaxies for four magnitude bins. 
Emission line galaxies have been excluded.
\addtolength{\baselineskip}{3pt}
}

The trends of the indices with galaxy luminosity are shown clearly in
Fig.~3, where it can be seen that there is a positive correlation
between the age-sensitive indices and R magnitude, and an
anticorrelation between the metallicity-sensitive indices and R
magnitude.\footnote{Similar relations are found for other
metallicity-sensitive and age-sensitive indices but not shown.}  It is
remarkable that these relations hold over the whole magnitude range
explored in this study (6.5 mag) down to $M_B\sim -14$.  The scatter
around these relations progressively increases for fainter galaxies
and a group of distinguished outliers -- mainly dwarfs -- can be
identified in most of these plots.  Note that the incidence of
outliers, especially in the Balmer-indices plots, is not symmetric
around the mean relation, and there are many more points lying {\it above}
the relation than below.  The remainder of this section deals with the
interpretation of the mean observed relations, the origin of the
scatter and the characteristics of the outliers\footnote{We note that
the majority of the outliers in the right-hand plots are not the same
objects that stand out as outliers in the left-hand plots.  This will
become clearer examining the index-index diagrams in the next
section. As an example, most of the $\rm H\beta$-strong galaxies with
$\rm Hbeta>4$ have $\rm Mg_2<0.2$.}.

We find that correlations similar to those of Fig.~3 exist also
between the indices and the R-band effective surface brightness
$\mu$. This is not surprising given the correlation between the R
magnitude and the effective surface brightness observed in this sample
(Mobasher et al. 2001, Paper II).  The slope B and the intercept A of
the linear fits to the data are given in Table~2 for the relations of
six indices with the R magnitude and with $\mu$.

Since there is no such a thing as a ``pure'' index that depends only
on age or only on metallicity, each of the correlations shown in
Fig.~3 implies either that the fainter galaxies are younger, or that
they are less metal rich, or both. In order to separate these effects
it is necessary to compare the positions of the galaxies in
index-index plots with the results of our models.

\hbox{~}
\centerline{\hspace{1.3in}\psfig{file=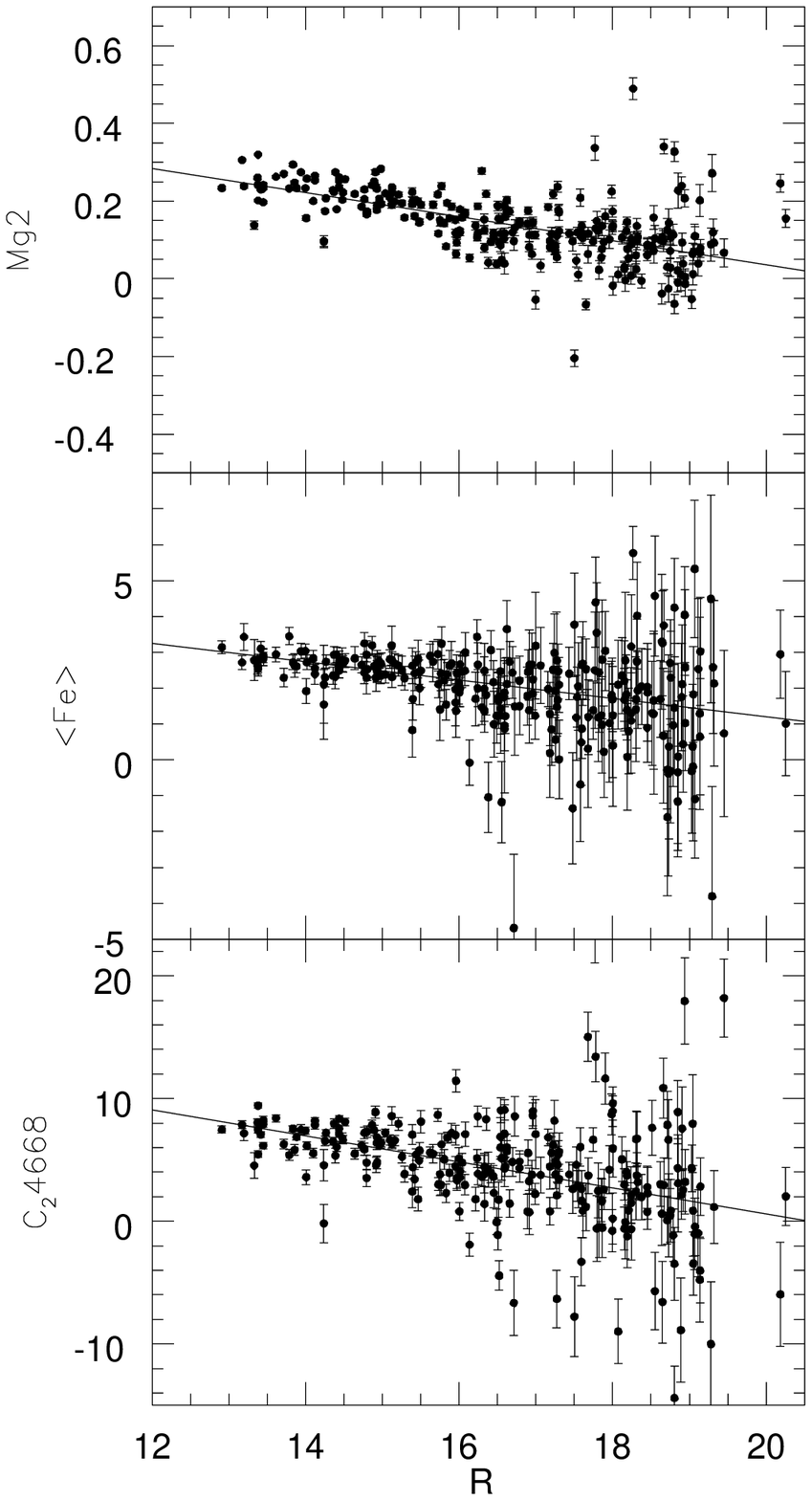,angle=0,width=5in}
\hspace{-2in}\psfig{file=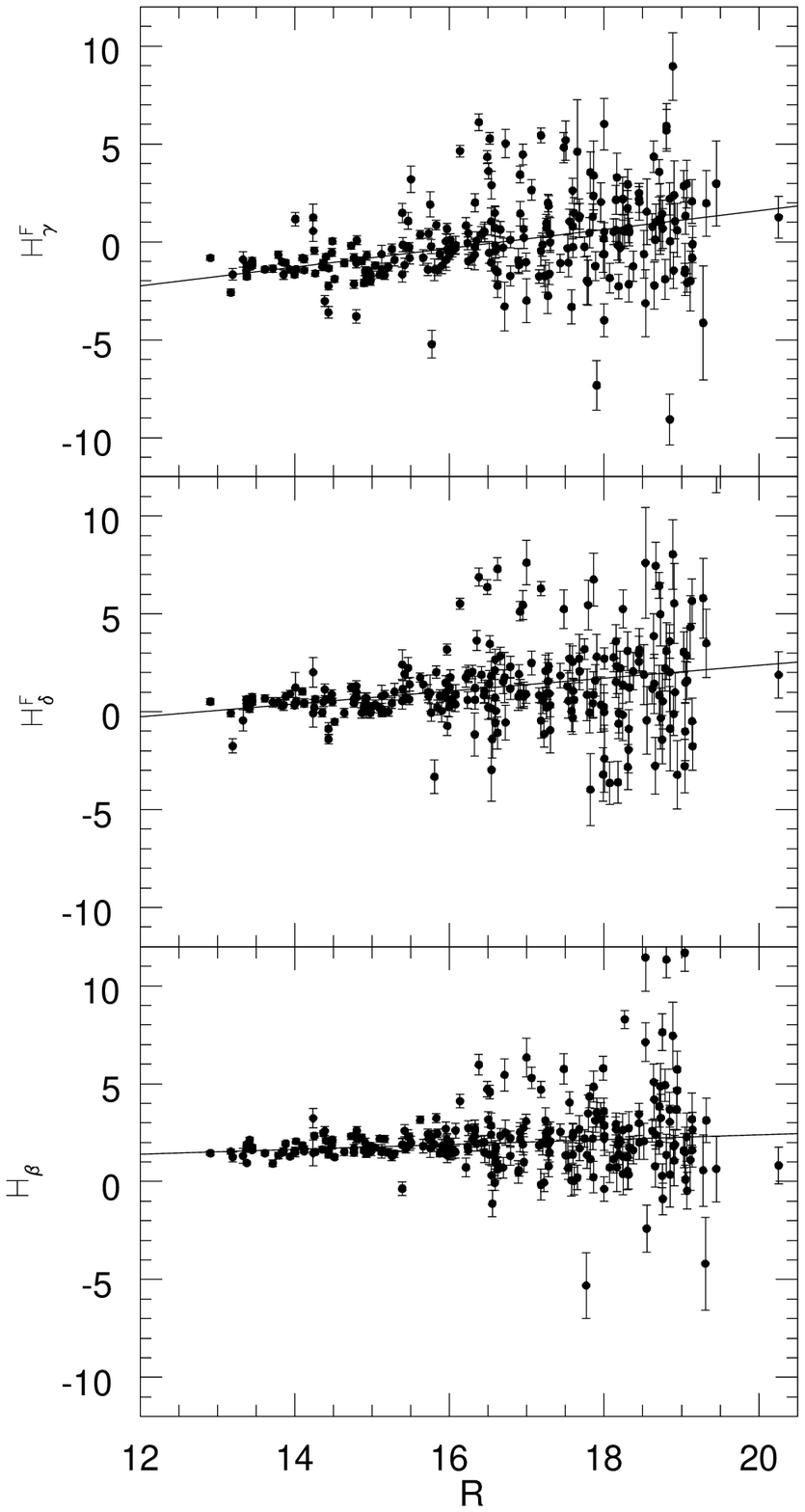,angle=0,width=5in}}
\noindent{\scriptsize
\addtolength{\baselineskip}{-3pt} 
\hspace*{0.1cm} Fig.~3.\ The relations between R magnitude
and the metallicity-sensitive 
indices (left), and the age-sensitive indices (right).
Emission-line galaxies have been omitted from this figure.
Both Coma1 and Coma3 galaxies are included.
The straight line in each panel is a robust 
linear fit to the data found by means of an M-estimate that minimizes
the absolute deviation; the slope B and the intercept A of these fits are
given in Table~2.
\addtolength{\baselineskip}{3pt}
}

\begin{table}
{\scriptsize
\begin{center}
\centerline{\sc Table 2}
\vspace{0.1cm}
\centerline{\sc Parameters of the fit $Y=A + B \times R \, (\rm or \, \mu)$}
\vspace{0.3cm}
\begin{tabular}{lcccc}
\hline\hline
\noalign{\smallskip}
Y  & $A_R$ & $B_R$ & $A_{\mu}$ & $B_{\mu}$ \cr
\hline
\noalign{\smallskip}
 $\rm Mg_2$        & 0.656 & -0.031 & 0.947 & -0.038 \cr 
 $<Fe>$            & 6.299 & -0.255 & 9.228 & -0.343 \cr
 $\rm C_2$4668     & 21.8  & -1.06  & 31.56 & -1.302 \cr
 $\rm {H\gamma}_F$ & -7.994 & 0.479 & -15.70& 0.753 \cr
 $\rm {H\delta}_F$ & -4.238 & 0.331 & -8.705 & 0.478 \cr
 $\rm H_\beta$     & -0.139 & 0.127 & -1.79 & 0.180\cr
&&&& \cr
Z(Hb) & 3.006 & -0.220 & & \cr
Z(Hg) & 1.944 & -0.144 & & \cr
\noalign{\smallskip}
\noalign{\hrule}
\noalign{\smallskip}
\noalign{\smallskip}
\end{tabular}
\end{center}
}
\vspace*{-0.8cm}
\end{table}

\subsection{Index-index diagrams}

In this section we wish to investigate whether the trends of the
indices with galaxy magnitude presented in \S4.1 are due to 
metallicity, age, or both.  To discriminate between these
effects it is necessary to analyse simultaneously an age-sensitive
index and a metallicity-sensitive index.  Comparing them with a
two-dimensional theoretical grid of single stellar populations (single
burst and single metallicity at a given age) yields an estimate of the
{\it luminosity-weighted} age and metallicity of the galaxy.

Figure~4 shows the index-index plots of the dwarf (filled circles) and
giant (empty circles) samples. Here we present the two most
age-sensitive indices according to Worthey \& Ottaviani (1997) ($\rm
H\beta$ and $\rm H{\gamma}_F$) versus two indices which are mostly
sensitive to metallicity ($\rm Mg_2$ and $<$Fe$>$)\footnote{Although
the $C_{2}4668$ index shows the strongest sensitivity to metallicity
(Worthey \& Ottaviani 1997, Worthey 1998), we prefered to use the
other metallicity indicators because the reliability of $C_{2}4668$
has been questioned (Jones, Smail \& Couch 2000) and it is unclear
what chemical species drives its enhancement with respect to iron in
the most luminous galaxies (Kuntschner 2000).}.  A subset of the model
grid is shown in more detail in Fig.~5.

\hbox{~}
\vspace{-1in}
\centerline{\hspace{-0.6in}\psfig{file=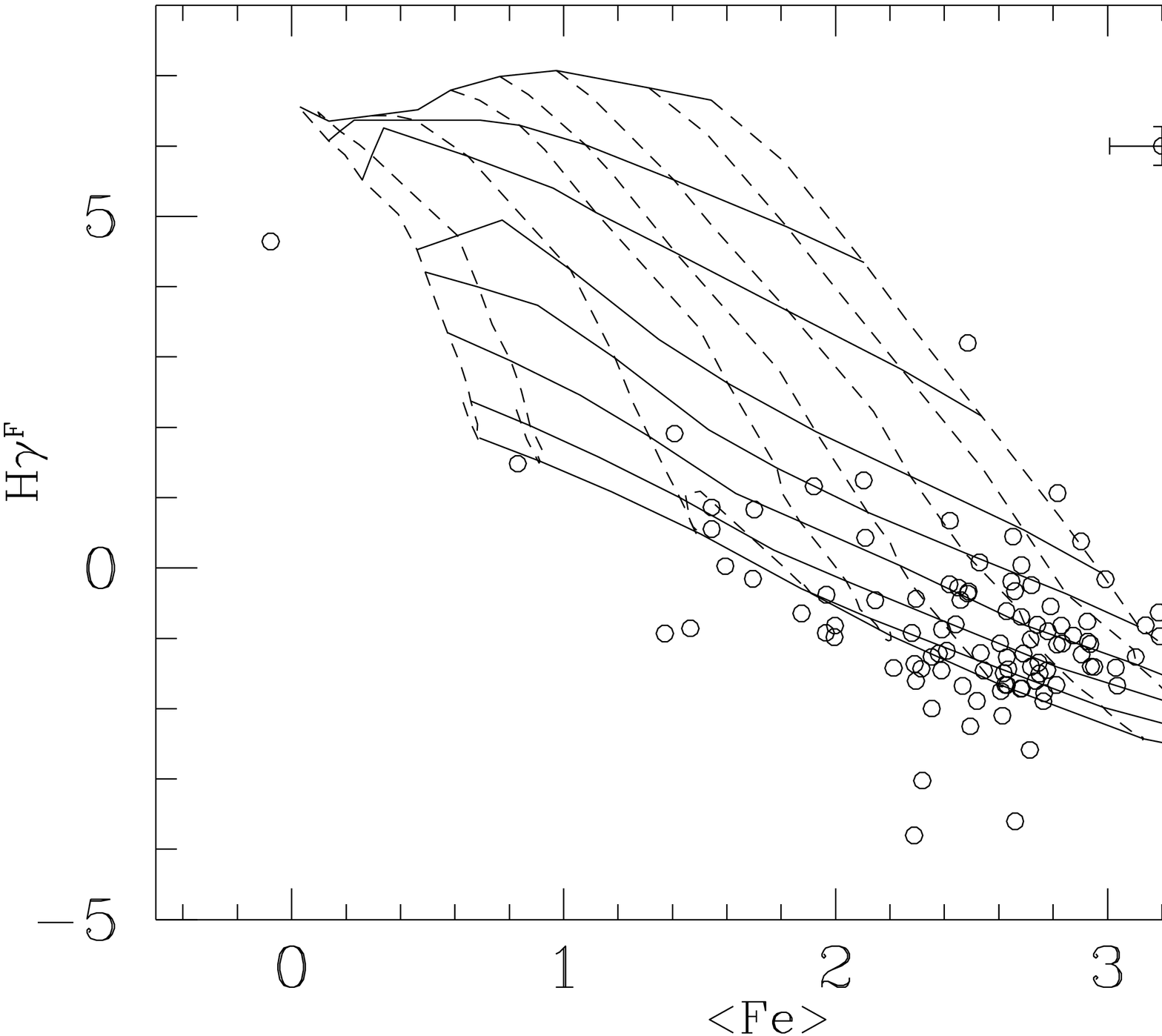,angle=0,width=2.3in} \hspace{0.6in}\psfig{file=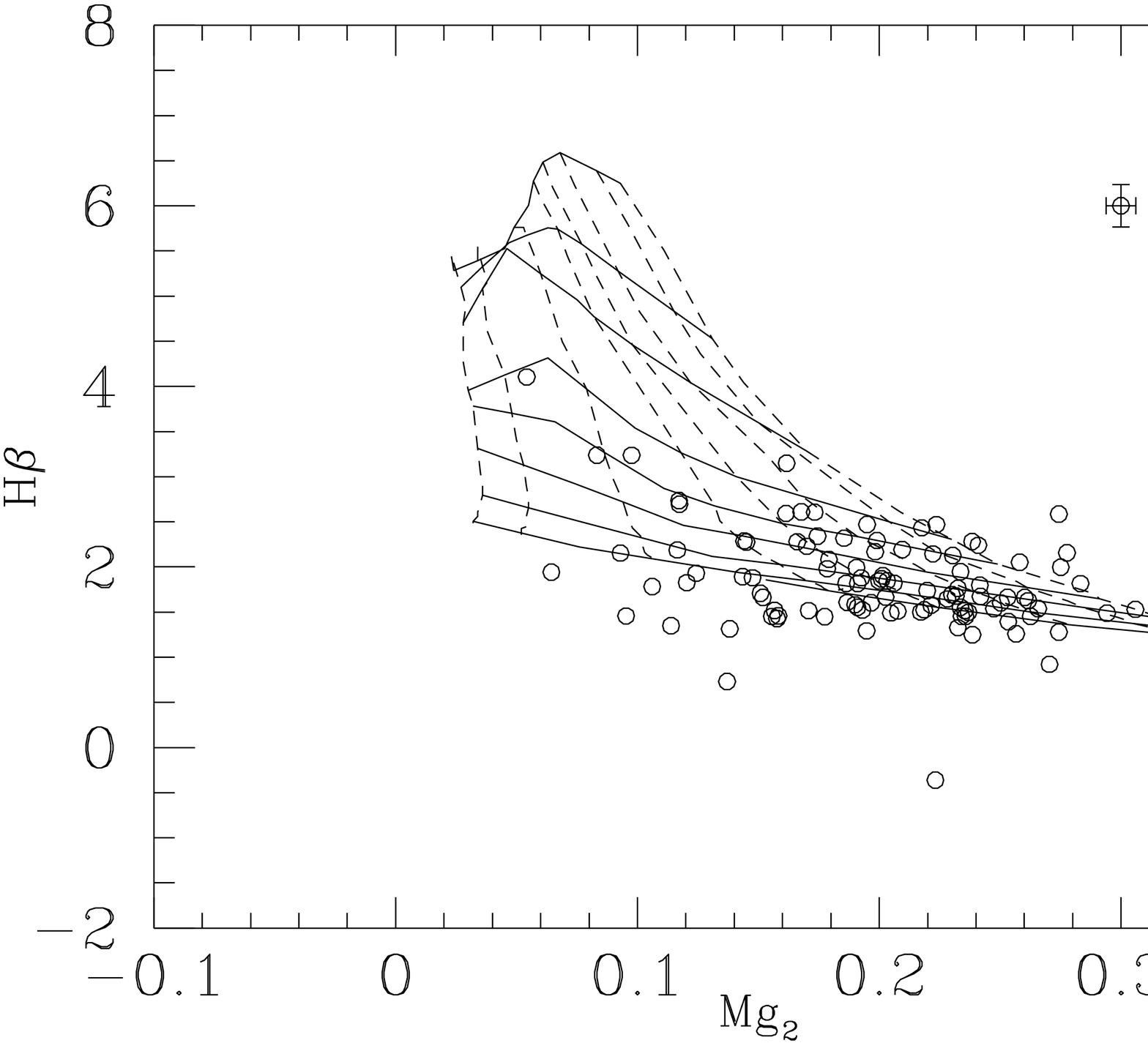,angle=0,width=2.3in}}
\vspace{-0.7in}
\centerline{\hspace{-0.6in}\psfig{file=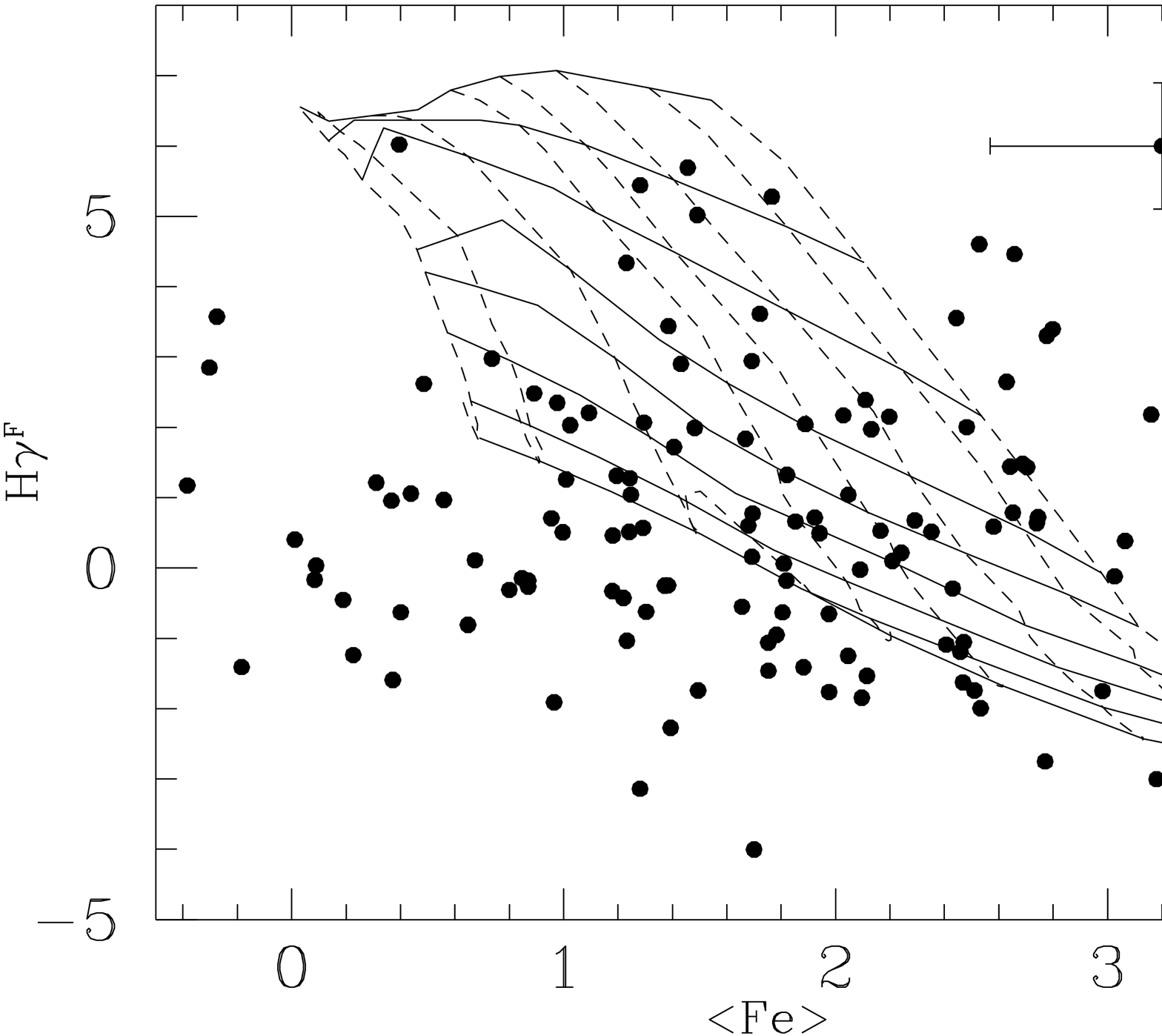,angle=0,width=2.3in} \hspace{0.6in}\psfig{file=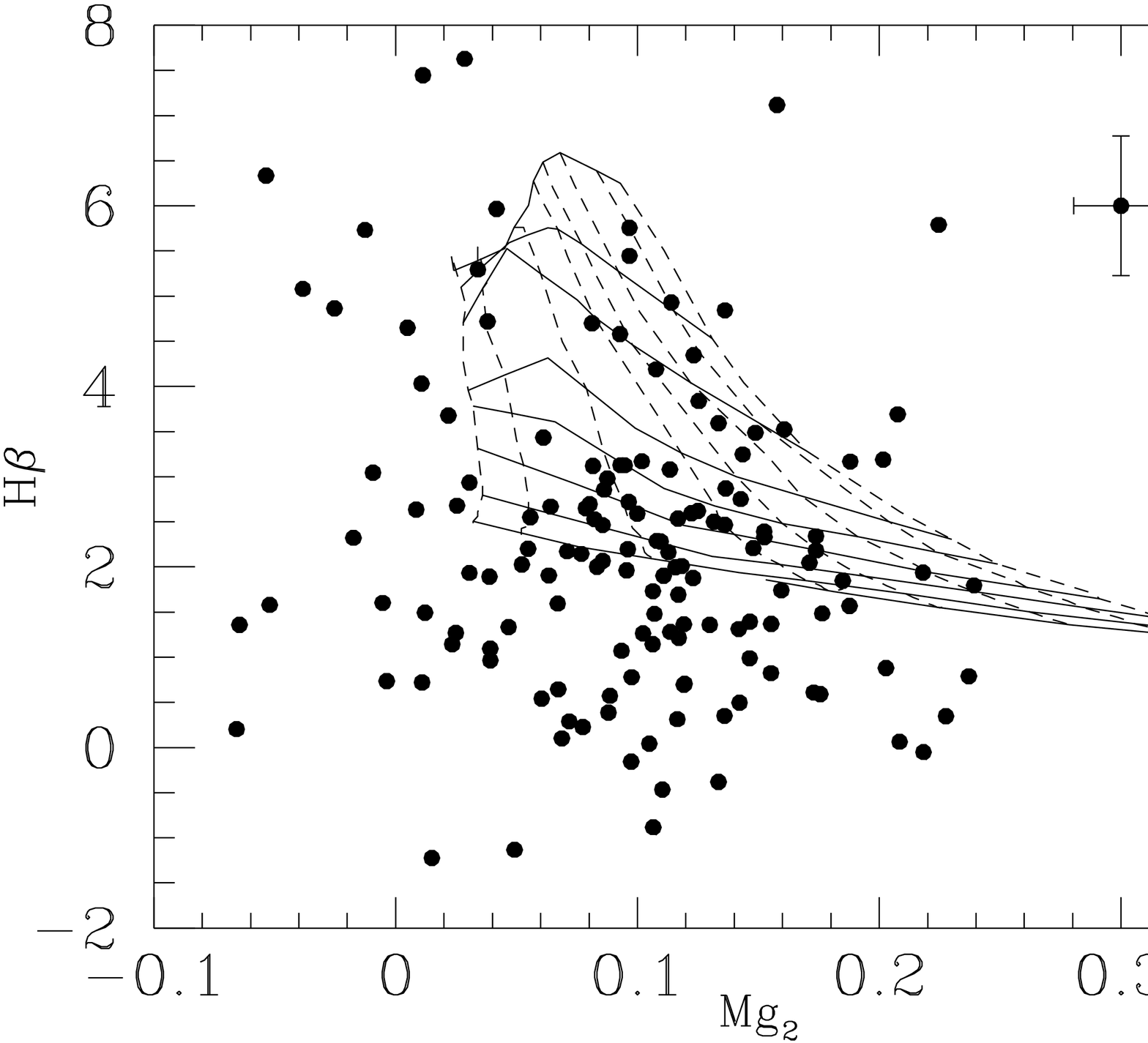,angle=0,width=2.3in}}
\vspace{-0.7in}
\centerline{\hspace{-0.6in}\psfig{file=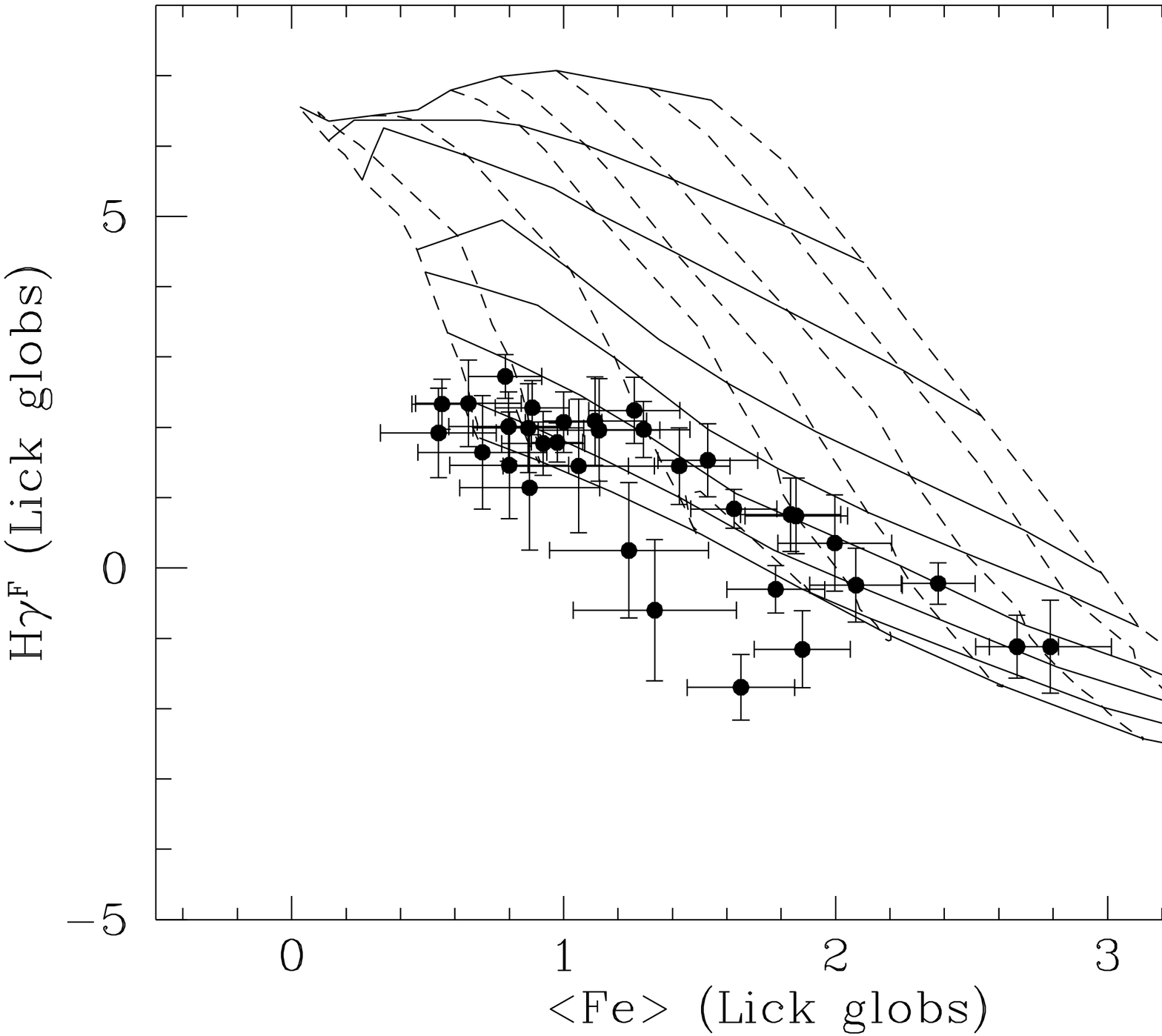,angle=0,width=2.3in} \hspace{0.6in}\psfig{file=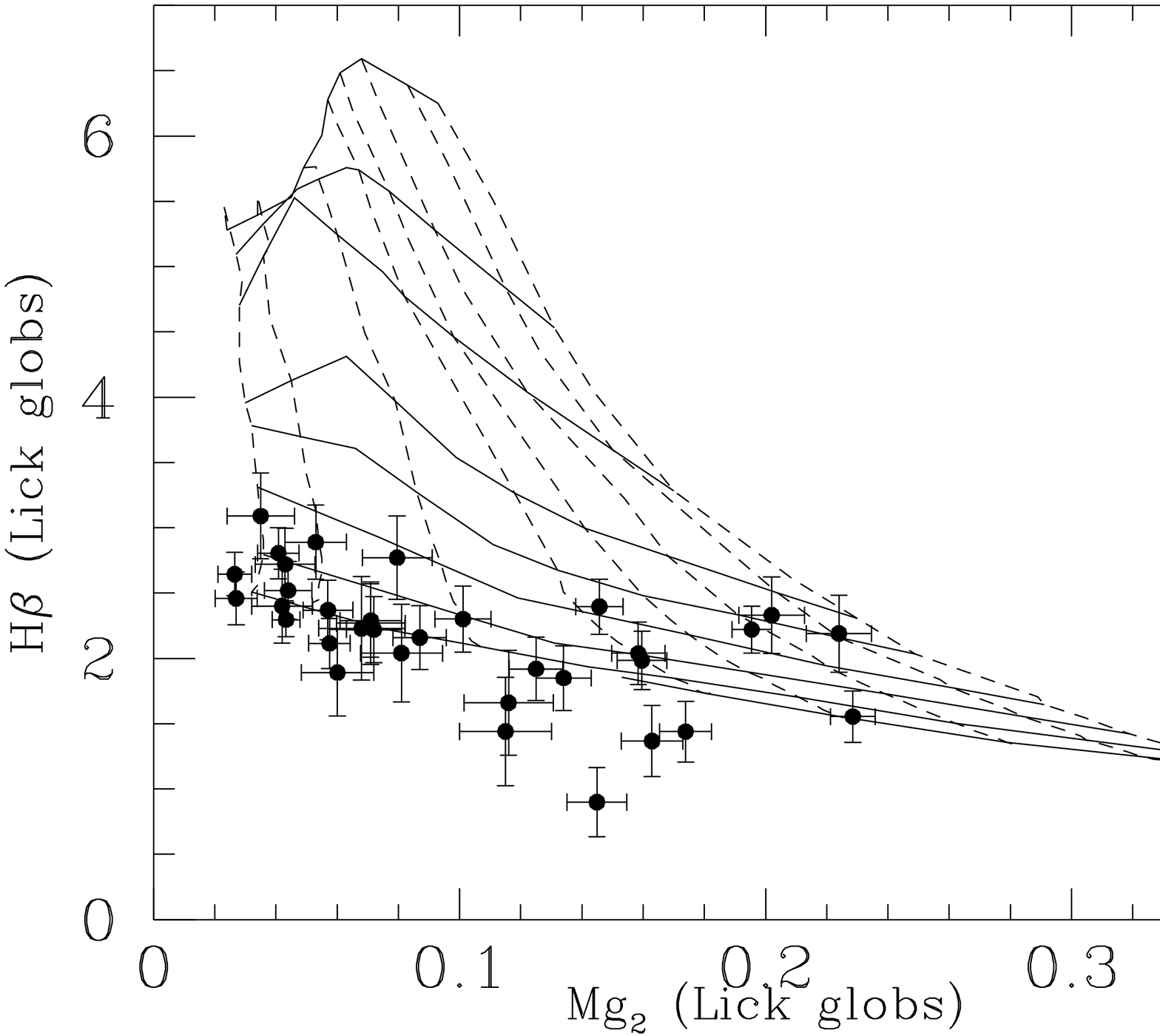,angle=0,width=2.3in}}
\noindent{\scriptsize
\addtolength{\baselineskip}{-3pt} 
\hspace*{0.1cm} Fig.~4.\ The two best age-sensitive indices 
($\rm {H\gamma}_F$ and $\rm H\beta$) are plotted against two 
metallicity-sensitive indices ($<$Fe$>$ and $\rm Mg_2$). 
Top and middle panels: giant (empty symbols) and dwarf (filled symbols)
galaxies, 
respectively. For clarity, only the mean errorbars for dwarfs and giants 
are shown in the top right corner of the plot. Overplotted are some Padova
models by Worthey (see \S3) of single stellar populations (see Fig.~5). 
Bottom panels: results for Lick globular clusters, kindly provided
by G.~Worthey ahead of publication.
\addtolength{\baselineskip}{3pt}
}

\hbox{~}
\vspace{-1in}
\centerline{\hspace{-0.6in}\psfig{file=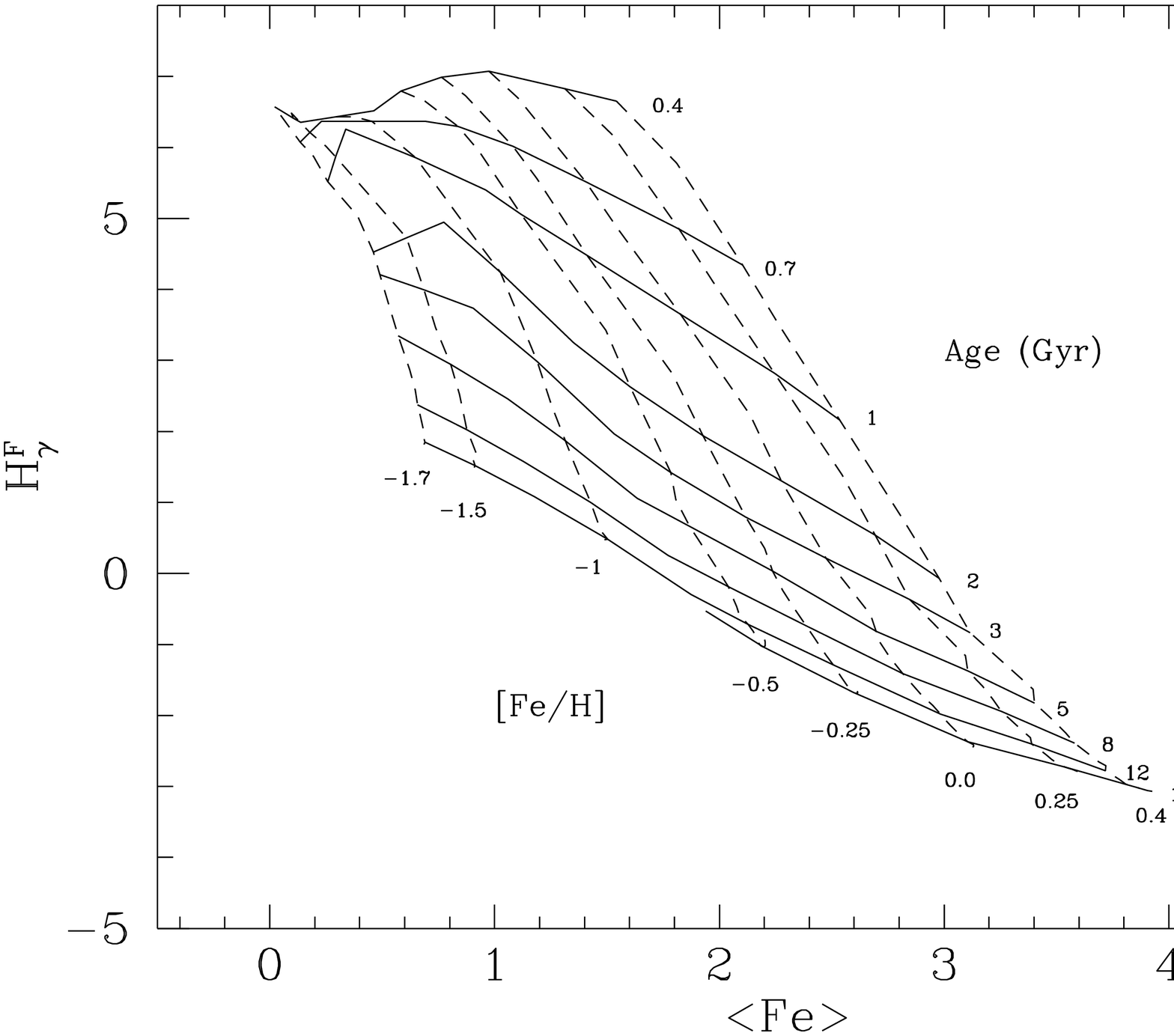,angle=0,width=2.3in} \hspace{0.6in}\psfig{file=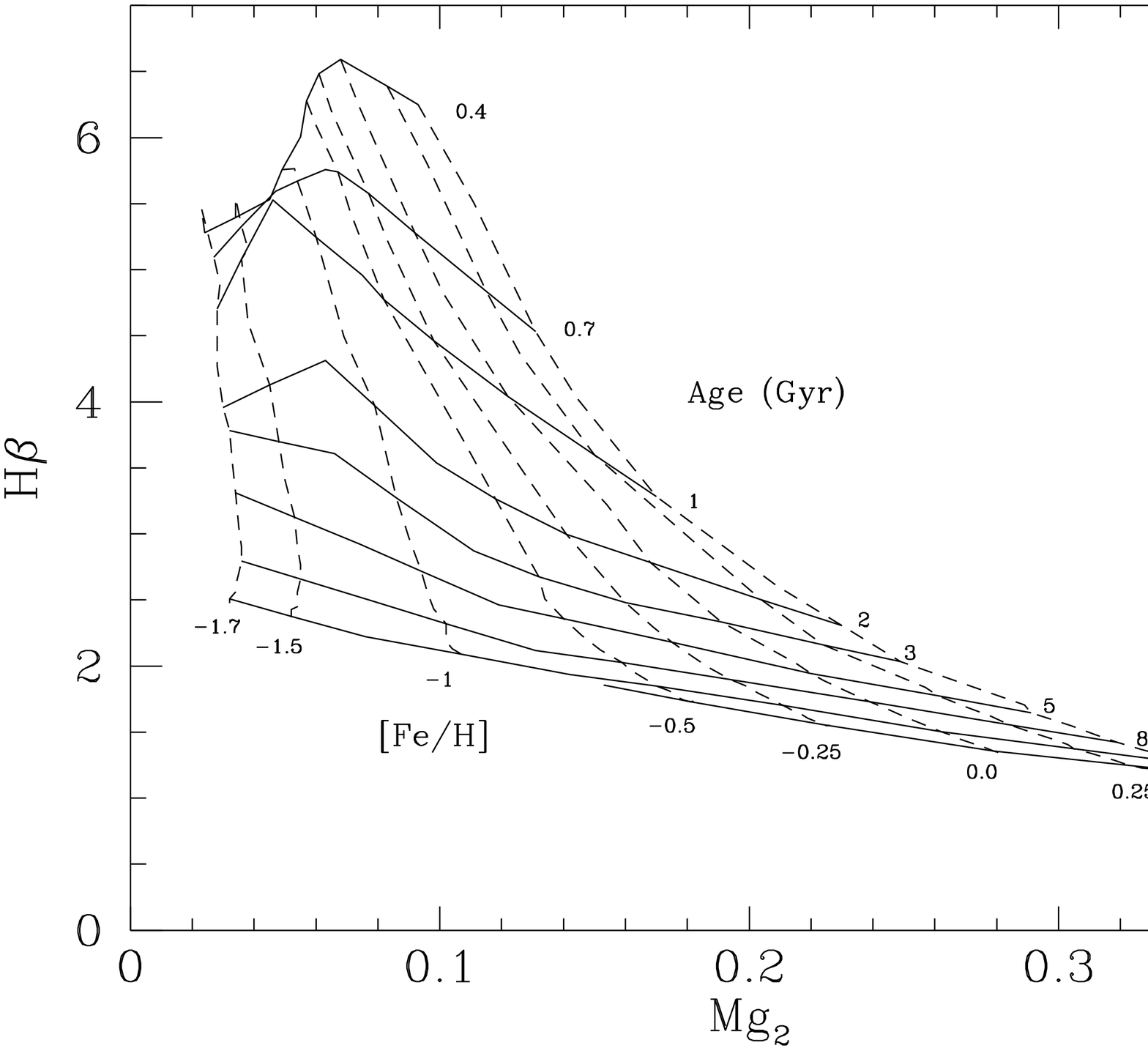,angle=0,width=2.3in}}
\noindent{\scriptsize
\addtolength{\baselineskip}{-3pt} 
\hspace*{0.1cm} Fig.~5.\ Model grids as in Fig.~4. For clarity only a 
subset of the model set is displayed (\S3). Note that the age resolution
decreases with age and that the
model grid is practically degenerate at ages greater than $\sim 9-10$ Gyr.
\addtolength{\baselineskip}{3pt}
}

A first, visual inspection of these diagrams allows us to draw some qualitative
conclusions and reveals that:

1) Most of the giant galaxies lie in the high-metallicity and old age corner 
of the model grid ([Fe/H]$>-0.25, $$t>5$ Gyr).

2) Dwarf galaxies occupy a larger region than giants in the diagrams, 
covering the whole range of metallicity but mostly lying at [Fe/H]$<-0.25$. 
They have a broad range in ages, from very old to young ($<1$ Gyr), but the 
majority of them reside in a region where the models are older than 3 Gyr.
Their metallicity-index range overlaps with that of the Lick globular clusters 
(kindly provided by G.~Worthey ahead of publication, bottom
panels in Fig.~4), while their age-index range extends to ages
much lower than those of the globular clusters.

3) The errorbars on the indices of the giant galaxies are quite small
(e.g. the mean ${\sigma}_{\rm H\beta}=0.16$ and ${\sigma}_{\rm
Mg_2}=0.004$ for $R<15$, \, mean ${\sigma}_{\rm H\beta}=0.22$ and
${\sigma}_{\rm Mg_2}=0.006$ for $R<16.3$). For comparison, the mean
errors in the Lick/IDS sample of galaxies are ${\sigma}_{\rm
H\beta}=0.31$ and ${\sigma}_{\rm Mg_2}=0.009$.  On the contrary, the
index errors for fainter galaxies in our sample are large, with a mean
${\sigma}_{\rm H\beta}=0.68$ and ${\sigma}_{\rm Mg_2}=0.020$ for the
whole dwarf subsample ($R>16.3$). This reflects the variation of mean
signal-to-noise with magnitude. The mean S/N measured between
4800 and 5200 \AA $\,$ is 15.5, 13.8, 8.6 and 7.8 for
$R<15$, $15<R<16.3$, $16.3<R<18$ and $R>18$, respectively.
Large errors are a critical issue for the dwarf galaxies,
and allow us to obtain only broad general conclusions regarding
their ages and metallicities, as discussed in the next sections.

4) Quite a large number of datapoints (mostly dwarfs)
lie outside of the model grid. As explained in \S3, the grid
extends over the ranges [Fe/H]=-1.7/0.4 and t=0.4/20 Gyr.
The most notable group of outliers is located {\it below} the grid,
at very low Balmer indices. While some of these outliers
overlap with the grid if their errorbars are taken into account, 
many of them display Balmer indices which are far too low compared with 
those of any model. 

The same problem has been observed in the spectra of a number of
globular clusters (Rose 1994, Jones 1996, Cohen, Blakeslee \& Ryzhov
1998, Vazdekis \& Arimoto 1999) and it is also apparent in the bottom
panels of Fig.~4 showing the results of the Lick globular clusters.
The best-studied case is the metal-rich globular cluster 47 Tuc, whose
high-resolution, high-signal-to-noise spectra yield very low Balmer
indices, implying an embarrassing spectroscopic age $>$ 20 Gyr (Gibson
et al. 1999; Vazdekis et al. 2001) which is at odds with the age
derived from the resolved color-magnitude diagram (9-14 Gyr). A number
of early-type galaxies with $\rm H\beta$ indices below the model grid
have also been found by Kuntschner et al. (2001). Nebular emission
appears unlikely to be the source of the discrepancy (Gibson et
al. 1999; Vazdekis et al. 2001, Kuntschner et al. 2001); in our case,
if emission were the main cause of the low Balmer lines we would
expect this phenomenon to be much more prominent in $\rm H\beta$
(right panel in Fig.~4) than in $\rm H\gamma$ (left panel in Fig.~4)
and this is not the case.  It has been suggested that the reason might
be a problem in the zero point of the models which probably fail to
reach sufficiently low Balmer line strengths at old ages due to the
fact that the main sequence turnoff (TO) is cooler than current models
assume (Vazdekis et al. 2001; also Alexandre Vazdekis (1999) and
Scott Trager (2000), private communications).  In fact, there are
suggestions from recent developments of stellar evolution theory that
improving the input physics of the models (i.e. including the effects
of diffusion and the Coulomb correction to the equation of state)
produces a decrease of the turnoff temperature (Lee, Yoon \& Lee
2000 and references therein, Vazdekis et al. 2001). If this is the
cause of the mismatch between the models and the low Balmer indices
observed and if the metallicity estimate is sufficiently unaffected by
the TO temperature as argued by Vazdekis et al. (2001), then the
objects lying below the model grids in Fig.~4 are old stellar systems
(luminosity-weighted age $> 9$ Gyr) whose approximate metallicity can
be inferred from their metallicity-index. Adopting this as our working
hypothesis, we derived relative ages and metallicities as explained
below.

\subsection{Ages and metallicities}

The age-index versus Z-index plots of Fig.~4 were used to derive the
values of the luminosity-weighted stellar ages and metallicities.
Given a chosen pair of indices, we computed the age and abundance of
each datapoint lying within the model grid interpolating from the
closest model points. In the case of the low-Balmer points below the
grid, the abundance was estimated extrapolating the iso-Z model lines
(dashed lines in Fig.~4) and an old age, arbitrarily recorded as 30
Gyr to keep track of these objects, was assigned. Similarly, the age
of the few points lying to the right(left) of the model grid was
estimated extrapolating the iso-age model lines and an arbitrarily
high(low) metallicity was recorded ([Fe/H]=+3(-3)).  Errors were found
by perturbing the index values with their errorbar and recomputing the
age and Z.\footnote{We remind the reader that in this paper the
metallicity ``Z'' is $\rm Z=[M/H]=log(M/H)-log(M/H)_{\odot}$ where M
represents the ``metal'' content. In the case of solar abundance
ratios, [Fe/H]=[Mg/H]=[any other heavy element/H]=[$Z$/H] where $Z$
indicates the sum of all elements heavier than helium. Since variation
in abundance ratios might be present within our sample, it should be
kept in mind that Z indicates the total metallic content as determined
either by the magnesium or the iron indices.}

The luminosity-weighted metallicities derived from the two pairs of
indices are shown as a function of galaxy luminosity in Fig.~6.  In
both cases the mean metallicity increases towards brighter magnitudes.
A straight line fit to the Z($H\beta$) estimate (based on the $\rm
H\beta$/$\rm Mg_2$ model grid) yields somewhat lower metallicity
values at faint magnitudes than the $\rm {H\gamma}_F$-based estimate,
possibly due to varying abundance ratios.  The parameters of the best
fit Z-R relations are given in Table~2.

\hbox{~}
\centerline{\hspace{0.7in}\psfig{file=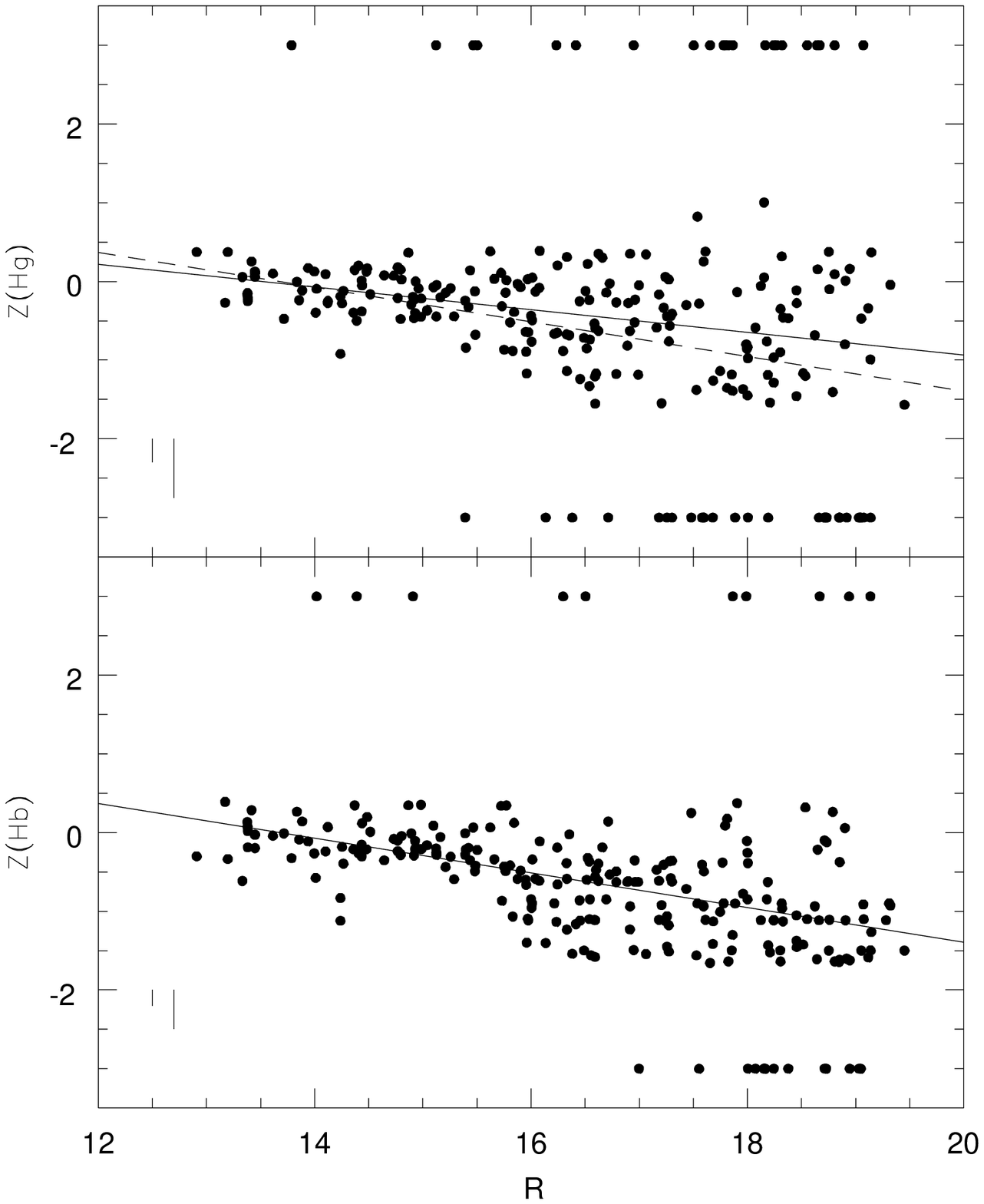,angle=0,width=5.0in}}
\noindent{\scriptsize
\addtolength{\baselineskip}{-3pt} 
\hspace*{0.1cm} Fig.~6.\ Metallicities derived from the comparison
with the model grids $\rm {H\gamma}_F$/$<$Fe$>$ (top panel)
and $\rm H\beta$/$\rm Mg_2$ (bottom panel) are plotted against
the R band magnitude. Best fit relations are shown as solid lines. 
The best-fit line of the bottom panel (Z(Hb) vs. R) is shown as
the long dashed line in the top panel for comparison. 
The typical errorbars for giants and dwarfs are found in the left bottom 
corner of each panel. Note that, while 
the mean Z(Hg) shows a steady decline with R, the mean Z(Hb)-R
relation appears to be almost flat for giants and drops to lower values for
the dwarfs, with a break at around R=16. It is unclear whether this effect
is real, or is an artifact due to the combination of the errors and the 
shape of the model grid.
\addtolength{\baselineskip}{3pt}
}

Hence, a correlation between metallicity and luminosity exists for
galaxies in our sample.  The spread in metallicity is much larger at
faint magnitudes, as clearly shown by the Z distributions presented in
Fig.~8. In this figure the arrows represent the mean metallicity in
each magnitude bin of Coma1 (solid arrow) and Coma3 (dotted arrow)
galaxies. It is beyond the scope of this paper to discuss the
variation of the galactic properties as a function of the position in
the cluster, but we note here that Coma1 galaxies tend to be on
average more metal-rich than Coma3 galaxies in any given magnitude
interval.  This finding could be consistent with the result of Secker
(1996), who detected a color gradient in the projected radial
distribution of Coma dwarf ellipticals which he interpreted as an
increase in metallicity closer to the cluster centre.

In contrast to metallicity, age does not correlate with R in a
simple way (Fig.~7). However, we are going to show that there are
systematic trends in the proportion of galaxies of different ages with
luminosity.  First of all, in Fig.~7 it is possible to notice two
voids in the left bottom corner and in the right top corner of the
diagram, corresponding to a lack of young bright galaxies and of old
faint galaxies. The top right void disappears if one takes into
account the numerous points lying at age=30 Gyr. The corresponding age
distributions are presented in Fig.~8 for Coma1 and Coma3 galaxies in
each magnitude bin. The mean ages found from $\rm {H\gamma}_F$ are
typically 1 Gyr younger than the corresponding mean $\rm H\beta$ ages
of the same magnitude bin. In Coma1, the mean age slightly decreases
towards fainter galaxies, ranging from 9.3 ($R<15$) to 7.6 Gyr ($R>18$)
for $\rm{H\gamma}_F$ and from 10.0 to 8.4 Gyr for $\rm H\beta$.
There is no obvious trend in the mean age of Coma3 galaxies for
different magnitude bins, possibly due to small number statistics, and
we note that bright galaxies in Coma3 are on average younger than in
Coma1.

\hbox{~}
\centerline{\hspace{0.7in}\psfig{file=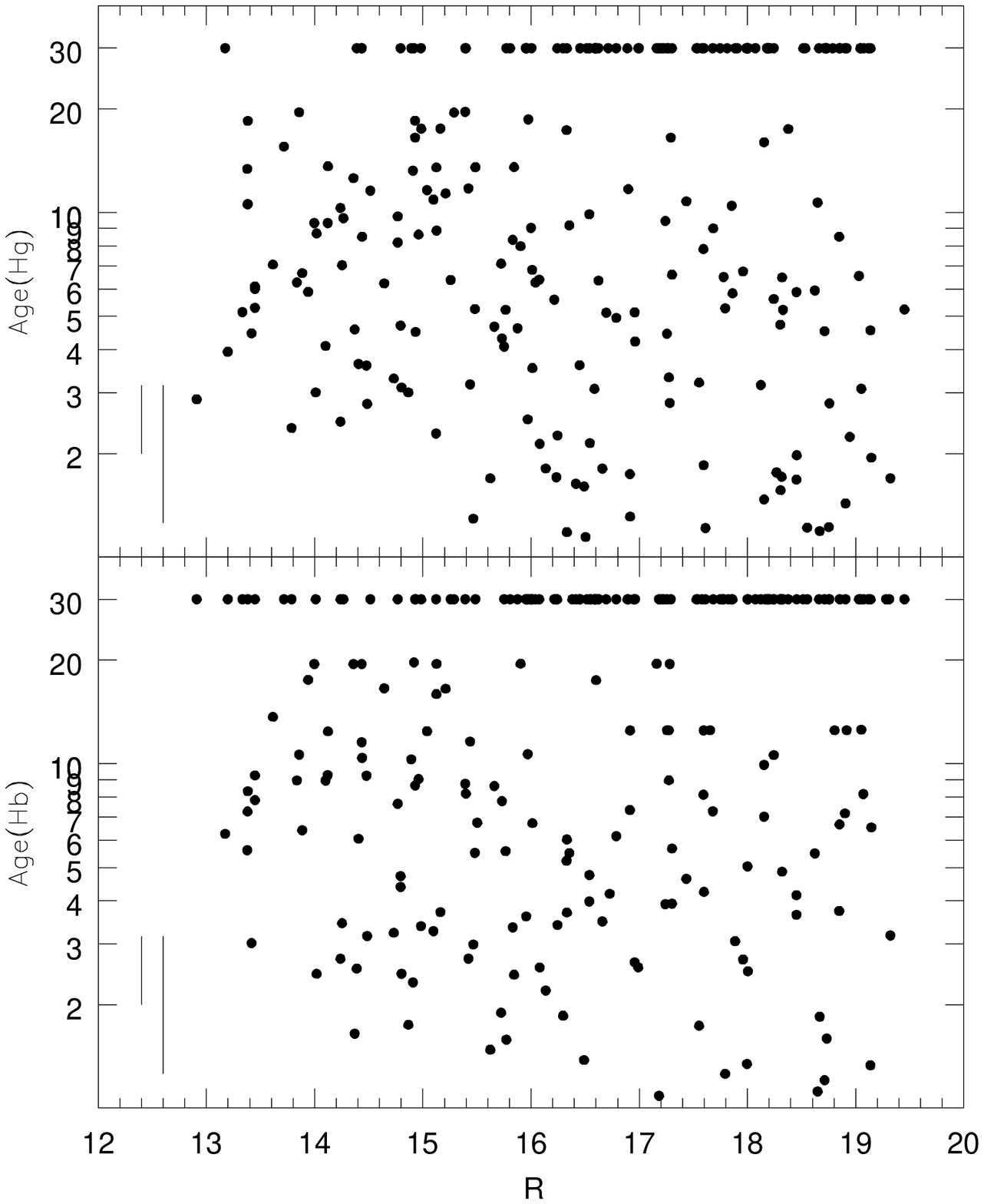,angle=0,width=5.0in}}
\noindent{\scriptsize
\addtolength{\baselineskip}{-3pt} 
\hspace*{0.1cm} Fig.~7.\ 
Ages derived from the model grids $\rm{H\gamma}-F$/$<$Fe$>$ (top 
panel) and $\rm H\beta$/$\rm Mg_2$ (bottom panel) plotted against R 
magnitude. The typical
errorbars of dwarfs and giants are shown in the left bottom corner.
The {\it length} of the errors is approximately constant 
throughout the diagram and corresponds to an uncertainty
of 1-2 Gyr at ages $<4$ and to several Gyrs at older ages.
This is because the age resolution decreases roughly logarithmically for 
older populations. 
\addtolength{\baselineskip}{3pt}
}

Since the errorbars of the indices of faint galaxies are too large to
look for subtle differences (such as for example 1 Gyr at old ages or
0.1 dex in Z), we have grouped all galaxies into three broad age and
metallicity classes: old (age$>9$ Gyr), intermediate ($3<\rm age<9$
Gyr), young (age$<3$ Gyr) and metal-rich (Z $>-0.15$), intermediate
$-1<\rm Z <-0.15$, and metal-poor Z $< -1$. The ranges chosen for these
classes are broader than the uncertainty of even the faintest
galaxies, thus it is safe to draw conclusions on the basis of this
classification. Repeated spectra of 9 dwarf galaxies confirm the 
robustness of this broad division in classes and show that this is
actually a conservative choice.

The proportions of young, intermediate-age and old galaxies among
Coma1 galaxies of each magnitude bin are shown in
Fig.~9. Interestingly, the fraction of old galaxies is approximately
constant within the errors regardless of the magnitude, while the
fraction of young galaxies increases towards {\it fainter} magnitudes
and the fraction of intermediate-age galaxies increases towards {\it
brighter} magnitudes.\footnote{The trends might be similar for Coma3
galaxies but for them the Poissonian errorbars are too large to reach
any firm conclusions.}  The absolute values of these fractions depend
on which index-index diagram is considered (compare left and right
panels in Fig.~9), but the trends are the same in both cases.  This is
a striking result, if one considers that a look-back time of 3 Gyr
corresponds to z$\sim$0.35 ($H_0=65 \, \rm km \, s^{-1} \, Mpc^{-1}$,
$q_0=0.05$), 5 Gyr to z$\sim$0.65 and 9 Gyr to z$\sim$2. If the
luminosity-weighted age can be considered to be approximately
representative of the epoch of the last episode of significant star
formation\footnote{In general, the luminosity-weighted age will be an
upper limit to the time elapsed since the last star formation
activity, because it is diluted by the light of all previous stellar
populations.} (Trager et al. 2000b), taking
Fig.~9 at face value implies that among the galaxies without active
star formation at z=0 in the central region of Coma:

a) about 50 to 60\% of them did not have significant star formation
activity since $z\sim 2$ (old galaxies). Remarkably, these figures are
valid for both giants and dwarfs over 6 magnitudes, regardless of the
luminosity. Conversely, this means that about 40-50\% of the
present-day non-active cluster population -- at all magnitudes -- {\it
did} have significant star formation at redshifts below 2.

b) the fraction of (present-day) {\it luminous} galaxies that have had
significant star formation in the intermediate-redshift regime between
z=2 and z=0.35 (intermediate-age galaxies) was higher than the
fraction of (present-day) {\it dwarfs} that were active at that epoch;

c) at low-redshift ($z<0.35$), the star formation activity has involved
a higher proportion of faint galaxies than bright ones.

These fractions are found simply considering the ages derived from the
$\rm H\beta-Mg_2$ diagram and ignoring the effects of the errors on
the index measurements.  The errors are relatively small and will not
significantly affect the proportions of young, intermediate-age and
old galaxies in the two brightest magnitude bins of Fig.~9 (giant
galaxies, see \S4.2), but they are large and can potentially have
systematic effects on the fractions of dwarfs (faintest two bins in
Fig.~9).  To assess how relevant this can be, we have recomputed the
fractions of dwarfs excluding from the ``old'' and ``young'' classes
all those galaxies whose errorbars are large enough to possibly move
them into a different age class. In this way, we derive a ``minimum''
fraction of ``secure'' old dwarfs $\sim$ 30\% and of ``secure'' young
dwarfs $\sim$ 20\%. Assuming the error estimates are realistic, these 
are lower limits for the proportions of old and young dwarfs.  The
fraction in the intermediate-age class is obviously the most uncertain
one, and could represent as much as 45\% of the dwarf population, in
the most pessimistic case that {\it all} those galaxies with large
errorbars classified as old and young would instead fall into the
intermediate class.  In conclusion, it is important to keep in mind
that a realistic uncertainty on the fractions for the two faintest
bins presented in Fig.~9 is surely larger than the Poissonian
errorbars shown.  However, even taking the errors into account, 
two results appear to be solid: 1) the fraction
of young dwarfs is higher ($\geq 20$\%) than the 
fraction of young giants, and 2) a significant fraction of the dwarfs (at
least $\sim$1/3 of them) have old luminosity weighted ages.

It is also worth recalling that the results discussed above 
refer to the central region of each
galaxy (\S2). If significant radial age gradients are present within a
galaxy, these results cannot be considered representative of the whole
star formation history.  Since we cannot exclude that some SF activity
occurred in the outer regions of the ``old'' galaxies, the values for
young and intermediate-age galaxies of Fig.~9 are lower limits to the
fractions of galaxies that experienced star formation at $z<0.35$ and
$0.35<z<2$.

These results provide some constraints for theories of galaxy
evolution in clusters and especially dwarf galaxy evolution. They seem
to point to a ``down-sizing effect'', suggesting that the last star
formation activity (possibly related to the epoch of accretion onto
the cluster) occurs on average at lower redshifts for progressively fainter
galaxies. This would imply that the ``typical mass'' of a star-forming
galaxy in Coma decreases towards lower redshifts.
Obviously, these findings will need to be compared with the
accretion history of clusters predicted by hierarchical models. If
these conclusions are representative of the evolutionary history of
rich clusters in general, they represent a relatively
precise local (z=0) assessment of the star formation history of the
cluster galactic population to be compared with distant cluster
studies. In a qualitative manner, these results agree with the
photometric and spectroscopic studies of the Butcher-Oemler effect at
intermediate redshifts which are mostly limited to luminous galaxies
and find current or recently halted star formation activity in a large
fraction of cluster galaxies at z=0.3-0.6 (Butcher \& Oemler 1978,
1984; Dressler \& Gunn 1982, 1983, 1992; Couch \& Sharples 1987;
Barger et al. 1996; Dressler et al. 1999; Poggianti et
al. 1999). Totally passive (k-type) galaxies with no recent star
formation (no SF at redshifts below 0.8 in the cosmology adopted here)
were found to be typically 50\% of the cluster population of 10
clusters at $z\sim 0.5$ (Poggianti et al. 1999).

Finally, it is extremely interesting that regardless of the variation
in the epoch of the last SF episode in galaxies in this sample
(Fig.~8), a general relation between the metallicity and the
luminosity exists (Fig.~6), although with a significant
scatter.\footnote{Another example of this are the dwarf galaxies of
the Local Group, where galaxies with very diverse star formation
histories follow the same global relations between absolute magnitude
and mean metallicity (Caldwell 1999, Grebel \& Guhathakurta
1999).}  This could be suggesting that the relatively recent star
formation we observe is due to secondary episodes that do not affect
the chemical enrichment history of the galaxy sufficiently to demolish
the luminosity-weighted Z versus luminosity relation which was likely
established at earlier epochs.

\hbox{~}
\centerline{\psfig{file=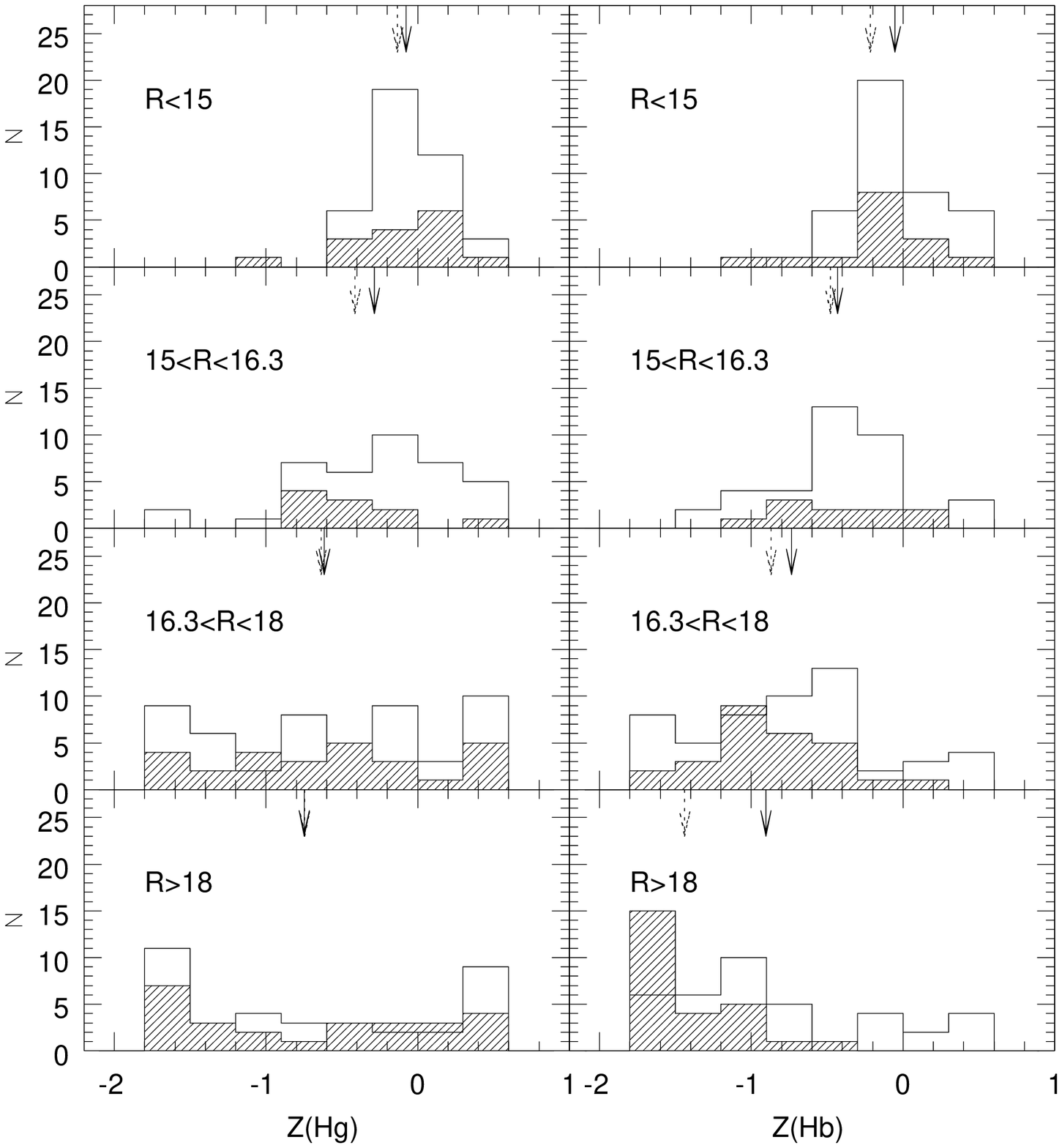,angle=0,width=4.5in}\hspace{-0.6in}
\psfig{file=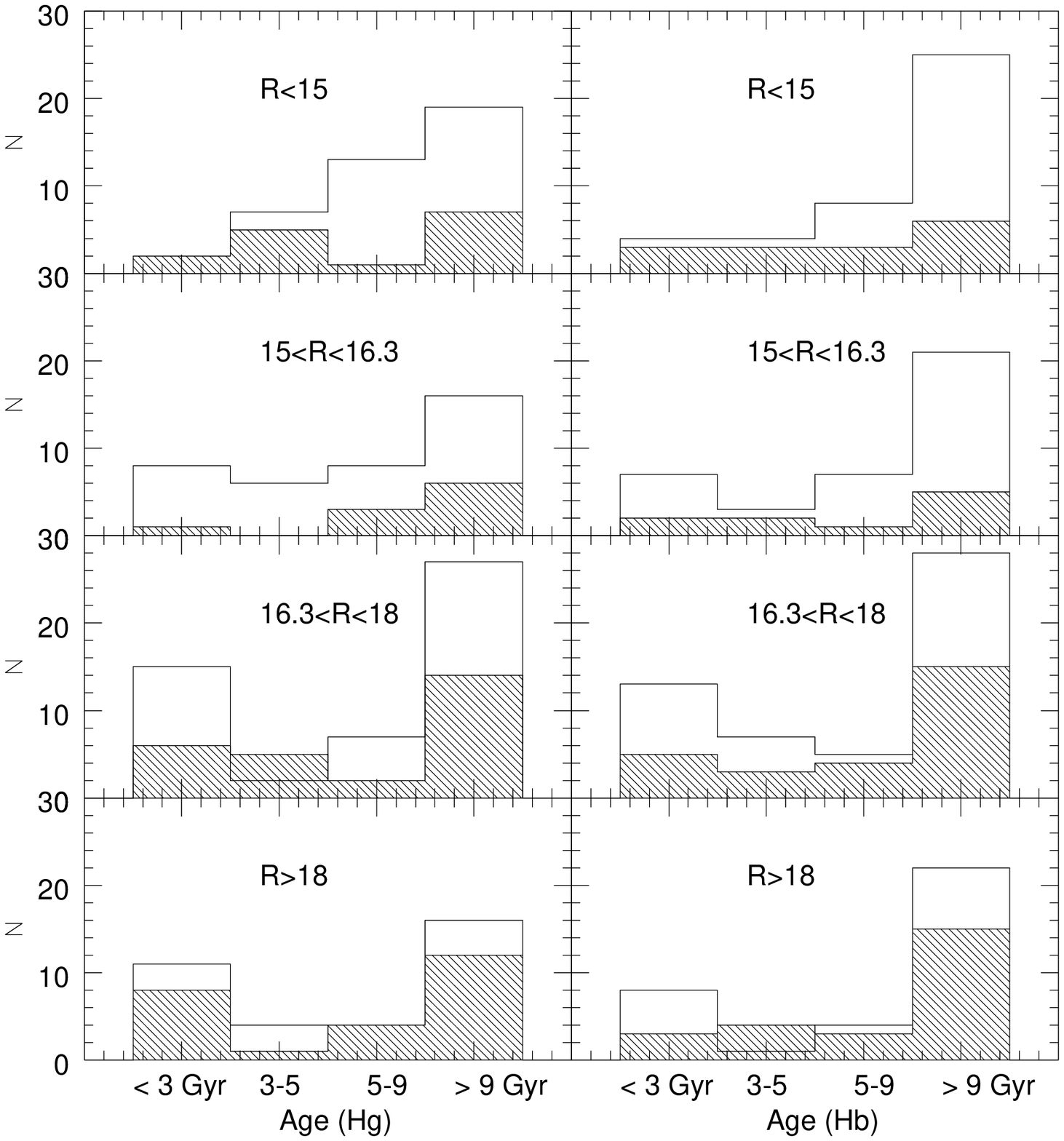,angle=0,width=4.5in}}
\noindent{\scriptsize
\addtolength{\baselineskip}{-3pt} 
\hspace*{0.1cm} Fig.~8\ Distributions of Z (left) and ages (right)
for galaxies in the four magnitude bins. Empty and shaded
histograms refer to Coma1 and Coma3 galaxies respectively. 
The arrows in the left panel indicate the mean value of Z
in Coma1 (solid arrow) and Coma3 (dotted arrow) in each magnitude
bin. In these diagrams the outliers (Z=-3,+3) have been
assigned to Z=0.5 and Z=-1.8 respectively.
The width of the bins has been chosen to be comparable to the uncertainties
in age and Z for the faintest galaxies.
\addtolength{\baselineskip}{3pt}
}

\hbox{~}
\vspace{-2in}
\centerline{\psfig{file=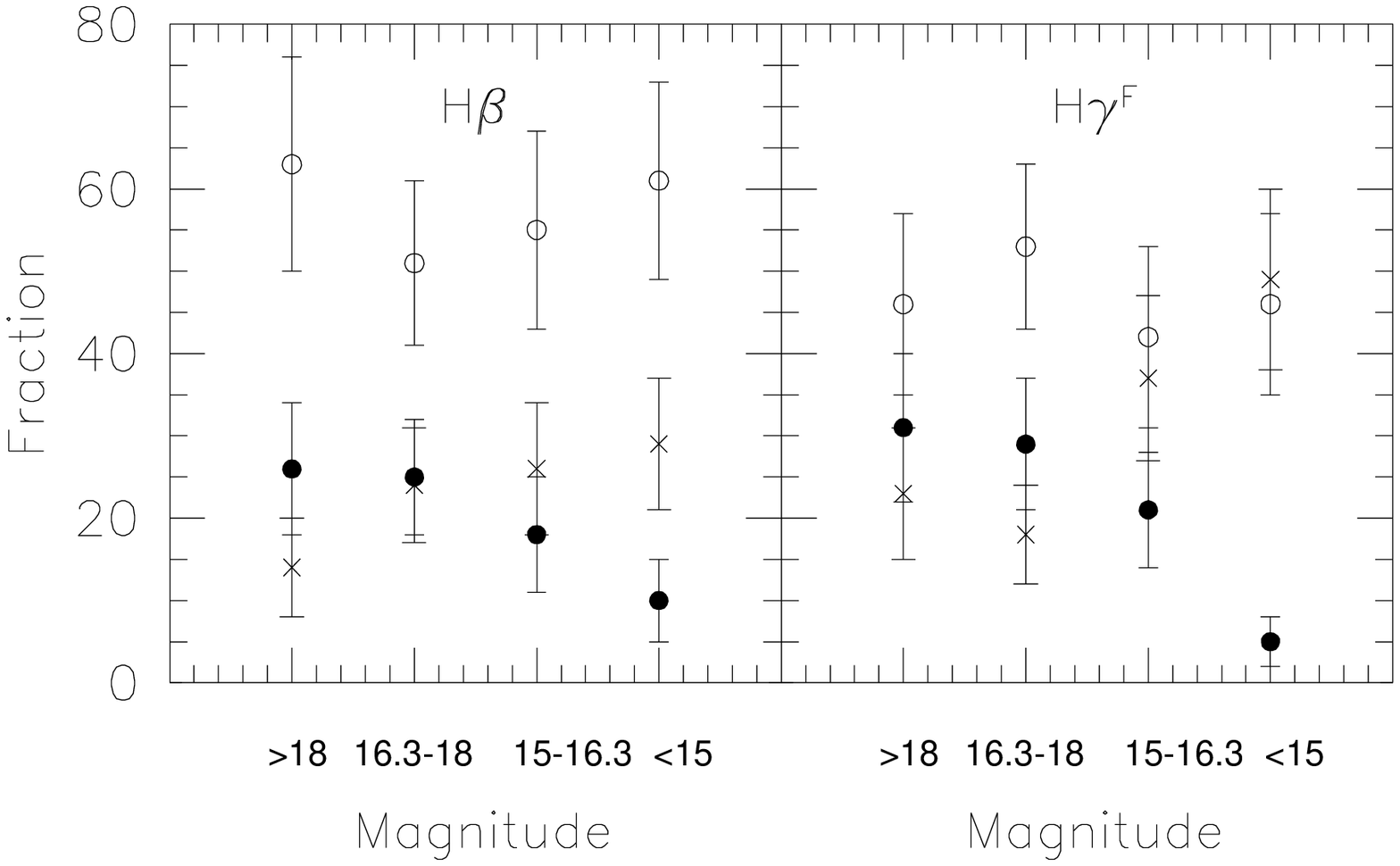,angle=0,width=6.0in}}
\noindent{\scriptsize
\addtolength{\baselineskip}{-3pt} 
\hspace*{0.1cm} Fig.~9.\ Fraction of young (filled dots, age $<3$ Gyr), 
intermediate-age (crosses, 3 to 9 Gyr) and old (empty dots, age $>9$ Gyr) 
Coma1 galaxies within each magnitude bin as
derived from the $\rm H\beta$/$\rm Mg_2$ diagram (left) and the 
$\rm{H\gamma}_F$/$<$Fe$>$ diagram (right). The errorbars
are Poissonian.
\addtolength{\baselineskip}{3pt}
}

\subsection{The composite nature of the faint galaxy population}

We will now discuss the main properties of the faint galaxies
in our sample, especially concentrating on the young ones.
We stress that, given the large errors on individual measurements,
conclusions for faint galaxies are reliable only in a statistical sense,
not on a one-to-one basis. We have seen in \S4.3 that
at faint magnitudes ($R>16.3$), more than half of the galaxies
are old, $\sim$20\% have an intermediate age and $\sim$25\% are young.
We remind the reader once again that terms such as ``old'' and ``young'' refer
to the luminosity-weighted integrated age and that these
proportions are found having assumed that galaxies with very low Balmer
indices (those lying below the model grid in Fig.~4) are indeed old.
If the $\rm H\beta$ index in some of these galaxies is actually contaminated 
by emission, then these are ``young'' galaxies that have been misclassified.
For this reason and for the possible aperture effects discussed in \S4.3, 
a 25\% fraction of young galaxies should be considered as a lower limit.

\begin{table}
{\scriptsize
\begin{center}
\centerline{\sc Table 3}
\vspace{0.1cm}
\centerline{\sc Indices measured on the coadded spectra}
\vspace{0.3cm}
\begin{tabular}{lrrc}
\hline\hline
\noalign{\smallskip}
Index Bandpass & Metal-poor & Metal-rich & MR$>$MP ? \cr
\hline
\noalign{\smallskip}
$\rm CN_1$    & -0.123$\pm$0.017  & -0.054$\pm$0.015 & + \cr  
$\rm CN_2$    & -0.091$\pm$0.020  & -0.027$\pm$0.016 & + \cr
Ca4227        &  0.667$\pm$0.143  &  0.588$\pm$0.133 & - \cr
G4300         &  1.576$\pm$0.497  &  5.864$\pm$0.378 & + \cr
Fe4383        &  0.848$\pm$0.544  &  3.119$\pm$0.455 & + \cr
Ca4455        &  0.675$\pm$0.232  &  1.596$\pm$0.236 & + \cr
Fe4531        &  1.377$\pm$0.329  &  2.366$\pm$0.221 & + \cr
$\rm C_2$4668 &  1.103$\pm$0.574  &  1.222$\pm$0.652 & + \cr
Fe5015        &  3.581$\pm$0.827  &  4.128$\pm$0.227 & + \cr
$\rm Mg_1$    &  0.010$\pm$0.018  &  0.049$\pm$0.007 & + \cr
$\rm Mg_2$    &  0.003$\pm$0.007  &  0.160$\pm$0.004 & + \cr
Mg$b$         &  0.790$\pm$0.191  &  2.735$\pm$0.123 & + \cr
Fe5270        &  2.162$\pm$0.358  &  1.491$\pm$0.063 & - \cr
Fe5335        &  1.600$\pm$0.228  &  1.984$\pm$0.098 & + \cr
Na D          & -0.507$\pm$0.318  &  1.325$\pm$0.125 & + \cr
\noalign{\smallskip}
\noalign{\hrule}
\noalign{\smallskip}
\multispan4{Sky-lines contamination (OI5577,NaD5890/6,OI6302)
affects Fe5406,\hfil}\cr 
\multispan4{5709,5782,TiO1 and TiO2, 
which have been omitted from this table. \hfil}\cr
\multispan4{Errors have been computed from the index differences 
of various \hfil}\cr 
\multispan4{``Monte-carlo'' combinations of metal-poor and metal-rich
spectra. \hfil}\cr
\noalign{\smallskip}
\end{tabular}
\end{center}
}
\vspace*{-0.8cm}
\end{table}

The metallicity distribution of the young, intermediate-age and old
faint galaxies are shown in Fig.~10 (top panels) together with the
distribution of the residuals (bottom panels) from the best fit of the
relation Z($\rm H\beta$)-R of Fig.~6.  Old, faint galaxies have
metallicities lower than -0.15, and many of them (the majority at the
faintest magnitudes) have $Z<-1$. None of them is metal-rich and their
distribution of residuals has a Gaussian shape centered at
Z(meas.)-Z(fit)=-0.15. This is what one expects if the best-fit
relation is mostly driven by these old metal-poor galaxies and it is
only slightly modified by the existence of more metal-rich (young and
intermediate-galaxies) galaxies in the same magnitude range.

\hbox{~}
\centerline{\hspace{2.4in}\psfig{file=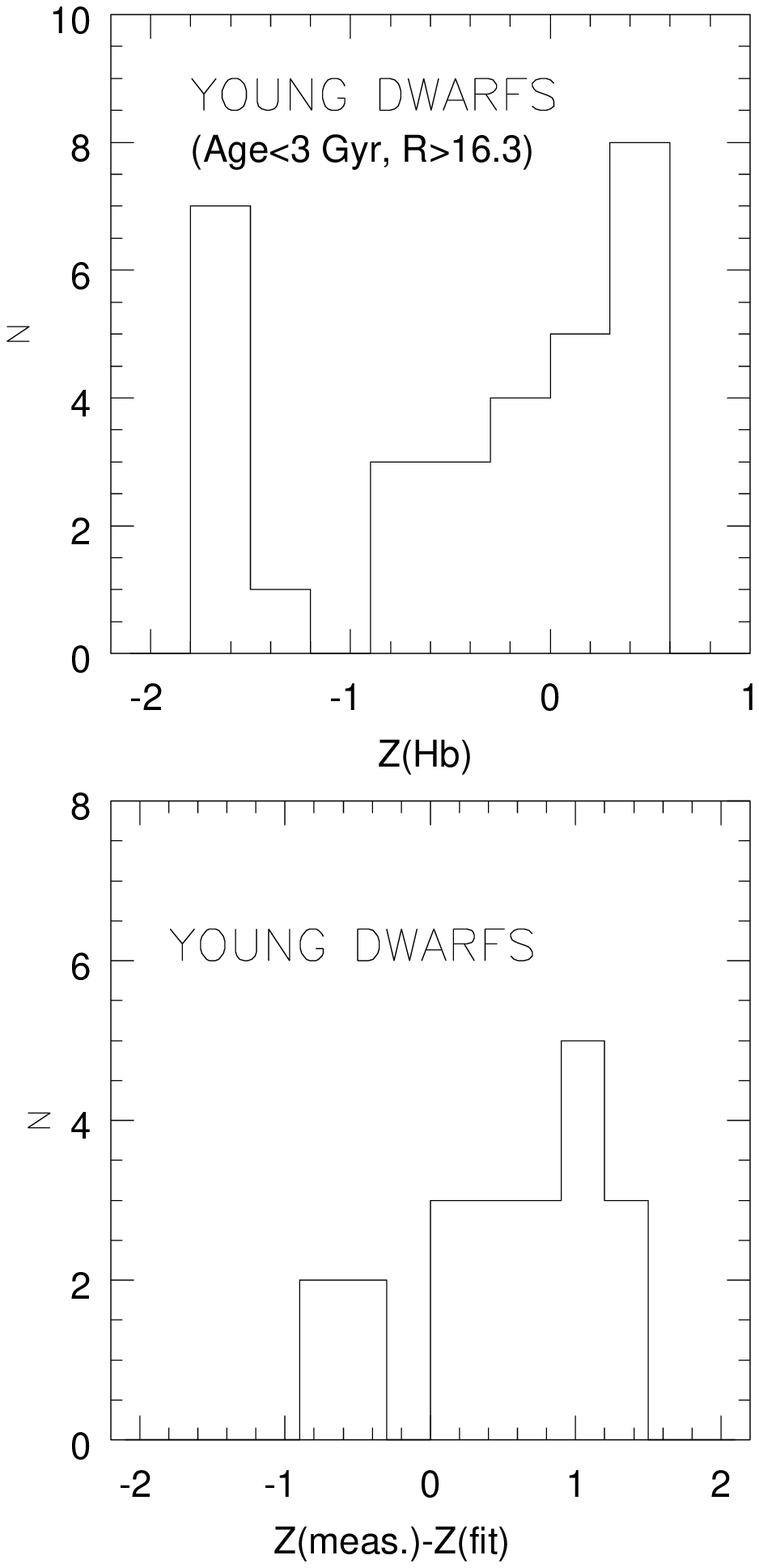,angle=0,width=5.0in}\hspace{-2.6in}\psfig{file=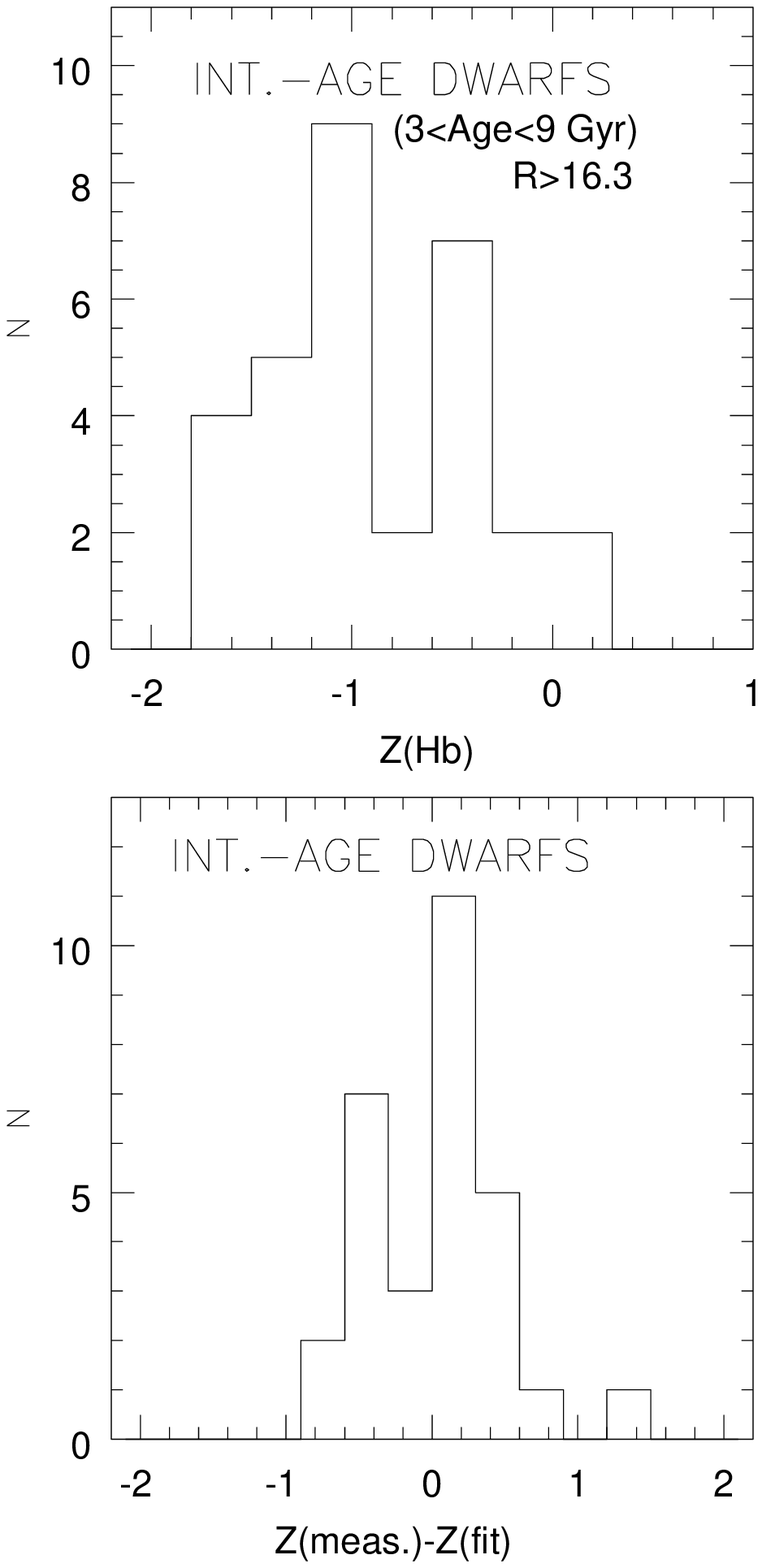,angle=0,width=5.0in}\hspace{-2.6in}\psfig{file=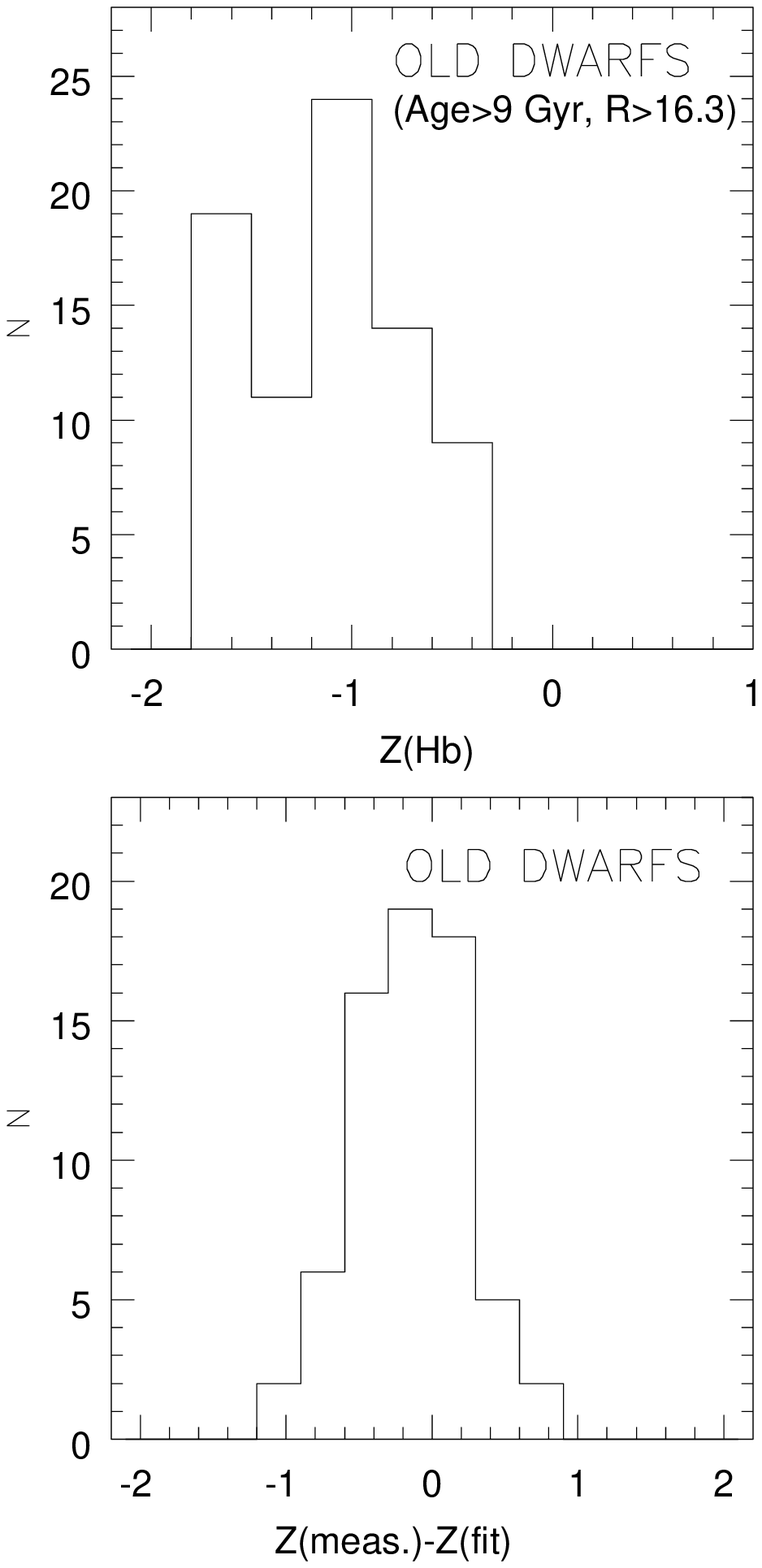,angle=0,width=5.0in}}
\noindent{\scriptsize
\addtolength{\baselineskip}{-3pt} 
\hspace*{0.1cm} Fig.~10.\ Metallicity distribution (top panels) and
distribution of the residuals from the best-fit line of the relation
Z($\rm H\beta$)-R of Fig.~6 (bottom panels).  Distributions for young
faint galaxies are shown in the left panels, for intermediate-age
faint galaxies in the middle panels and for old faint galaxies in the
right panels.  Galaxies with Z=-3/+3 have been assigned to the bins
Z=-1.65 and Z=+1.65 respectively, and have been omitted in the plot of
the residuals. 
\addtolength{\baselineskip}{3pt} }

In contrast, the young faint galaxies show a bimodal metallicity
distribution, a group of them being metal-rich and another group being
metal-poor.  The metal-rich young dwarfs are generally ``too
metal-rich'' for their luminosity and the metal-poor ones are ``too
metal-poor'', as shown by the residuals plot.  
Being among the faintest galaxies with the lowest
signal-to-noise spectra, especially below R=18, this result could be
affected by the large errorbars in the indices, although a significant
fraction of these galaxies are confirmed to be young and metal-rich
also by the $\rm {H\gamma}_F$ analysis. 

\hbox{~}
\centerline{\psfig{file=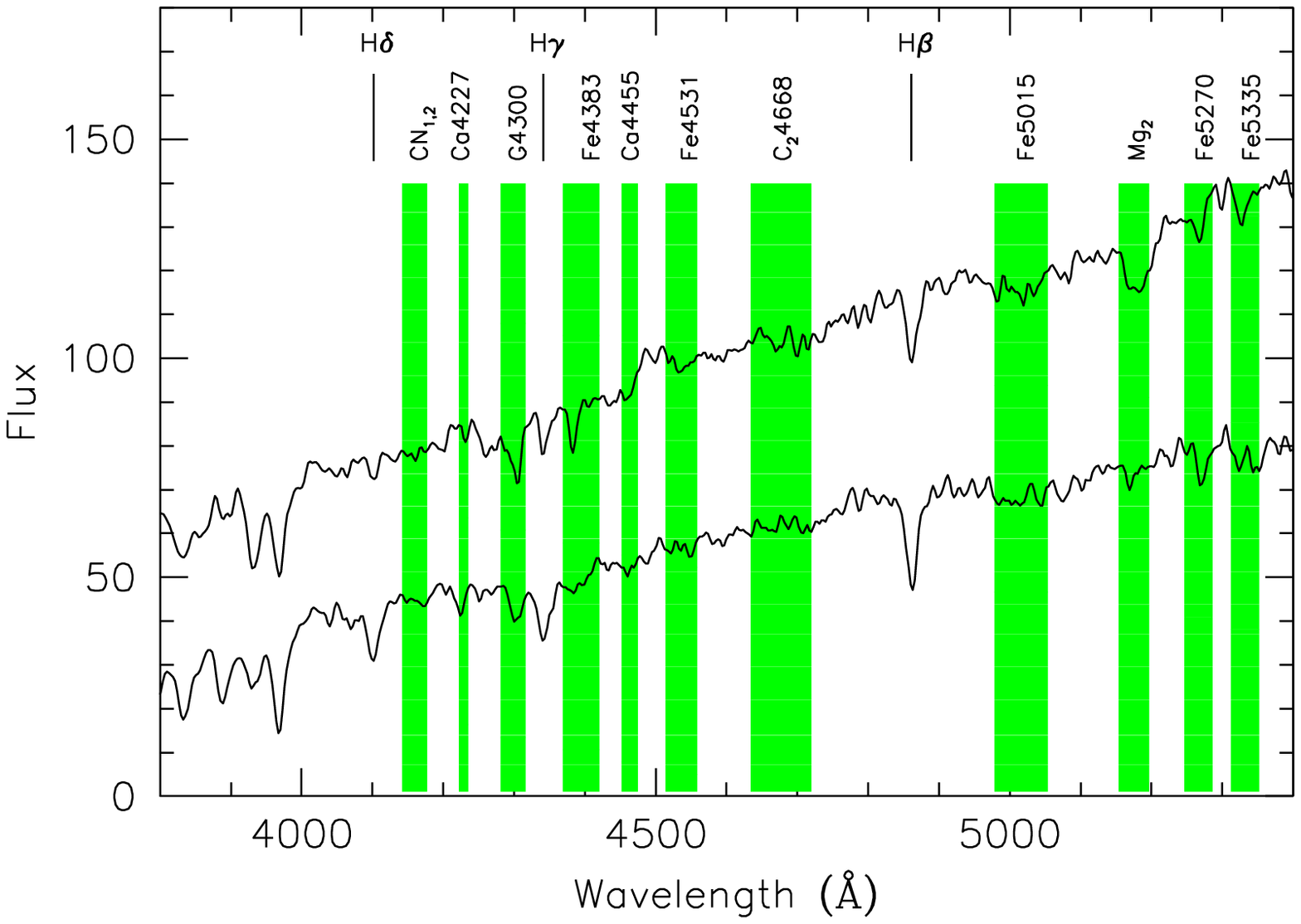,angle=0,width=5.0in}}
\vspace{1.0in}
\noindent{\scriptsize
\addtolength{\baselineskip}{-3pt} 
\hspace*{0.1cm} Fig.~11.\ Average spectra of metal-poor (lower) and metal-rich
(upper) faint young galaxies as found by coadding the spectra
of 7 metal-poor and 16 metal-rich dwarfs respectively 
(Z$<-1$ and Z$>-0.25$ in the top left panel of Fig.~10).
{\it All but one of the metal-rich galaxies are found in Coma1, while
4 out of 7 metal-poor galaxies are in Coma3.}
\addtolength{\baselineskip}{3pt}
}

In order to check the reality of these two metallicity
classes, we have coadded the spectra of metal-poor and metal-rich
young faint galaxies separately. The higher S/N, coadded spectra are
shown in Fig.~11 and the indices measured from them are presented in
Table~3. Besides the magnesium indices (on which the metal-rich vs
metal-poor division is based), most of the other
metallicity-sensitive indices are indeed much stronger in the
average spectrum of the metal-rich young dwarfs as compared to the
metal-poor coadded spectrum.  This strongly supports the 
hypothesis of a metallicity bimodality based
on the analysis of the individual lower S/N spectra (Fig.~10).
However, we believe that higher signal-to-noise spectra of
these galaxies will be necessary to draw definite conclusions.  If
confirmed, this dichotomy could point to two different formation
scenarios for the faint young galaxies.

Galaxies which have been recently accreted into the cluster and have
suffered from harassment (Moore et al. 1996, 1998), having lost a
significant fraction of their stars, are expected to be young and
unusually metal-rich for their post-harassment luminosity. Regarding
this, it is interesting to note that only one out of the fifteen
metal-rich (residual$>0$), 
young dwarfs is found in the Coma3 field, while all the
others are Coma1 galaxies.  Another viable mechanism to form faint
galaxies lying above the standard metallicity-luminosity relation is
the formation of tidal dwarfs created from the tidal debris of
collisions between metal-rich, massive galaxies (Duc \& Mirabel 1999).

Alternatively, the existence of metal-rich Balmer-strong galaxies
could be related to late evolutionary phases of {\it old} metal-rich
populations (such as AGB-manqu\'e stars) which are believed to be
responsible for the UV-upturn below 2000 \AA $\,$ in luminous
ellipticals (Greggio \& Renzini 1990).  Since this effect is not
included in the stellar evolutionary models employed
here\footnote{This effect must not be confused with the old {\it
metal-poor} populations with a blue Horizontal Branch that are
discussed in \S3 and are already included in our models.}, in
principle we cannot exclude that the Balmer-strong $\rm Mg_2$-strong
spectra of the dwarf galaxies which we classify as young and
metal-rich are instead {\it old} populations whose $\rm H\beta$ index
is enhanced due to old stars with metallicity much greater than
solar. However, on the basis of the current models, even extremely
high metallicities are not expected to be able to explain the $\rm
H\beta$ values observed in our ``young'' metal-rich dwarfs: models of
old SSPs (ages $>3$ Gyr) with $Z=1$ including hot-HB, AGB-manque' and
Post-AGB stars only reach $\rm H\beta=3$ (Bressan, Chiosi \& Tantalo
1996), while we observe $\rm H\beta=3-6$.

Finally, the metallicity distribution of the intermediate-age dwarfs
is somewhat intermediate between that of the old and that of the young
galaxies, with a broad metallicity range and possibly a composite
population as well. Overall, the distributions presented in Fig.~10
testify that the mean metallicity {\it increases} from old to
progressively younger galaxies. The existence and reality of such an
age-metallicity anti-correlation for galaxies of all luminosities are
the subject of the next section.

\subsection{The age-metallicity anticorrelation at all magnitudes}

Galaxies in our sample follow a general trend linking age and metallicity:
as displayed in Fig.~12, within a given magnitude bin younger galaxies
tend to be slightly {\it more metal-rich} than older galaxies.
The effect is more evident for bright galaxies whose relations
show less scatter. 

Several previous studies found evidence for a similar age-metallicity
anti-correlation (Faber et al. 1995, 
Worthey et al. 1995, Gorgas et al. 1997, Trager 1997, Worthey
1997, Kuntschner \& Davies 1998, Trager et al. 1998, Colless et
al. 1999, Ferreras, Charlot \& Silk 1999, Jorgensen 1999, Saglia et
al. 1999, Terlevich et al. 1999, Longhetti et al. 2000, Trager et
al. 2000b, Kuntschner 2000, Kuntschner et al. 2001, Rakos et al.
2001) and it has been suggested that this ``conspiracy'' could
contribute to the tightness of the global relations (e.g. the
index-magnitude relations) such that deviations due to age can be
counter-balanced by metallicity effects (e.g. Faber et al. 1995,
Worthey et al. 1995, Trager 1997, Jorgensen 1999, Trager et al. 2000b).

The reality of such an age-Z anti-correlation is hard to establish,
due to the fact that errors in the Z-index versus age-index diagram
are correlated: as shown by Trager et al. (2000b) and Kuntschner et
al. (2001), an error in a Z-index, for example, produces an error in
both the age and the metallicity estimate, in the sense that if the
age is underestimated then the metallicity is overestimated, and
vice-versa. As a result, a spurious anti-correlation between age and
metallicity is produced.  Index measurements with small errors
(i.e. high S/N spectra) are required to feel confident that the
anticorrelation is not just due to correlated errors (Trager et
al. 2000b); for the youngest five galaxies in their sample, Kuntschner
et al. (2001) conclude that they are indeed more metal-rich than the
average galaxy and therefore at least in these cases the age-Z
anticorrelation is real.

Although we believe at the moment this is an open issue, we note that
if correlated errors were the cause then the effect would be expected
to be more evident in the lowest S/N subset of our sample, i.e. at the
faintest magnitudes. This is not the case, on the contrary the
anticorrelation is best observed among the brightest subset of
galaxies (Fig.~12), whose mean error on $\rm H\beta$ is only 0.16 \AA.
This mild anticorrelation corresponds to clearly visible differences
in the metallicity distribution of bright galaxies of different ages
(Fig.~13).  Comparing this with the Z distributions of the dwarfs
(Fig.~10), it is the small group of {\it metal-poor} young dwarfs
which is ``anomalous'', and deviates from the general trend.

\hbox{~}
\centerline{\psfig{file=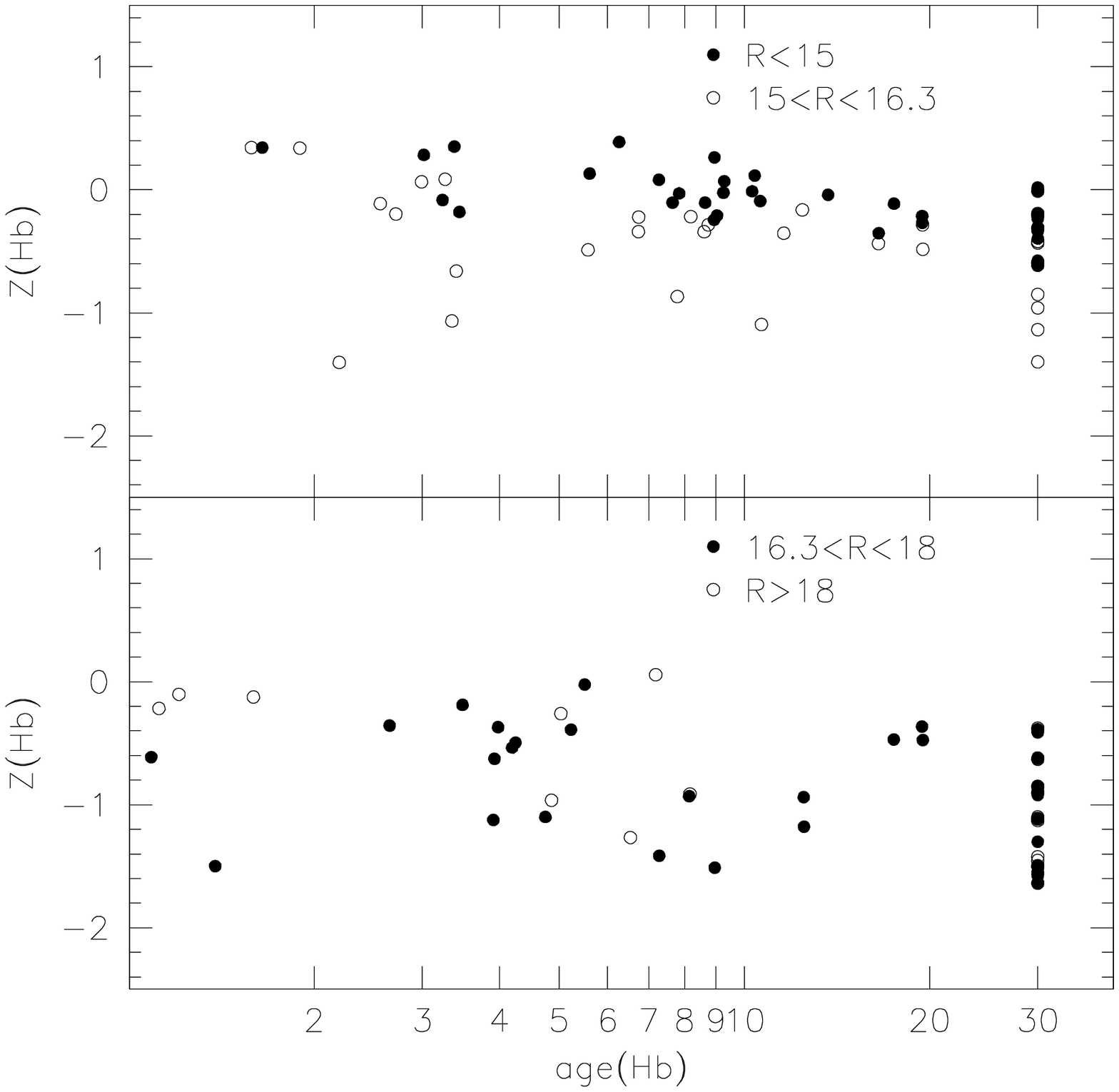,angle=0,width=5.0in}}
\noindent{\scriptsize
\addtolength{\baselineskip}{-3pt} 
\hspace*{0.1cm} Fig.~12.\ Ages versus metallicities of Coma1 galaxies
as derived from the $\rm H\beta$/$\rm Mg_2$ diagram. The best-fit relations
in the top panels are Z=0.445 $-$ 0.461 $\times$ log(age) ($R<15$) and
Z=0.051 $-$ 0.4 $\times$ log(age) ($15<R<16.3$).
\addtolength{\baselineskip}{3pt}
}

\hbox{~}
\centerline{\psfig{file=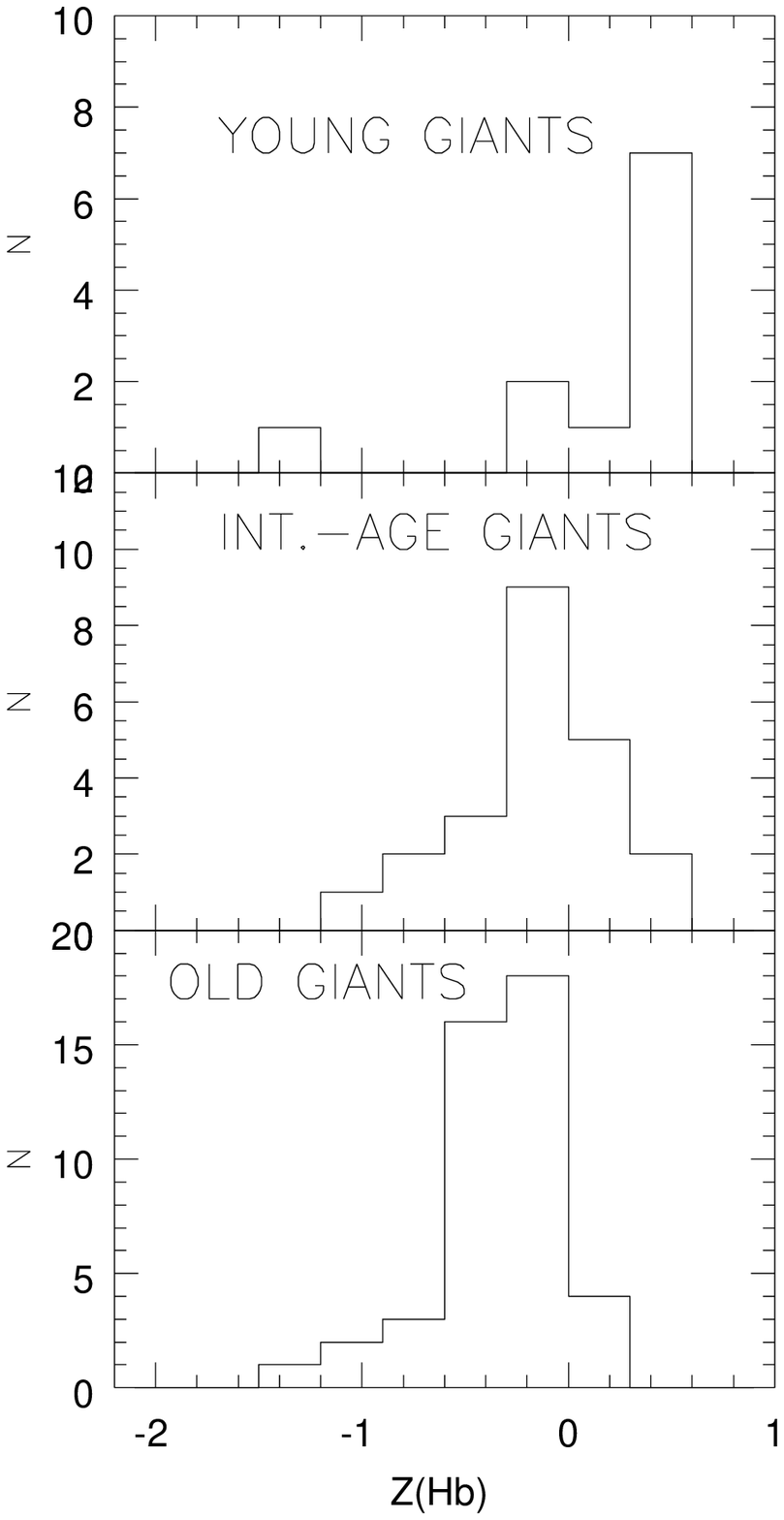,angle=0,width=5.0in}}
\noindent{\scriptsize
\addtolength{\baselineskip}{-3pt} 
\hspace*{0.1cm} Fig.~13.\ Metallicity distributions of 
young, intermediate-age and old giants ($R<16.3$)
in Coma1.
\addtolength{\baselineskip}{3pt}
}

\subsection{The slope, the scatter and the outliers of the index-magnitude 
relations}

Once we have determined the ages and metallicities of each galaxy as
described above, we can investigate the origin of the index-magnitude
relations (Fig.~3). In particular, we wish to understand what
determines the slope and the scatter of these relations and what are
the properties of the outliers that lie above the general
trend. Hereafter the discussion will be limited to ages and
metallicities derived from the $\rm H\beta$-$\rm Mg_2$ diagram, which
have smaller errorbars than $\rm {H\gamma}_F$-$<$Fe$>$.\footnote{Conclusions 
similar to those presented in this section are
reached using the $\rm{H\gamma}_F$ index instead of $\rm H\beta$ and
$<$Fe$>$ instead of $\rm Mg_2$.}

The location in the index-R diagram of the various age and metallicity
classes defined in \S4.3 is shown in Figs.~14 and 15.  We first note
that, if the measured errors are not significantly underestimating
the true errors, the fact that the observed scatter is larger at fainter
magnitudes should be only partly due to the increasing uncertainties in the
index measurements: the {\it intrinsic} scatter is shown by the
vertical segments for four magnitude bins and it is larger for fainter
galaxies.

In Fig.~14a, it is evident how the old, intermediate-age and young
galaxies segregate in the $\rm H\beta$-R plot. Considering for the
moment only old galaxies, their upper envelope 
traces a correlation between the strength of
the $\rm H\beta$ index and magnitude. The existence of this
relation is a metallicity effect not related to age.  In fact, the
value of the $\rm H\beta$ index {\it of the old galaxies} lying 
on the index-magnitude relation is
anticorrelated with Z and has no correlation with age (Fig.~16). 
The presence of intermediate-age
and young galaxies in this plot produces a thickening in the $\rm H\beta$-R
relation at bright magnitudes and a slight shift in the zero point
towards stronger $\rm H\beta$, while young galaxies are responsible
for the outliers lying above the relation, mostly at faint magnitudes.
However, it can seen in Fig.~15a that a relation between $\rm H\beta$
and R persists even when only galaxies in a certain metallicity range
are considered. For example, the existence of the relation
$\rm H\beta$-R for metal-rich galaxies is an age effect, with
fainter galaxies being on average younger.
No obvious age segregation is instead visible in the $\rm
Mg_2$-R plot (Fig.~14b).

Turning to the location of galaxies with different metallicities, they
largely segregate in the $\rm Mg_2$-R diagram (Fig.~15b), going from
higher indices (the metal-rich galaxies) to progressively lower
indices (intermediate-Z and metal-poor galaxies), with some overlap in
the index range of the three classes. Considering one metallicity
class at a time, for example the metal-rich galaxies, we observe
that they still display a trend with R: such a relation is an age effect
(fainter galaxies being on average younger than bright galaxies within
the same metallicity class), as proved by the fact that the $\rm Mg_2$
index {\it of the metal-rich galaxies} -- or, equivalently, another
metallicity class -- is correlated with age and not with Z
(Fig.~16). The $\rm Mg_2$-R relation traced by the metal-rich
galaxies has a slope similar to that of the {\it global} relation
produced by all three metallicity classes together, which is shown
as a solid line in Fig.~15b.  If the trend of decreasing mean age with
magnitude did not exist, the $\rm Mg_2$-R relation of the metal-rich
galaxies would be flat. However, a global $\rm Mg_2$-R relation would
still exist, due to the different index and magnitude
distributions of metal-rich, intermediate and metal-poor galaxies.
Strictly speaking, the relation would exist -- with 
a larger scatter -- even if only one of the trends with magnitude, either 
that of the mean age or that of the mean Z, were in place.  The fact
that both trends are present makes the relation {\it tighter},
especially at faint magnitudes.

To conclude, the characteristics of the index-magnitude relations
(slope, zero-point, scatter and outliers) are the result of both age
and metallicity trends with magnitude which are intertwined in an
intricate manner.  Similarly to the color-magnitude relation of
early-type galaxies (Bower, Lucey \& Ellis 1992; Kodama \& Arimoto
1997; Kauffmann \& Charlot 1998; Kodama et al. 1998; Terlevich et
al. 1999), a dominant factor appears to be metallicity, which is able
to largely account for the slope of the indices-R relations, hence would be
sufficient to produce index-R relations similar to the ones observed
even if the trend of decreasing mean age with magnitude did not
exist. Nevertheless, the spread in ages cannot be neglected if one
wants to understand in detail the characteristics of these relations
(slopes, zero-point, scatter) and the outliers.
Finally, it is natural to wonder how the variety of star 
formation histories observed in the present sample and presented in \S4.3
can give rise to relatively tight index-magnitude relations. 
In a sense this is not surprising, since
metallicity turns out to be the dominant effect in determining the slope
of the relations, and given the existence of the Z-luminosity relation.
Hence, the correct way to address this question is to aim at
understanding the reason for the Z dependence on L even in galaxies with
different star formation histories, as discussed in \S4.3. 

\hbox{~}
\vspace{-0.5in}
\centerline{\psfig{file=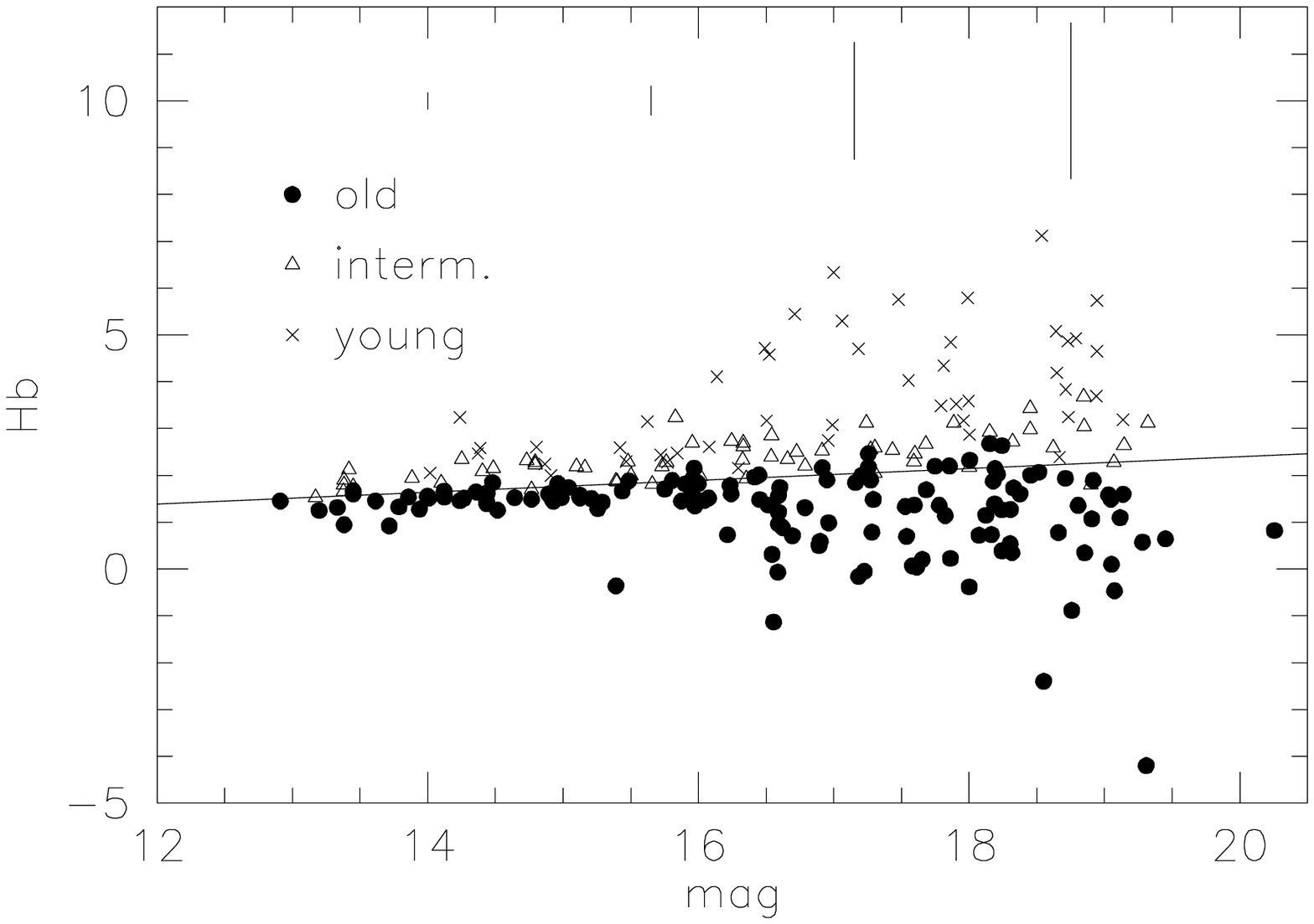,angle=0,width=4.0in}\hspace{-0.3in}
\psfig{file=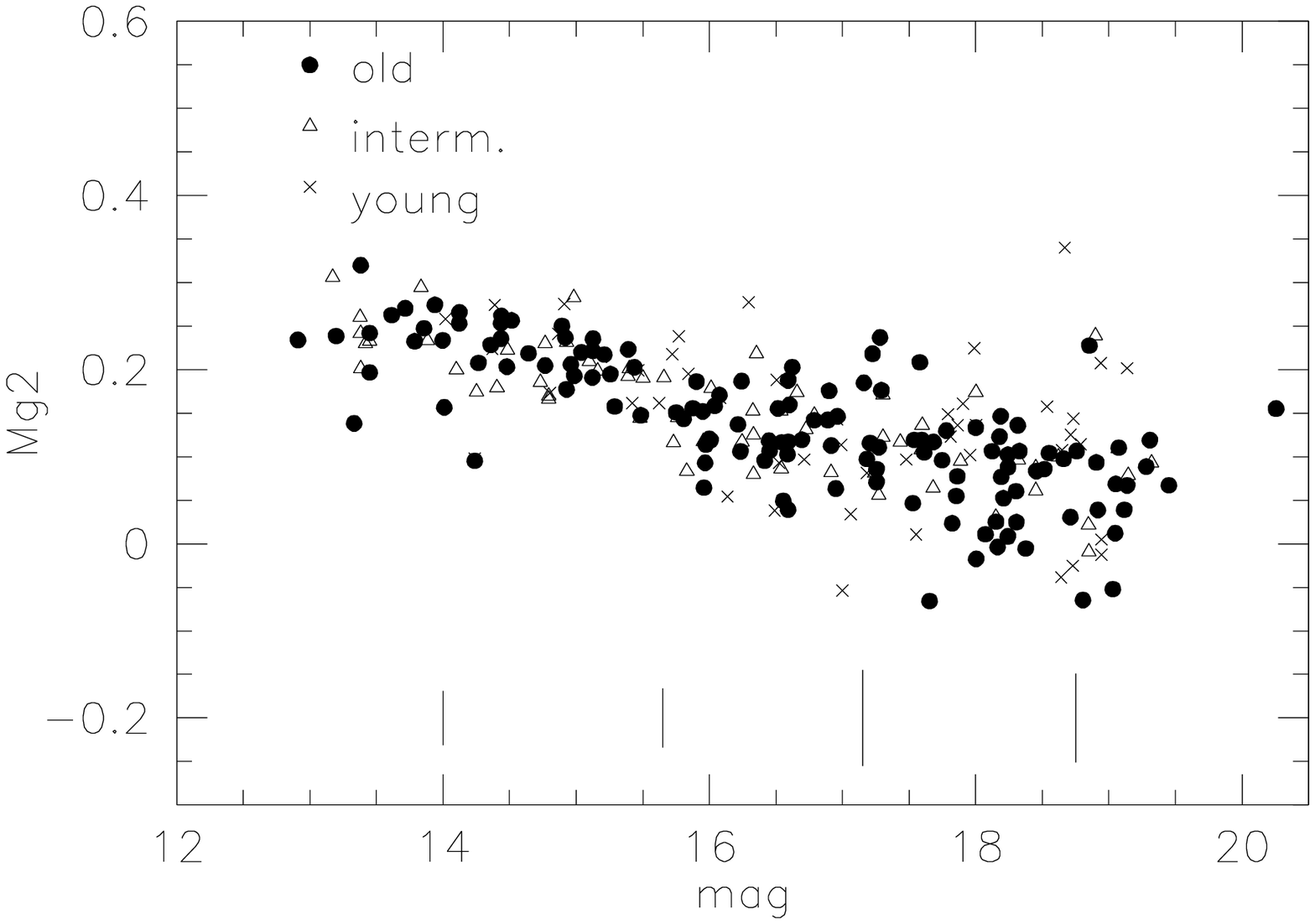,angle=0,width=4.0in}}
\noindent{\scriptsize
\addtolength{\baselineskip}{-3pt} 
\hspace*{0.1cm} Fig.~14.\ Index-magnitude relations as in Fig.~3 for 
old (age$>9$ Gyr), intermediate-age (between 3 and 9 Gyr) and young
(age$<3$ Gyr) galaxies. Both Coma1 and Coma3 galaxies are included.
The straight line in the left panel is the best fit given in Table~2
and shown also in Fig.~3.
The values of the intrinsic scatter in four magnitude bins are shown
as vertical segments. The intrinsic scatter has been computed 
as ${\sigma}^2_{intr.}=(y-B\times R-A)^2-{\sigma}^2_y$, where $y=A-B\times R$
is the observed best-fit relation between the index $y$ and the magnitude $R$
(Fig.~3) and ${\sigma}^2_y$ is the measured error on the index.
\addtolength{\baselineskip}{3pt}
}

\hbox{~}
\vspace{-1.2in}
\centerline{\psfig{file=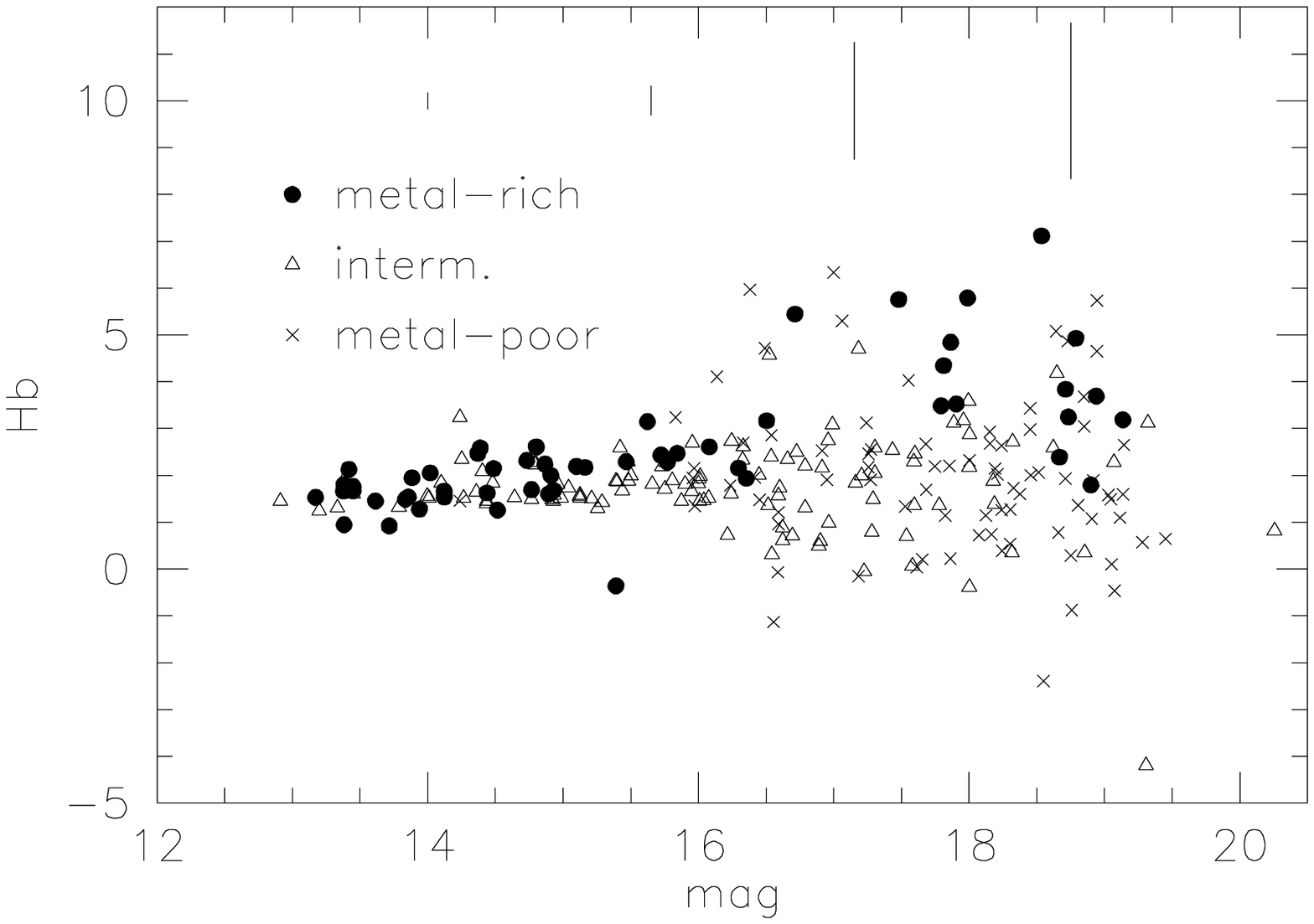,angle=0,width=4.0in}\hspace{-0.3in}
\psfig{file=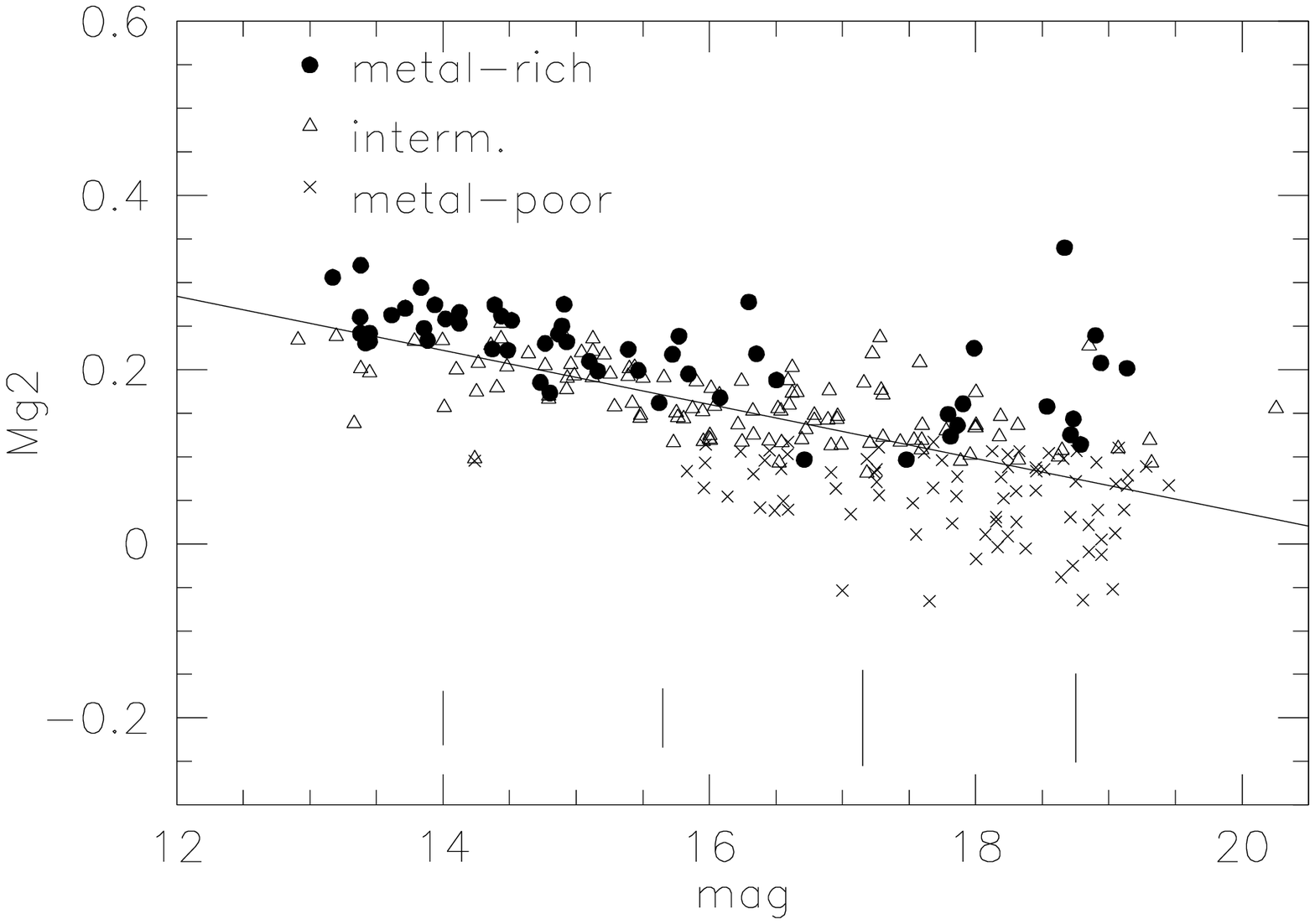,angle=0,width=4.0in}}
\noindent{\scriptsize
\addtolength{\baselineskip}{-3pt} 
\hspace*{0.1cm} Fig.~15.\ Index-magnitude relations as in Fig.~3 for 
metal-rich (Z $>-0.15$), intermediate-Z ($-$1 $<$ Z $<$ $-$0.15) and 
metal-poor (Z $<-1$) galaxies. Both Coma1 and Coma3 galaxies are included.
The straight line in the right panel is the best fit given in Table~2
and shown also in Fig.~3.
The values of the intrinsic scatter in four magnitude bins are shown
as vertical segments.
\addtolength{\baselineskip}{3pt}
}

\hbox{~}
\centerline{\psfig{file=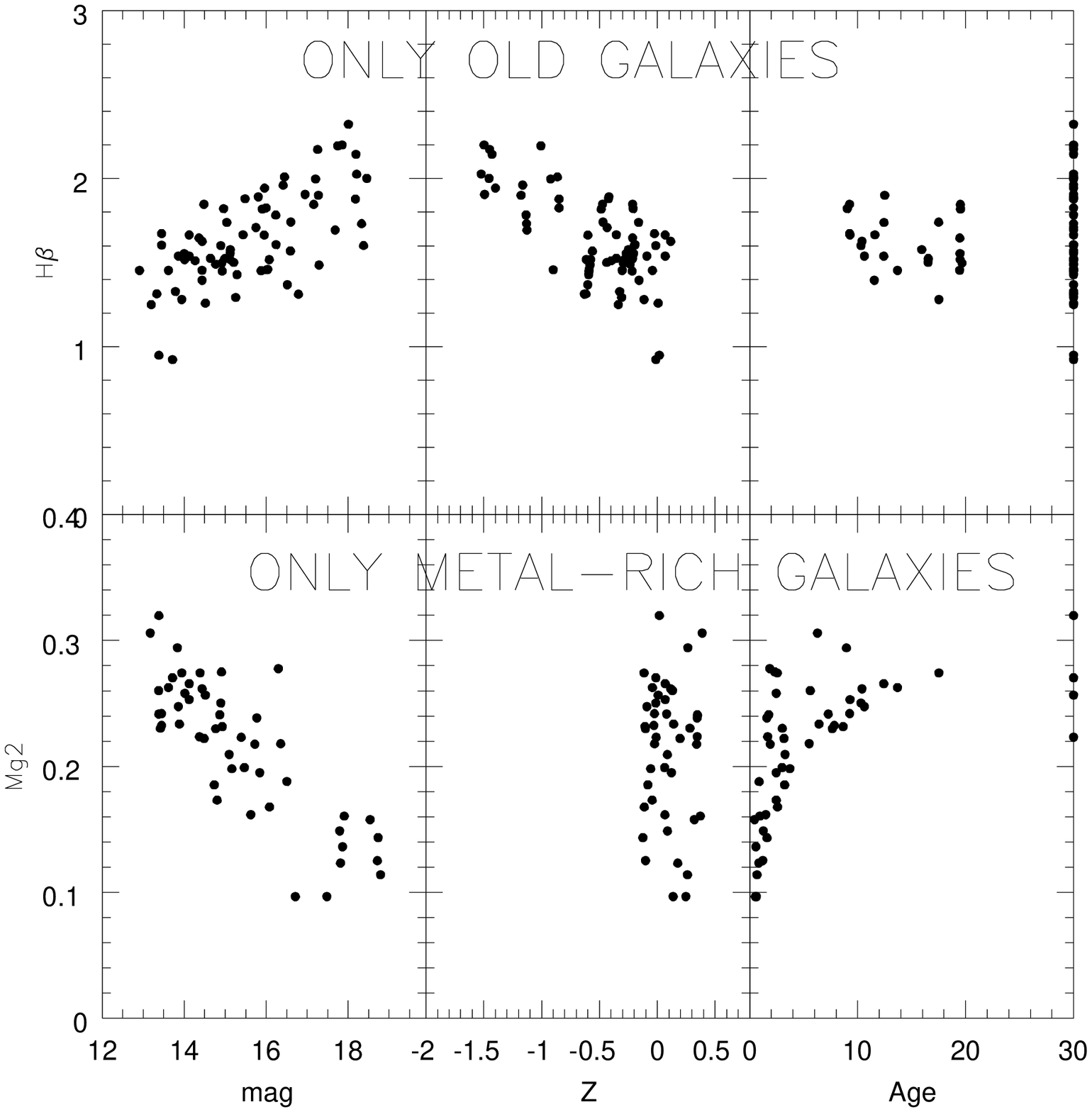,angle=0,width=4.0in}}
\noindent{\scriptsize
\addtolength{\baselineskip}{-3pt} 
\hspace*{0.1cm} Fig.~16.\ Top panels. $\rm H\beta$ index versus magnitude,
versus metallicity and versus age for old galaxies lying on the index-magnitude relation only.
Bottom panels. $\rm Mg_2$ index versus magnitude,
versus metallicity and versus age for metal-rich galaxies only.
\addtolength{\baselineskip}{3pt}
}

\section{Conclusions}

Analysis of the Lick indices of a large sample of Coma galaxies
with $-$20.5 $<$ M$_B$ $<$ $-$14 has shown systematic trends of the
index values with both galaxy luminosity and environment (Fig.~2).  In
this paper we have investigated the properties of the
non-emission-line galaxies and focussed on the relations between
index strength and luminosity, with the aim of understanding
the origin of the mean relations, the cause of their scatter and the
properties of the outlying points. The derived distributions of
luminosity-weighted ages and metallicities provide some important
indications about the evolutionary history of cluster galaxies as a
function of galactic luminosity.

The main results can be summarized as follows:

a) In our sample all the Lick indices sensitive to metallicity $\rm
Mg_2$, $<$Fe$>$, $C_24668$ etc.) display a correlation with 
luminosity, and all the age-sensitive indicators (Balmer indices)
anticorrelate with luminosity.  For the first time, we find that all these
relations hold also for faint, non-emission line galaxies, down to
$M_B\sim -14$.

b) By comparing the observed indices with model grids based on the
Padova isochrones, we derive luminosity-weighted ages and
metallicities.  In doing this, we find a large number of faint
galaxies whose Balmer indices are too low to be explained by the
current models. Interestingly, this phenomenon has also been
previously observed in the spectra of several globular clusters, even
at high signal-to-noise.  Moreover, on the basis of the random errors
and given the spectral characteristics of this sample, observational
errors and/or eventual emission filling of the Balmer lines do not
seem sufficient to explain the observed population of faint galaxies
with low Balmer lines.  A possible reason for the discrepancy between
the indices measured and the model grids is an overestimate of the
turn-off temperature in the models, hence we tentatively interpret the
galaxies with very low Balmer indices as old stellar systems (age $>9$
Gyr).

c) The mean metallicity increases with galaxy luminosity.  The spread
in metallicity at a given magnitude is large, especially for galaxies
fainter than $M_B \sim -18$. This spread is larger than the nominal
scatter due to the random errors in the index measurements, although it
should be kept in mind that additional sources of errors (such as
cosmic rays and sky residuals) have not been included.

d) A broad range of ages, from younger than 3 Gyr to older than 9 Gyr,
is found in galaxies of any magnitude but a large fraction of the
currently non-star-forming galaxies of {\it any} luminosity (dwarfs
and giants) are devoid of signs of star formation occurred at $z<2$.
This fraction is found to be $\sim 50$\% at all magnitudes, but it is
more uncertain and could be as low as 30\% for the dwarfs.  Although
no clear correlation is observed between age and magnitude, there are
systematic trends between the ``luminosity-weighted'' epoch of the
most recent star formation episode and luminosity.  At $z<0.35$, the
star formation activity has involved a higher proportion of faint
galaxies (at least 20\% of the present-day population) than bright
ones ($\sim$5-10\%).  Using the derived ages at face value, the
fraction of present-day luminous galaxies with significant star
formation at intermediate redshifts ($0.35<z<2$) is higher than the
fraction of present-day faint galaxies that were active at that epoch
(30-50\% against 15-25\%), but this last result could be 
affected by the large errorbars on the indices of the dwarfs.

e) The metallicity distributions of faint galaxies with young
luminosity-weighted ages ($<3$ Gyr) and, possibly, those with
intermediate-age (3-9 Gyr) are bimodal, with a group being ``too
metal-rich'' and another group being ``too metal-poor'' with respect
to the mean metallicity at their magnitude.  This bimodality is also
supported by the index strengths obtained coadding the spectra of the
two groups of galaxies separately. This is suggestive of two separate
mechanisms for the formation of the faint galaxy population in
clusters, but we stress that this result needs to be confirmed by
higher signal-to-noise spectra of a large number of galaxies at the
faintest magnitudes.
 
f) An anticorrelation between age and metallicity is found among
galaxies in any given luminosity bin. This could be
spuriously produced by the fact that the errors in the derivation of
the age and the metallicity are correlated, but the fact that the
age-metallicity anticorrelation is best observed in the subset of
galaxies with the highest signal-to-noise spectra seems to suggest
that it is real.

g) An interpretation of the observed index-magnitude relations is
presented. Both age and metallicity variations as a function of
luminosity play a role in determining the characteristics of the
index-magnitude relations (slope, scatter, zero-point and outliers).
The relations are
the consequence of both age and metallicity trends with luminosity:
each such trend on its own would be sufficient to produce
relations qualitatively similar (and not too different in slope) to
those observed.

\section*{Acknowledgements} 
We are sincerely grateful to Herve' Aussel who wrote for us the IDL 
program that convolves the spectra with a wavelength-dependent Gaussian.
We thank the referee, Dr. Guy Worthey, for his useful comments
which gave us an opportunity to improve the paper.
We also thank Matthew Colless for providing us with his spectroscopic catalog
of Coma galaxies and for assistance throughout this project.
This research has greatly benefited from the availability of the
PhD theses of Nicolas Cardiel Lopez (University of Madrid), 
Harald Kuntschner (University of Durham), Marcella Longhetti (University 
of Milano) and Scott C. Trager (University of California, Santa Cruz).
We acknowledge helpful discussions with and/or comments about this
manuscript from Cesare Chiosi, Fabio Governato,
George Hau, Enrico Held, Harald Kuntschner, Ian Smail,
Scott Trager, Alexandre Vazdekis and Guy Worthey, and we thank
the latter two colleagues for maintaining very useful WEB pages of their
models. This research has made use of the NASA/IPAC
Extragalactic Database (NED) which is operated by the Jet Propulsion
Laboratory, Caltech, under contract with the National Aeronautics and
Space Administration. 

\section*{Appendix A: Measurements of the Lick indices}

a) {\it Spectral resolution.}  The Lick/IDS spectral resolution varies
with wavelength as described in WO97 (see also Fig.~A1).  The
resolution of our spectra has been estimated from the arc exposures
and has been found to vary only slightly with wavelength; at each
given wavelength, the resolution stays approximately constant within
the ``internal fibres'' (fibre numbers 30-110) and worsens
progressively in the ``external fibres'' ($<30$ and $>110$) towards
the edges of the chip.  Fig.~A1 presents the mean FWHM\footnote{We
have verified that the variation of the resolution within the group of
external fibres is small enough not to affect the results.} of the
internal and external fibres found as a polynomial interpolation of
the FWHM at 4158, 4695, 5606 and 6416 \AA. We have broadened our
internal and external spectra with a Gaussian of wavelength dependent
width, in such a way to match the Lick/IDS resolution {\it at the
observed $\lambda$ corresponding to the rest-frame $\lambda$ of the
Lick/IDS system} (${\lambda}_{\rm obser.}=(1+z) \, {\lambda}_{\rm
Lick}$).  For cluster members this was done assuming a common redshift
$z=0.023$; the fore/background galaxies were divided into redshift
bins and the Gaussian convolution was done for each redshift bin at
the appropriate wavelength. In this way, we have degraded our spectra
to a spectral resolution which is the same of the Lick spectra at each
spectral index.  After Gaussian convolving and deredshifting the
spectra, the indices were measured using the IRAF-{\it sbands} tool
(see Appendix B).

\hbox{~}
\centerline{\psfig{file=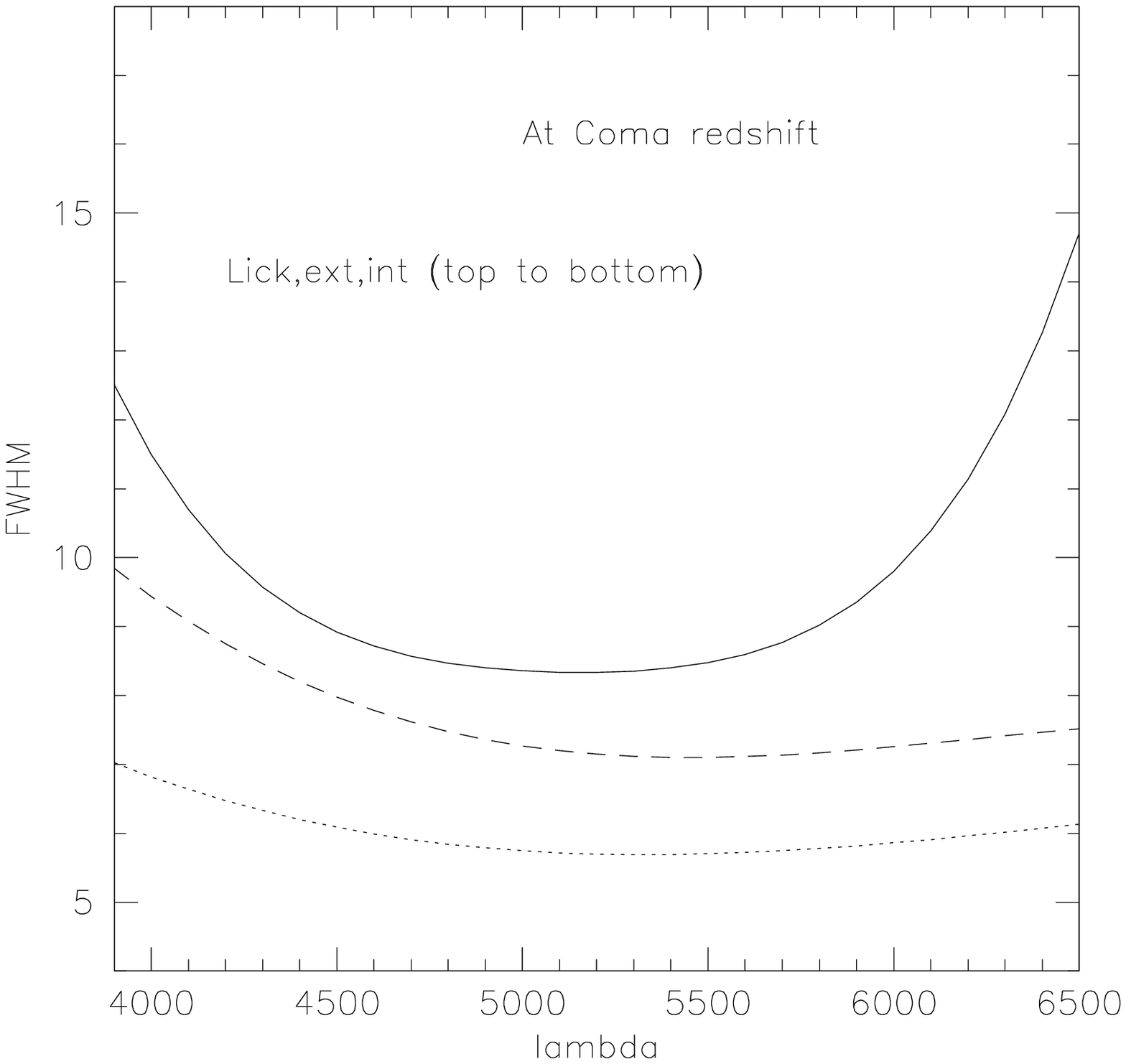,angle=0,width=4.0in}}
\noindent{\scriptsize
\addtolength{\baselineskip}{-3pt} 
\hspace*{0.1cm} Fig.~A1.\ Spectral resolution of the Lick spectra (solid line),
of our external-fibre spectra (long dashed line) and of our internal-fibre
spectra (dotted line) at the wavelengths observed at the Coma redshift.
\addtolength{\baselineskip}{3pt}
}

b) {\it Velocity dispersion correction.}  The line-of-sight velocities
of the stars in a galaxy broaden the spectral features of its
integrated spectrum; in contrast, the model integrated spectrum simply
adds up the contribution of all stars, hence in order to compare with
the Lick models it is necessary to correct the values of the observed
indices to a zero-velocity dispersion.

Velocity dispersions for 23 giant galaxies in our sample are available
from Lucey et al. (1997). Following Terlevich et al. (1999), we
constructed the relation between these velocity dispersions and our R
band magnitudes and used the fit to this relation to estimate the
velocity dispersion of the rest of the galaxies.  This relation is
expected to be valid for luminous galaxies, while the behaviour of the
$\sigma$-Luminosity relation for dwarf galaxies is not well
determined. Hence, a $\sigma = 50 \, \rm km \, s^{-1}$ was assigned to
those faint galaxies for which the relation described above yields
$\sigma < 50 \, \rm km \, s^{-1}$.  Given our spectral resolution, the
velocity dispersion correction for dwarf galaxies is negligible.

At the telescope we obtained 12 usable spectra of 5 different stars of
the Lick/IDS stellar library (HR2002, HR2600, HR3262, HR3427, HR3905).
After having convolved the flux-calibrated stellar spectra with a
wavelength-dependent Gaussian as described at point a), we further
broadened the spectra with a Gaussian filter to mimic a ``galaxy
velocity dispersion'' in the range $\sigma = 30-310 \, \rm km s^{-1}$
in steps of 20 $\rm km \, s^{-1}$. We have then measured the indices
for each star and each $\sigma$ and found the mean correction factors
$C^I(\sigma)=\rm Index(0)/Index(\sigma)$ (atomic indices) and
$C^I(\sigma)=\rm Index(0)-Index(\sigma)$ (molecular indices) that are
presented in Fig.~A2.  The indices correction factors of each galaxy
were computed from the galaxy velocity dispersion linearly
interpolating between points in Fig.~A2.

\hbox{~}
\centerline{\psfig{file=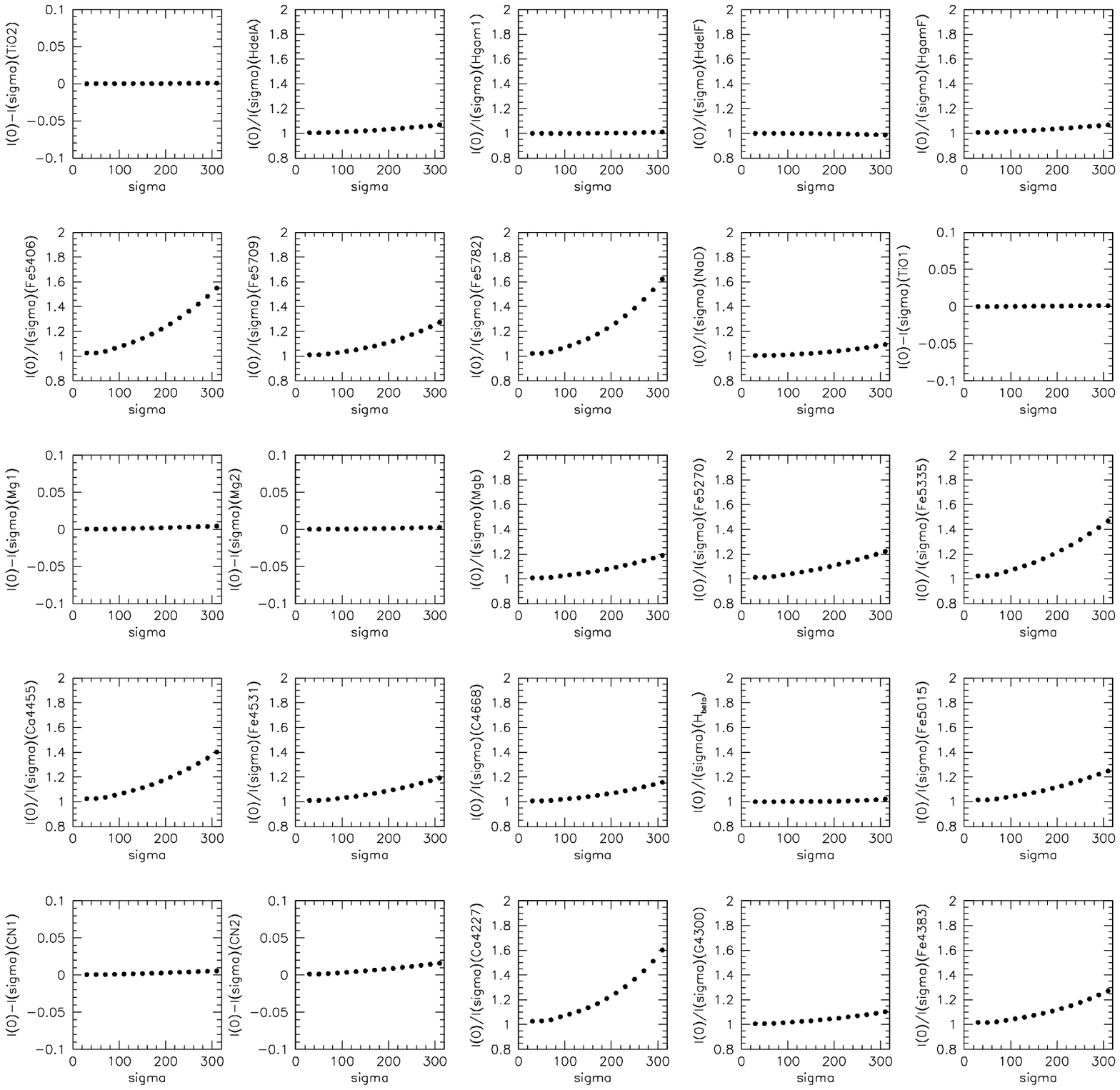,angle=0,width=4.0in}}
\noindent{\scriptsize
\addtolength{\baselineskip}{-3pt} 
\hspace*{0.1cm} Fig.~A2.\ Correction factors as a function of the
galaxy velocity dispersion. Corrections are highest for narrow atomic
indices, such as the Fe indices.
\addtolength{\baselineskip}{3pt}
}

c) {\it Offsets.}  The Lick spectra were not flux calibrated, but
normalized to a quartz-iodide tungsten lamp, therefore we need to
correct for the small systematic offsets produced by the different
continuum shape.  We have measured the Lick indices of the
flux-calibrated stellar spectra (after performing the
wavelength-dependent Gaussian convolution described at point a) and
compared them with the values given in the Lick papers (WO97). The
mean offsets between our and Lick measurements are listed in Table~A1;
our index values have been corrected for these offsets only if the
offset was bigger than its $1\sigma$ error.

\begin{table}
{\scriptsize
\begin{center}
\centerline{\sc Table A1}
\vspace{0.1cm}
\centerline{\sc Systematic offsets}
\vspace{0.3cm}
\begin{tabular}{lc}
\hline\hline
\noalign{\smallskip}
Index  & Offset (This work-Lick) \cr
\hline
\noalign{\smallskip}
 $\rm CN_1$    &  0.018$\pm$0.011 \cr
 $\rm CN_2$    &  0.016$\pm$0.016 \cr
 Ca4227        & -0.028$\pm$0.109 \cr
 G4300         &  0.123$\pm$0.291 \cr
 Fe4383        & -0.316$\pm$0.172 \cr
 Ca4455        & -0.260$\pm$0.136 \cr
 Fe4531        &  0.088$\pm$0.334 \cr
 $\rm C_2$4668 &  0.014$\pm$0.349 \cr
 $\rm H_\beta$ &  0.038$\pm$0.090 \cr
 Fe5015        & -0.362$\pm$0.313 \cr
 $\rm Mg_1$    &  0.015$\pm$0.028 \cr
 $\rm Mg_2$    &  0.011$\pm$0.012 \cr
 Mg$b$         & -0.017$\pm$0.109 \cr
 Fe5270        & -0.288$\pm$0.149 \cr
 Fe5335        & -0.119$\pm$0.245 \cr
 Fe5406        & -0.055$\pm$0.155 \cr
 Fe5709        & -0.071$\pm$0.135 \cr
 Fe5782        & -0.073$\pm$0.095 \cr
 Na D          & -0.176$\pm$0.105 \cr
 $\rm TiO_1$   &  0.003$\pm$0.003 \cr
 $\rm TiO_2$   &  0.015$\pm$0.011 \cr
$\rm {H\delta}_A$ & 0.330$\pm$0.371 \cr
$\rm {H\gamma}_A$ &-0.137$\pm$0.332 \cr
$\rm {H\delta}_F$ & 0.107$\pm$0.157 \cr
$\rm {H\gamma}_F$ &-0.038$\pm$0.126 \cr
\noalign{\smallskip}
\noalign{\hrule}
\noalign{\smallskip}
\noalign{\smallskip}
\end{tabular}
\end{center}
}
\vspace*{-0.8cm}
\end{table}

\section*{Appendix B: Tests of the Lick/IDS calibration}

The comparison of our measured indices with model predictions 
relies on a careful calibration onto the Lick/IDS system. 
We have assessed the accuracy of this calibration in a number of ways:

a) {\it Index measuring routine.}  we have verified that the method we
adopted for measuring the indices (the IRAF-sbands tool) reproduces
the index values measured by the Lick team. For allowing colleagues
to test their measuring routine, Guy Worthey provides on his WEB page
the spectra and Lick indices of 8 test-stars; the agreement between
our measurements and those of Lick is excellent and gives a mean
relative error of $0.5\pm0.4$\%.

b) {\it Comparison with the Lick index measurements.}  The Lick/IDS
galaxy library includes 5 Coma galaxies in common with our sample
(Trager et al. 1998, Worthey in prep.). The galaxy area covered by our
fibre diameter (2.7 arcsec) matches quite well the Lick/IDS standard
aperture ($1.4\times 4$ arcsec), hence we check the consistency with
the Lick/IDS system comparing directly our and their measurements in
Fig.~B1.  The figure shows a good general agreement within the errors.

\hbox{~}
\centerline{\psfig{file=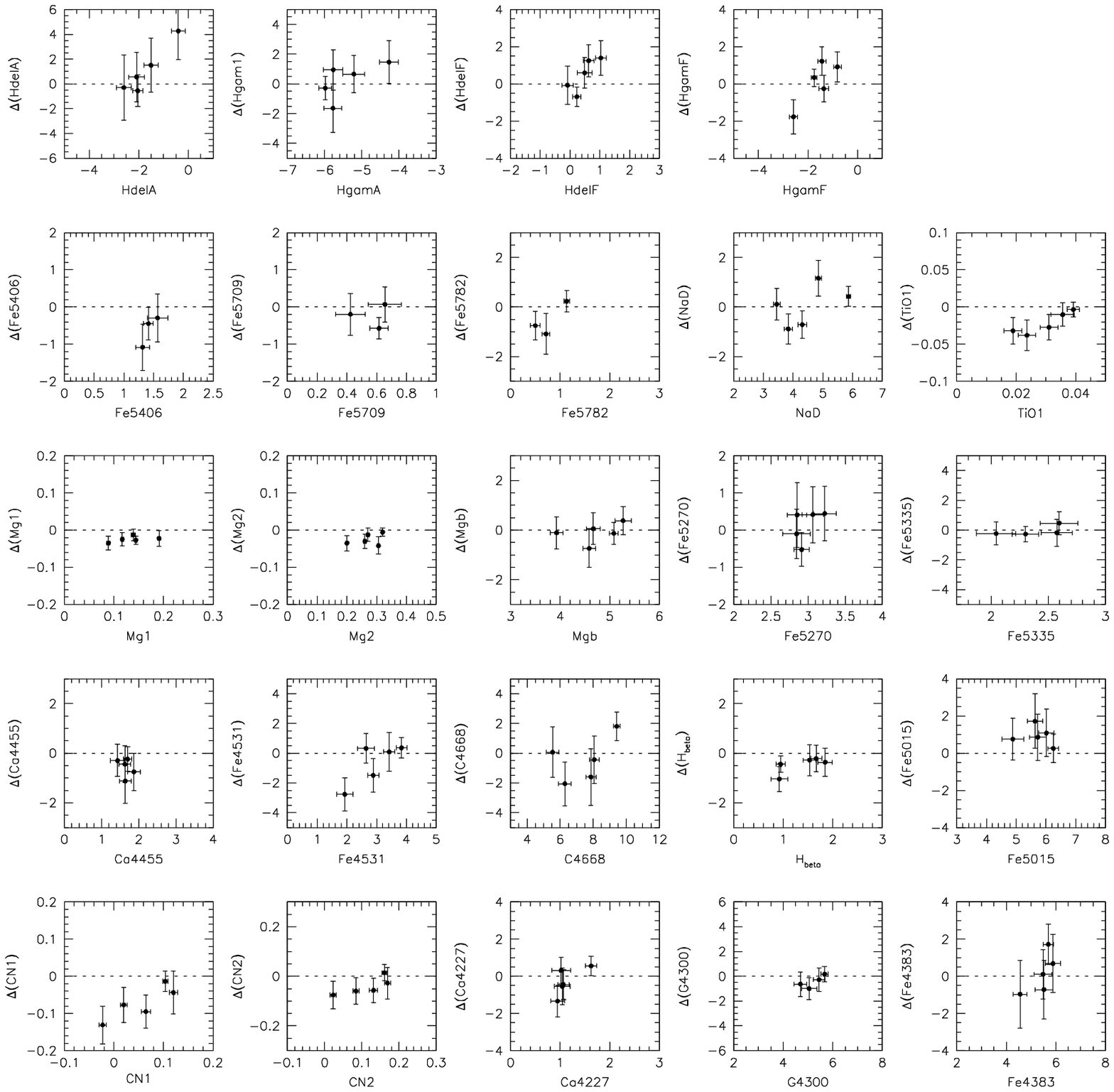,angle=0,width=5.0in}}
\noindent{\scriptsize
\addtolength{\baselineskip}{-3pt} 
\hspace*{0.1cm} Fig.~B1.\ Differences $\Delta$=(This Work - Lick) 
between our and Lick (Trager et al. 1998,
Worthey in prep.) index measurements for the 5 giant galaxies in common
versus our index measurements.
Some of the plots contain less than 5 galaxies when a measurement was missing
in the Lick list. The $\rm TiO_2$ plot is missing because this index 
was not measured by Lick for any of these galaxies.
\addtolength{\baselineskip}{3pt}
}

c) {\it Comparison with Jorgensen's index measurements.}

Recently, Jorgensen (1999, J99) published $\rm H\beta$, $\rm Mg_1$,
$\rm Mg_2$, Mgb and $\rm <Fe>$ indices of a large sample of E and S0
galaxies in Coma.  The index measurements of the 16 galaxies in common
are compared in Fig.~B2 (circles).  Jorgensen's spectroscopic
parameters are centrally measured values corrected to a circular
aperture of diameter 1.19 $\rm h^{-1}$ kpc (see J99 for details),
therefore we expect at least part of the differences to be due to
aperture effects. The agreement in Fig.~B2 is satisfactory, but we
note a possible anticorrelation between the index differences and the
index strength for $\rm Mg_2$ and - to a minor extent - $\rm Mg_1$ and
$\rm H\beta$.  To investigate the origin of this behaviour, in Fig.~B2
we also plot as crosses the differences between Jorgensen's and
Trager's measurements of the 5 common galaxies: as expected from the
agreement between Trager's and our measurements found in Fig.~B1, the
differences J-T follow the same trend as the differences between J99
and this work.  The source of this systematic trend could be related
to the dissimilar apertures, but a detailed investigation of the
discrepancy is beyond the scope of this work and we do not discuss it
further.

\hbox{~}
\centerline{\psfig{file=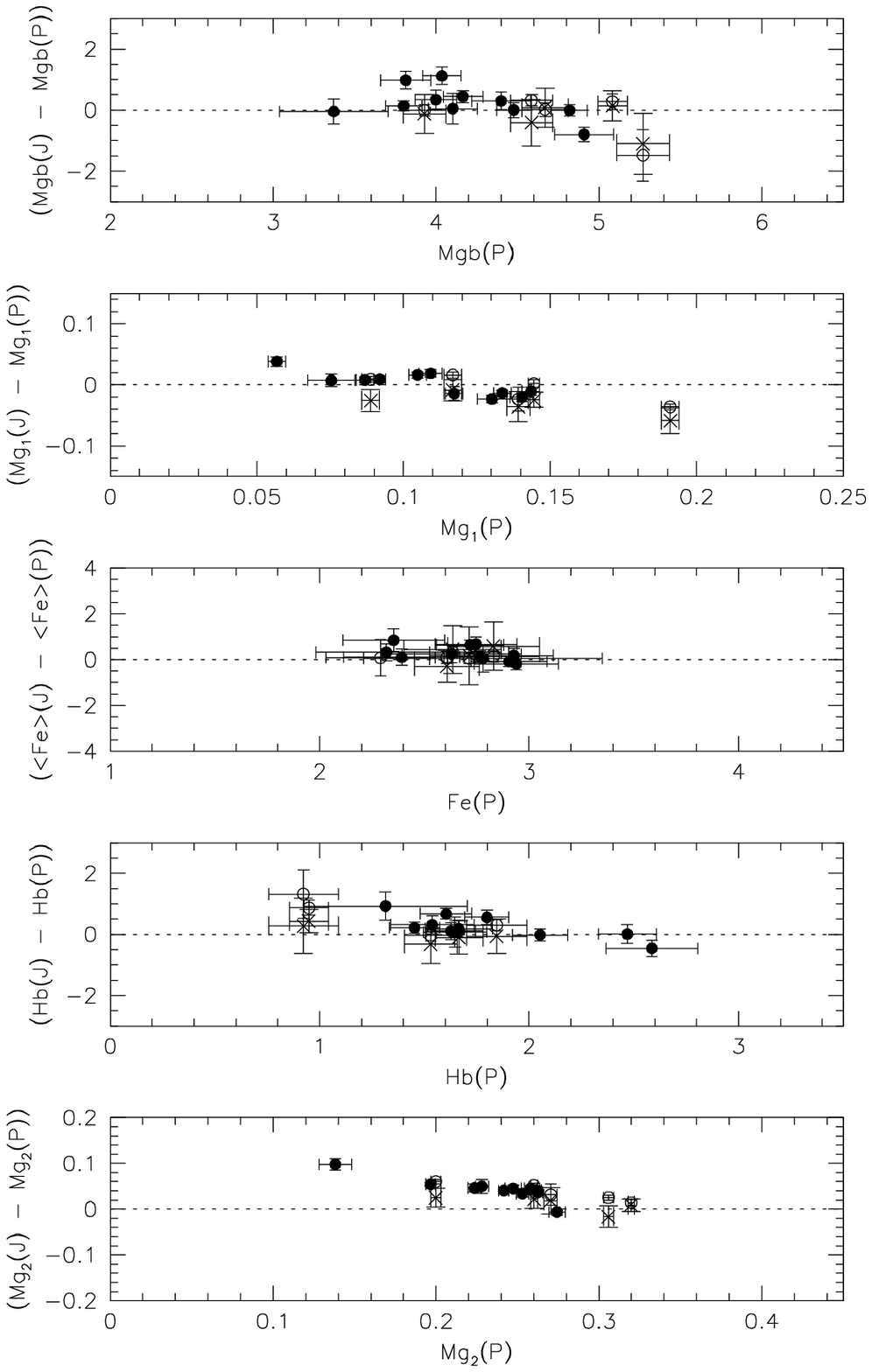,angle=0,width=4.0in}}
\noindent{\scriptsize
\addtolength{\baselineskip}{-3pt} 
\hspace*{0.1cm} Fig.~B2.\ Differences between Jorgensen's (1999)
and our (P) measurements for the 16 giant galaxies in common.
Empty circles identify those 5 galaxies in common among Trager,
Jorgensen and this work. The crosses mark the differences between
Jorgensen and Trager measurements of the common galaxies (see text
for details).
\addtolength{\baselineskip}{3pt}
}

d) {\it Index-index plots for indices dominated by similar species.}

The consistency of the calibration onto the Lick system and with the model
predictions can be investigated by means of index-index plots which
present indices that are sensitive to the same chemical species (Kuntschner 
2000).
In these plots the model predictions should closely follow the relation
traced by the data.
There are three main families of indices (Trager et al. 1998):
1) $\alpha$-element indices (CN, Mg, NaD, TiO2);
2) Fe-like indices (Ca, G band, TiO1, all Fe indices);
\footnote{$C_{2}4668$ has been shown by Trager et al. (1998) to be
somewhat intermediate between the $\alpha$ and Fe groups.}  3)
Balmer indices.

Here we only consider those indices that are used in this paper for
deriving ages and metallicities: their index-index plots are presented
in Fig.~B3 (lower panel of each plot). For comparison, the upper panel
of each plot displays the data of the Lick galaxies (dots) and
globular clusters (crosses). Note that the errorbars for our giant sample
are considerably smaller than the mean errors
in the Lick/IDS sample of galaxies (see \S4.2),
while our dwarf galaxy errors are larger.
The position of our giant galaxies in
the diagrams overlap with the positions of the Lick (giant) galaxies,
but our sample does not reach the highest metallicity-indices values
observed in the Lick sample, because of the different galaxy
luminosity ranges.  Dwarf galaxies span a wide range in all indices
and they also occupy the region of the diagrams where the Lick
globular clusters are found.  The models generally trace well the
relations where the data lie, with the well known exception of the Mg
indices due to the enhanced Mg/Fe ratio of the luminous galaxies.
These model deviations are observed in both the lower and the upper
panels of Fig.~B3, as well as in any other sample (e.g. Kuntschner
2000).  

\vfill\eject

\hbox{~}
\centerline{\hspace{1.3in}\psfig{file=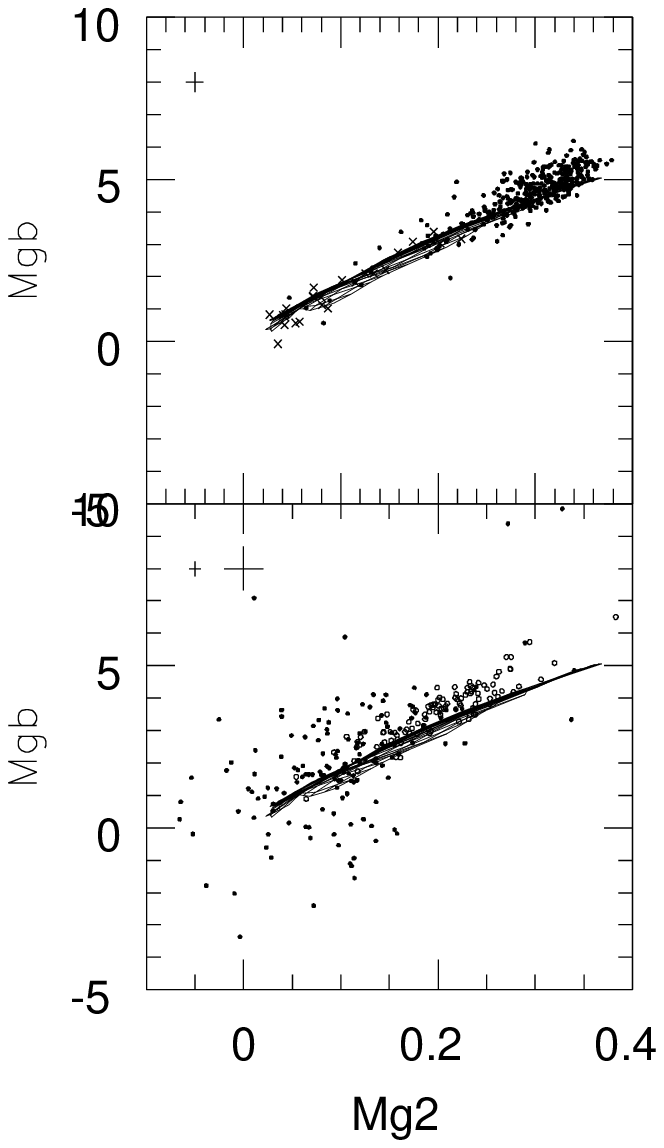,angle=0,width=5in}
\hspace{-3.4in}\psfig{file=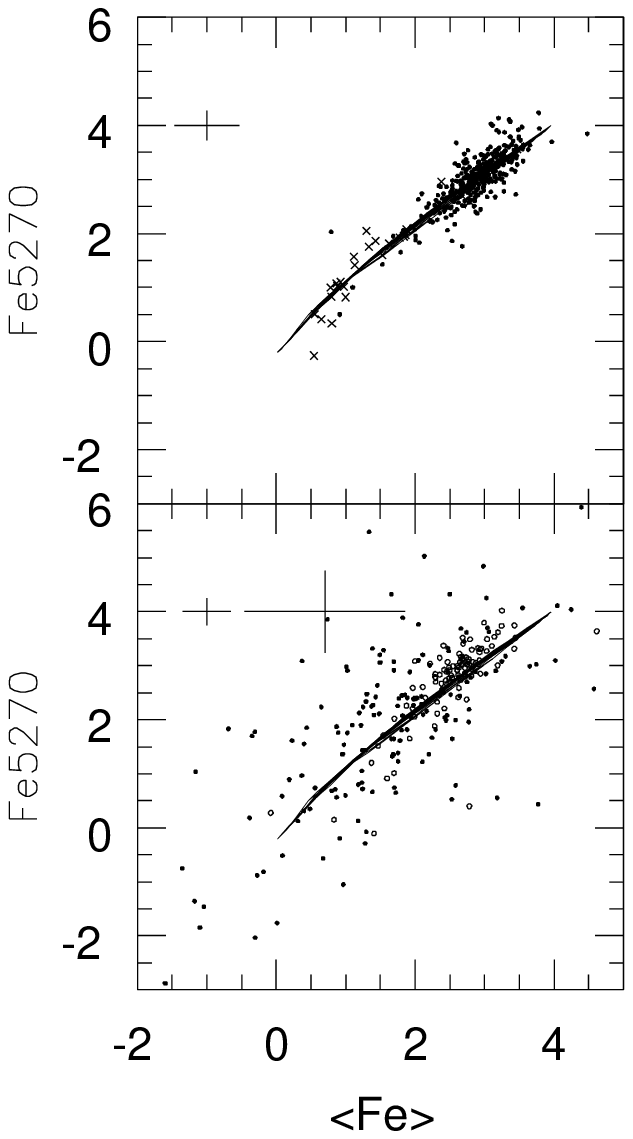,angle=0,width=5in}
\hspace{-3.4in}\psfig{file=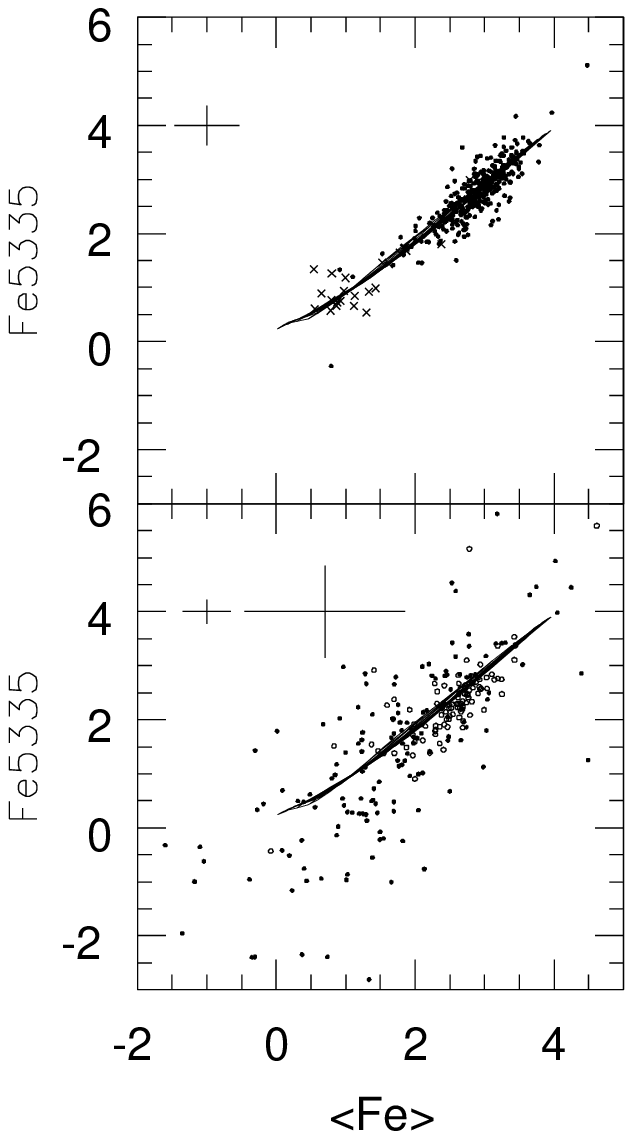,angle=0,width=5in}
\hspace{-3.3in}\psfig{file=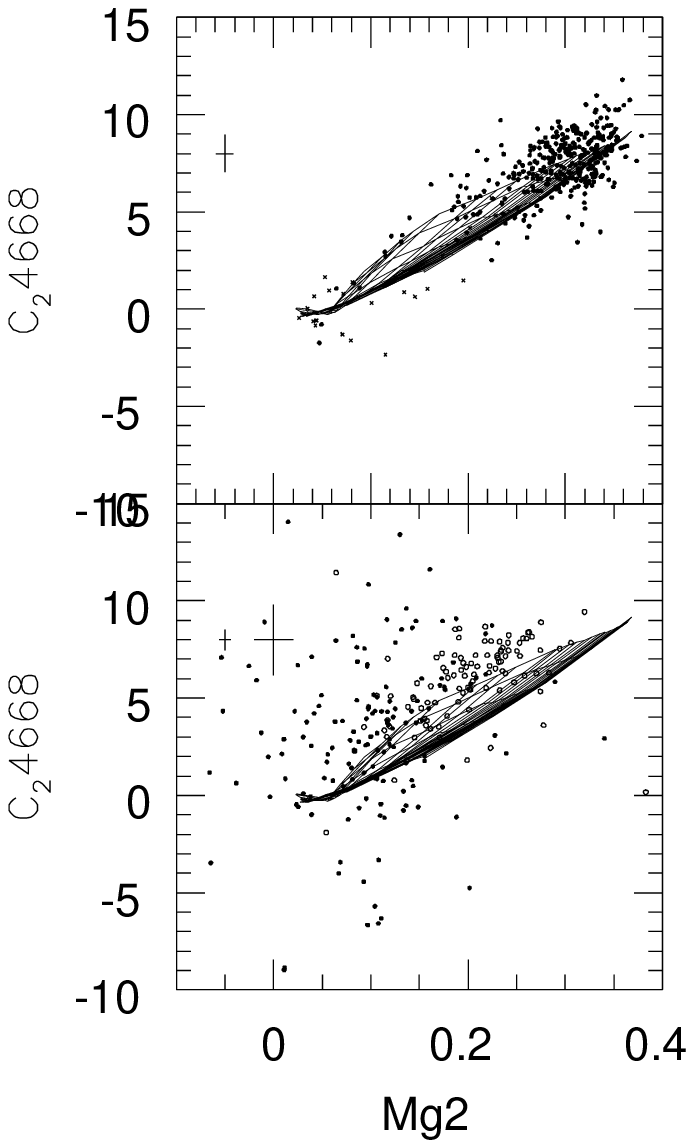,angle=0,width=5in}}
\vspace{-2in}
\centerline{\hspace{1.3in}\psfig{file=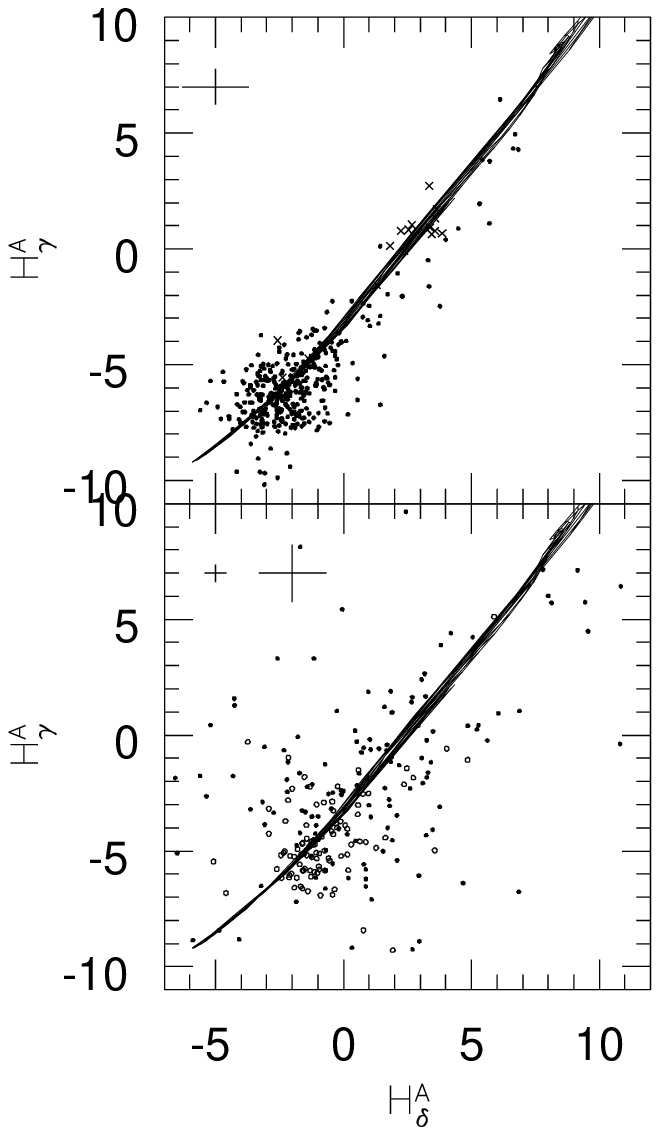,angle=0,width=5in}
\hspace{-3.4in}\psfig{file=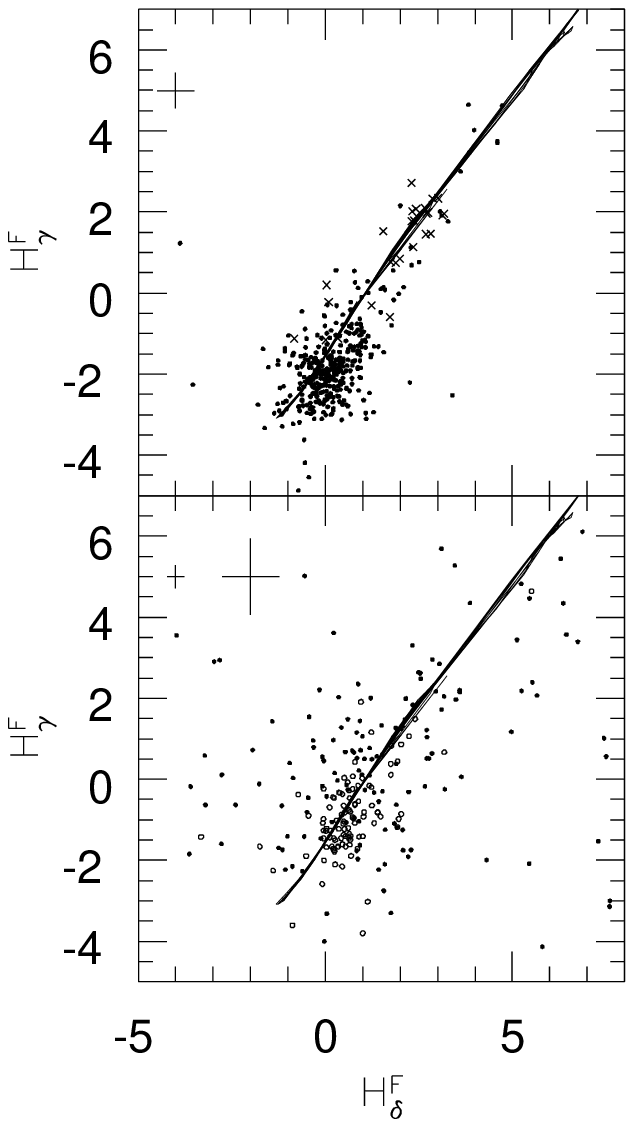,angle=0,width=5in}
\hspace{-3.4in}\psfig{file=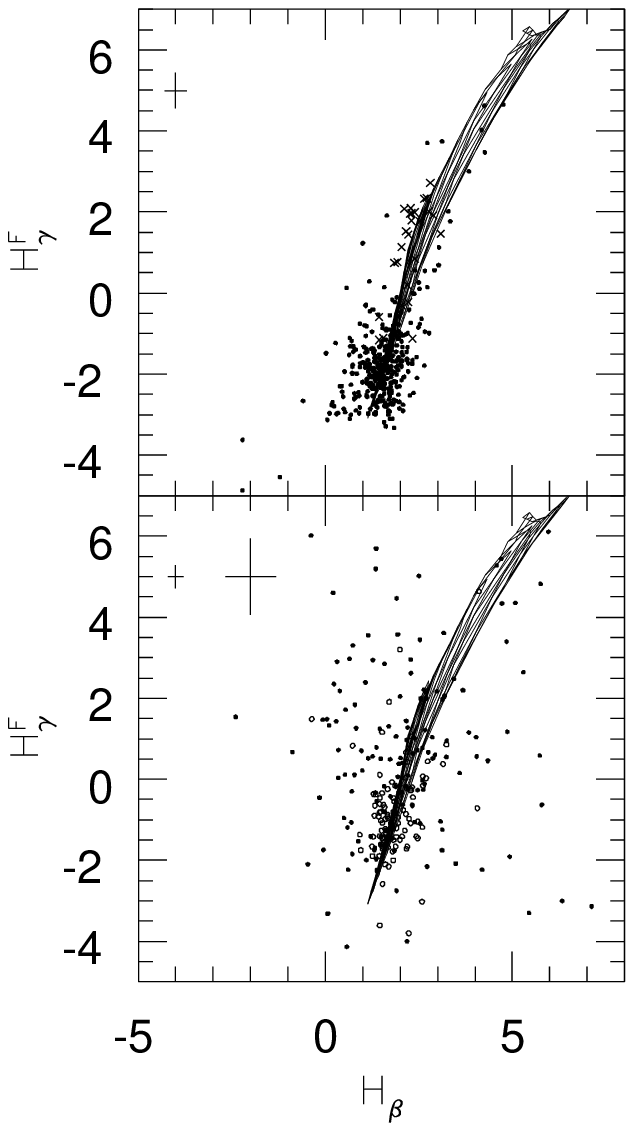,angle=0,width=5in}
\hspace{-3.4in}\psfig{file=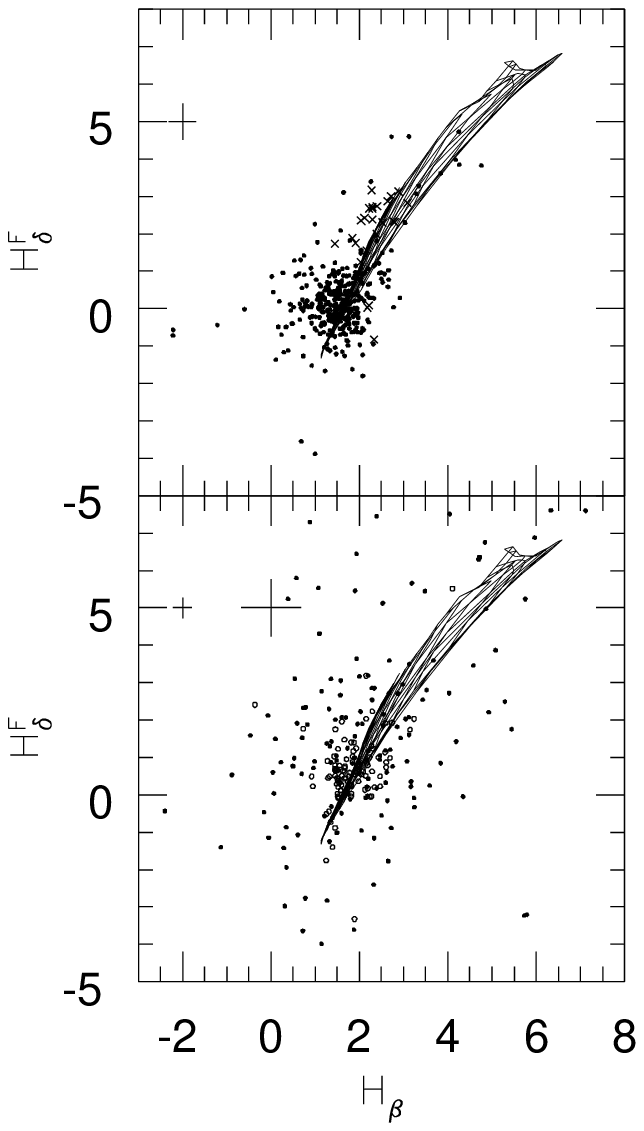,angle=0,width=5in}}
\noindent{\scriptsize
\addtolength{\baselineskip}{-3pt} 
\hspace*{0.1cm} Fig.~B3.\ Index versus index plots of indices dominated by
the same chemical species.
Our whole galaxy sample is shown in the lower panel of each plot:
filled dots are dwarfs and empty dots are giants.
Galaxies with emission lines have been excluded from this figure. 
The upper panel of each plot presents the Lick/IDS sample
(Trager et al. 1998, Worthey in prep.)
of galaxies (filled dots) and globular clusters (crosses).
The median errorbars of our giant (left) and dwarf (right) galaxies
are shown in the upper left corner of the lower panels. The median errorbar
of the Lick galaxies is in the upper left corner of the upper panels.
Overplotted are Worthey's models (see text).
\addtolength{\baselineskip}{3pt}
}

\vfill\eject

\smallskip

\end{document}